\documentstyle[epsf,12pt]{book}
\begin{document}
\setlength{\baselineskip}{18pt}
%%%%%%%%------macros-----------------------------------------------------
\jot=8pt
\gdef\journal#1,#2,#3,#4.{{\it #1~}{\bf #2} (#4) #3}
\def\prd{\journal Phys. Rev. D,}
\def\prl{\journal Phys. Rev. Lett.,}
\def\npb{\journal Nucl. Phys. B,}
\def\plb{\journal Phys. Lett. B,}
\def\apj{\journal Ap. J.,}
\def\apjl{\journal Ap. J. Lett.,}
\def\MNRAS{\journal MNRAS,}
%%%%%%%%%%%%%%%%%%%%%%%%%%%%%%%%%%%%%%%%%%%%%%%%%%%%%%%%%%%%%%%%%%%%%%%%%%%%%%
\def\be{\begin{equation}}\def\bea{\begin{eqnarray}}\def\beaa{\begin{eqnarray*}}
  \def\ee{\end{equation}}  \def\eea{\end{eqnarray}}  \def\eeaa{\end{eqnarray*}}
\def\double{\baselineskip 24pt \lineskip 10pt}
\def\re#1{{[\ref{#1}]}}
\def\fun#1#2{\lower3.6pt\vbox{\baselineskip0pt\lineskip.9pt
        \ialign{$\mathsurround=0pt#1\hfill##\hfil$\crcr#2\crcr\sim\crcr}}}
%%%%%%%%%%%%%%%%%%%%%%%%%%%%%%%%%%%%%%%%%%%%%%%%%%%%%%%%%%%%%%%%%%%%%%%%%%%%
%\begin{document}
%\psdraft    % uncomment to get an empty box with the size of the figure
%%%%%%%%%%%%%%%%%%%%%%%%%%%%%%%%%%%%%%%%%%%%%%%%%%%%%%%%%%%%%%%%%%%%%%%%%%%%
\def\half{{\textstyle{ 1\over 2}}}
\def\frac#1#2{{\textstyle{#1\over #2}}}
\def\gsim{\mathrel{\raise.3ex\hbox{$>$\kern-.75em\lower1ex\hbox{$\sim$}}}}
\def\lsim{\mathrel{\raise.3ex\hbox{$<$\kern-.75em\lower1ex\hbox{$\sim$}}}}
%%%%%%%%%%%%%%%%%%%%%%%%%%%%%%%%%%%%%%%%%%%%%%%%%%%%%%%%%%%%%%%%%%%%%%%%%%%%
\def\la{\bigl\langle} \def\ra{\bigr\rangle}
\def\cd{\!\cdot\!}
\def\a{\hat a}      \def\b{\hat b}      \def\c{\hat c}
\def\ab{\a\cd\b}    \def\ac{\a\cd\c}    \def\bc{\b\cd\c}
\def\cg{\cos\gamma} \def\ca{\cos\alpha} \def\cb{\cos\beta}
\def\go{ \hat\gamma_1}  
\def\gt{\hat\gamma_2}  \def\gth{\hat\gamma_3}
\def\gf{\hat\gamma_4}
\def\got{ \hat\gamma_1\cd\hat\gamma_2} \def\ggo{ \hat\gamma\cd\hat\gamma_1}
\def\goth{\hat\gamma_1\cd\hat\gamma_3} \def\ggt{ \hat\gamma\cd\hat\gamma_2}
\def\gtth{\hat\gamma_2\cd\hat\gamma_3} \def\ggth{\hat\gamma\cd\hat\gamma_3}
\def\n{\hat n}       \def\no{\hat n_1}   \def\nt{\hat n_2}  \def\nth{\hat n_3}
\def\nont{\no\cd\nt} \def\nonth{\no\cd\nth} \def\ntnth{\nt\cd\nth}
\def\nogo{\no\cd\hat\gamma_1} \def\nogt{\no\cd\hat\gamma_2} 
\def\nogth{\no\cd\hat\gamma_3} 
\def\ntgo{\nt\cd\hat\gamma_1} \def\ntgt{\nt\cd\hat\gamma_2}
\def\ntgth{\nt\cd\hat\gamma_3} 
\def\nthgo{\nth\cd\hat\gamma_1} \def\nthgt{\nth\cd\hat\gamma_2} 
\def\nthgth{\nth\cd\hat\gamma_3} 
\def\D{ {\Delta T \over T} }   \def\dO{d\Omega}
\def\d{ {\delta T \over T} }  
\def\etal{{\sl et al.}}
\def\eg{{\sl e.g.}}
\def\ie{{\sl i.e.}}

\def\bbox#1{\hbox{\boldmath{$#1$}}} 
\def\edth{\;\raise1.5pt\hbox{$'$}\hskip-6pt\partial\;} 
\def\baredth{\;\overline{\raise1.5pt\hbox{$'$}\hskip-6pt \partial}\;}  

%%%%%%%%%%%%%%%%%%%%%%%%%%%%%%%%%%%%%%%%%%%%%%%%%%%%%%%%%%%%%%%%%%%%%%%%%%
%\include{bol_macros}
%%%%%%%%------end macros----------------------------------------------------

%%%%%%%%------title----------------------------------------------------
\pagestyle{empty}

\begin{center}
{\huge Topological Defects in Cosmology
\footnote{Lecture Notes for the First Bolivian School on Cosmology \\
          La Paz, 24--28 September, 2001 \\
          {\tt http://www.umsanet.edu.bo/fisica/cosmo2k1.html}}}

% In La Paz, you follow the rules (`soroche'):
% Caminar despacito,
% comer poquito y 
% dormir solito.

\vspace{1.0cm}

{\large Alejandro Gangui}

\vspace{1.0cm}

Instituto de Astronom\'{\i}a y F\'{\i}sica del Espacio, \\
Ciudad Universitaria, 1428 Buenos Aires, Argentina

and

Dept. de F\'{\i}sica, Universidad de Buenos Aires, \\
Ciudad Universitaria -- Pab. I, 
1428 Buenos Aires, Argentina  

\vspace{1.0cm}

{\large Abstract}

\end{center}                  

%\addcontentsline{toc}{chapter}{Abstract}

\noindent
Topological defects are ubiquitous in condensed--matter physics but
only hypothetical in the early universe. In spite of this, even an
indirect evidence for one of these cosmic objects would revolutionize
our vision of the cosmos.
We give here an introduction to the subject of cosmic topological
defects and their possible observable signatures.
Beginning with a review of the basics of general defect formation and
evolution, we then focus on mainly two topics in some detail:
conducting strings and vorton formation, and some specific imprints
in the cosmic microwave background radiation from simulated cosmic
strings.

%\end{center}                  

%\include{bol_title}
%%%%%%%%------end title----------------------------------------------------

%\clearpage
\pagestyle{plain}
\pagenumbering{roman}  

\tableofcontents

\clearpage
\pagenumbering{arabic}
\setcounter{page}{1}
\setcounter{chapter}{0}
\pagestyle{myheadings}

\chapter{Topological Defects in Cosmology}
\label{chap-ag}
%\markboth{Chapter 1. ~Topological Defects in Cosmology}{Chapter 1. ~xxx}

%%%%-----------------------------------------------------------
\section{Introduction}
\label{sec-intro}      

On a cold day, ice forms quickly on the surface of a pond. But it does
not grow as a smooth, featureless covering.  Instead, the water begins
to freeze in many places independently, and the growing plates of ice
join up in random fashion, leaving zig--zag boundaries between
them. These irregular margins are an example of what physicists call
``topological defects'' -- {\sl defects} because they are places where
the crystal structure of the ice is disrupted, and {\sl topological}
because an accurate description of them involves ideas of symmetry
embodied in topology, the branch of mathematics that focuses on the
study of continuous surfaces.

Current theories of particle physics likewise predict that a variety
of topological defects would almost certainly have formed during the
early evolution of the universe. Just as water turns to ice (a phase
transition) when the temperature drops, so the interactions between
elementary particles run through distinct phases as the typical energy
of those particles falls with the expansion of the universe. When
conditions favor the appearance of a new phase, it generally crops up
in many places at the same time, and when separate regions of the new
phase run into each other, topological defects are the result. The
detection of such structures in the modern universe would provide
precious information on events in the earliest instants after the Big
Bang. Their absence, on the other hand, would force a major revision
of current physical theories.

The aim of this set of Lectures is to introduce the reader to the
subject of topological defects in cosmology. We begin with a review of
the basics of defect formation and evolution, to get a grasp of the
overall picture.  We will see that defects are generically predicted
to exist in most interesting models of high energy physics trying to
describe the early universe.  The basic elements of the standard
cosmology, with its successes and shortcomings, are covered elsewhere
in this volume, so we will not devote much space to them here.  We
will then focus on some specific topics. We will first treat
conducting cosmic strings and one of their most important predictions
for cosmology, namely, the existence of equilibrium configurations of
string loops, dubbed vortons. We will then pass on to study some key
signatures that a network of defects would produce on the cosmic
microwave background (CMB) radiation, \eg, the CMB bispectrum of the
temperature anisotropies from a simulated model of cosmic strings.
Miscellaneous topics also reviewed below are, for example, the way in
which these cosmic entities lead to large--scale structure formation
and some astrophysical footprints left by the various defects, and we
will discuss the possibility of isolating their effects by
astrophysical observations.  Also, we will briefly consider
gravitational radiation from strings, as well as the relation of
cosmic defects to the well--known defects formed in condensed--matter
systems like liquid crystals, etc.

Many areas of modern research directly related to cosmic defects are
not covered in these notes. The subject has grown so wide, so fast,
that the best thing we can do is to refer the reader to some of the
excellent recent literature already available. So, have a look, for
example, to the report by Achucarro \& Vachaspati [2000] for a
treatment of semilocal and electroweak strings\footnote{Animations of 
semilocal and electroweak string formation and evolution
can be found at {\tt http://www.nersc.gov/\~{}borrill/}}, 
and to [Vachaspati, 2001] 
for a review of certain topological defects, like monopoles,
domain walls and, again, electroweak strings, virtually not covered
here.  For conducting defects, cosmic strings in particular, see for
example [Gangui \& Peter, 1998] for a brief overview of many different
astrophysical and cosmological phenomena, and the comprehensive
colorful lecture notes by Carter [1997] on the dynamics of branes
with applications to conducting cosmic strings and vortons.  If your
are in cosmological structure formation, Durrer [2000] presents a good
review of modern developments on global topological defects and their
relation to CMB anisotropies, while Magueijo \& Brandenberger [2000]
give a set of imaginative lectures with an update on local string
models of large-scale structure formation and also baryogenesis with
cosmic defects.

If you ever wondered whether you could have a pocket device, the size
of a cellular phone say, to produce ``topological defects'' on demand
[Chuang, 1994], then the proceedings of the school held {\sl aux}
Houches on topological defects and non-equilibrium dynamics, edited by
Bunkov \& Godfrin [2000], are for you; the ensemble of lectures in
this volume give an exhaustive illustration of the interdisciplinary
of topological defects and their relevance in various fields of
physics, like low--temperature condensed--matter, liquid crystals,
astrophysics and high--energy physics.

Finally, all of the above (and more) can be found in the concise
review by Hindmarsh \& Kibble [1995], particularly concerned with the
physics and cosmology of cosmic strings, and in the monograph by
Vilenkin \& Shellard [2000] on cosmic strings and other topological
defects.

%%-------------------------------------------------------------
\subsection{How defects form}
\label{subsec-howdefects}   

A central concept of particle physics theories attempting to unify all
the fundamental interactions is the concept of symmetry breaking.  As
the universe expanded and cooled, first the gravitational interaction,
and subsequently all other known forces would have begun adopting
their own identities.  In the context of the standard hot Big Bang
theory the spontaneous breaking of fundamental symmetries is realized
as a phase transition in the early universe.  Such phase transitions
have several exciting cosmological consequences and thus provide an
important link between particle physics and cosmology.

There are several symmetries which are expected to break down in the
course of time.  In each of these transitions the space--time gets
`oriented' by the presence of a hypothetical force field called the
`Higgs field', named for Peter Higgs, pervading all the space. This
field orientation signals the transition from a state of higher
symmetry to a final state where the system under consideration obeys a
smaller group of symmetry rules.  As an every--day analogy we may
consider the transition from liquid water to ice; the formation of the
crystal structure ice (where water molecules are arranged in a well
defined lattice), breaks the symmetry possessed when the system was in
the higher temperature liquid phase, when every direction in the
system was equivalent.  In the same way, it is precisely the
orientation in the Higgs field which breaks the highly symmetric state
between particles and forces.
                                                                          
Having built a model of elementary particles and forces, particle
physicists and cosmologists are today embarked on a difficult search
for a theory that unifies all the fundamental interactions. As we
mentioned, an essential ingredient in all major candidate theories is
the concept of symmetry breaking. Experiments have determined that
there are four physical forces in nature; in addition to gravity these
are called the strong, weak and electromagnetic forces. Close to the
singularity of the hot Big Bang, when energies were at their highest,
it is believed that these forces were unified in a single,
all--encompassing interaction. As the universe expanded and cooled,
first the gravitational interaction, then the strong interaction, and
lastly the weak and the electromagnetic forces would have broken out
of the unified scheme and adopted their present distinct identities in
a series of symmetry breakings.

Theoretical physicists are still struggling to understand how gravity
can be united with the other interactions, but for the unification of
the strong, weak and electromagnetic forces plausible theories
exist. Indeed, force--carrying particles whose existence demonstrated
the fundamental unification of the weak and electromagnetic forces
into a primordial ``electroweak'' force -- the W and Z bosons -- were
discovered at CERN, the European accelerator laboratory, in 1983.  In
the context of the standard Big Bang theory, cosmological phase
transitions are produced by the spontaneous breaking of a fundamental
symmetry, such as the electroweak force, as the universe cools. For
example, the electroweak interaction broke into the separate weak and
electromagnetic forces when the observable universe was $10^{-12}$ seconds old,
had a temperature of $10^{15}$ degrees Kelvin, and was only one part
in $10^{15}$ of its present size. There are also other phase
transitions besides those associated with the emergence of the
distinct forces. The quark-hadron confinement transition, for example,
took place when the universe was about a microsecond old. Before this
transition, quarks -- the particles that would become the constituents
of the atomic nucleus -- moved as free particles; afterward, they
became forever bound up in protons, neutrons, mesons and other
composite particles.

\begin{figure}[t]
  \begin{center}
    \leavevmode
    \epsfxsize = 10cm
    \epsffile{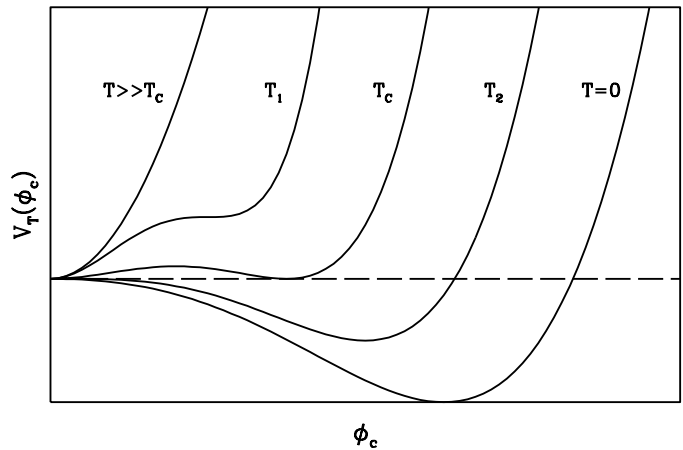}
  \end{center}
\caption{{\sl Temperature--dependent effective potential for a
first--order phase transition for the Higgs field. For very high
temperatures, well above the critical one $T_c$, the potential
possesses just one minimum for the vanishing value of the Higgs
field. Then, when the temperature decreases, a whole set of minima
develops (it may be two or more, discrete or continuous, depending of
the type of symmetry under consideration). Below $T_c$, the value
$\phi = 0$ stops being the global minimum and the system will
spontaneously choose a new (lower) one, say $\phi = \eta
\exp(i\theta)$ (for complex $\phi$) 
for some angle $\theta$ and nonvanishing $\eta$, amongst
the available ones.  This choice signals the breakdown of the symmetry
in a cosmic phase transition and the generation of random regions of
conflicting field orientations $\theta$. In a cosmological setting,
the merging of these domains gives rise to cosmic defects.}}
\label{fig-pot_phtrans}
\end{figure}

As we said, the standard mechanism for breaking a symmetry involves
the hypothetical Higgs field that pervades all space. As the universe
cools, the Higgs field can adopt different ground states, also
referred to as different vacuum states of the theory. In a symmetric
ground state, the Higgs field is zero everywhere. Symmetry breaks when
the Higgs field takes on a finite value (see Figure
\ref{fig-pot_phtrans}).

Kibble [1976] first saw the possibility of defect formation when he
realized that in a cooling universe phase transitions proceed by the
formation of uncorrelated domains that subsequently coalesce, leaving
behind relics in the form of defects.  
In the expanding universe, widely separated regions in space have not
had enough time to `communicate' amongst themselves and are therefore
not correlated, due to a lack of causal contact.  It is therefore
natural to suppose that different regions ended up having arbitrary
orientations of the Higgs field and that, when they merged together,
it was hard for domains with very different preferred directions to
adjust themselves and fit smoothly. In the interfaces of these
domains, defects form. 
Such relic `flaws' are unique examples of incredible amounts of energy
and this feature attracted the minds of many cosmologists.

%%-------------------------------------------------------------
\subsection{Phase transitions and finite temperature field theory}
\label{sec-PhaseTrans}      

Phase transitions are known to occur in the early universe.  Examples
we mentioned are the quark to hadron (confinement) transition, which
QCD predicts at an energy around 1 GeV, and the electroweak phase
transition at about 250 GeV.  Within grand unified theories (GUT), aiming to
describe the physics beyond the standard model, other phase
transitions are predicted to occur at energies of order $10^{15}$ GeV;
during these, the Higgs field tends to fall towards the minima of its
potential while the overall temperature of the universe decreases as a
consequence of the expansion.

A familiar theory to make a bit more quantitative the above
considerations is the $\lambda |\phi|^4$ theory, 
\be
\label{lambda4}
{\cal L} = {1\over 2} |\partial_\mu\phi|^2 + {1\over 2} m_0^2 |\phi|^2
- {\lambda \over 4!} |\phi|^4 ~, \ee with $m^2_0 > 0$.  
The second and third terms on the right hand side yield the usual
`Mexican hat' potential for the complex scalar field.  For energies
much larger than the critical temperature, $T_c$, the fields are in
the so--called `false' vacuum: a highly symmetric state characterized
by a vacuum expectation value $\la | \phi | \ra = 0$.  But when
energies decrease the symmetry is spontaneously broken: a new `true'
vacuum develops and the scalar field rolls down the potential and sits
onto one of the degenerate new minima. In this situation the vacuum
expectation value becomes $\la | \phi | \ra^2 = 6 m_0^2 / \lambda$.

Research done in the 1970's in finite--temperature field theory
[Weinberg, 1974; Dolan \& Jackiw, 1974; Kirzhnits \& Linde, 1974] has
led to the result that the temperature--dependent effective potential
can be written down as \be
\label{VfiniT}
V_T( | \phi | ) = -{1\over 2} m^2(T) |\phi|^2 + {\lambda \over 4!}
|\phi|^4 \ee 
with $T_c^2 = 24 m_0^2 / \lambda $, $m^2(T) = m_0^2 (1 -
T^2 / T_c^2)$, and $\la | \phi | \ra^2 = 6 m^2(T) / \lambda$.  We
easily see that when $T$ approaches $T_c$ from below the symmetry is
restored, and again we have $\la | \phi | \ra = 0$.  In
condensed--matter jargon, the transition described above is
second--order
%, or also first--order proceeding by spinoidal
%decomposition (caused by a rapid quench in the system) 
[Mermin, 1979].\footnote{In a first--order phase transition the order parameter
(\eg, $\la | \phi | \ra$ in our case) is not continuous.  It may
proceed by bubble nucleation [Callan \& Coleman, 1977; Linde, 1983b]
or by spinoidal decomposition [Langer, 1992].  Phase transitions can
also be continuous second--order processes. The `order' depends
sensitively on the ratio of the coupling constants appearing in the
Lagrangian.}

\subsection{The Kibble mechanism}
\label{sec-Kibbbb}                                 

The model described in the last subsection is an example in which the
transition may be second--order.  As we saw, for temperatures much
larger than the critical one the vacuum expectation value of the
scalar field vanishes at all points of space, whereas for $T < T_c$ it
evolves smoothly in time towards a non vanishing $\la | \phi | \ra$.
Both thermal and quantum fluctuations influence the new value taken by
$\la | \phi | \ra$ and therefore it has no reasons to be uniform in
space.  This leads to the existence of domains wherein the $\la | \phi
(\vec x) | \ra$ is coherent and regions where it is not.  The
consequences of this fact are the subject of this subsection.
                                 
Phase transitions can also be first--order proceeding via bubble
nucleation.  At very high energies the symmetry breaking potential has
$\la | \phi | \ra = 0$ as the only vacuum state. When the temperature
goes down to $T_c$ a set of vacua, degenerate to the previous one,
develops. However this time the transition is not smooth as before,
for a potential barrier separates the old (false) and the new (true)
vacua (see, \eg\ Figure \ref{fig-pot_phtrans}).  
Provided the barrier at this small temperature is high enough,
compared to the thermal energy present in the system, the field $\phi$
will remain trapped in the false vacuum state even for small ($< T_c$)
temperatures. Classically, this is the complete picture.  However,
quantum tunneling effects can liberate the field from the old vacuum
state, at least in some regions of space: there is a probability per
unit time and volume in space that at a point $\vec x$ a bubble of
true vacuum will nucleate.  The result is thus the formation of
bubbles of true vacuum with the value of the field in each bubble
being independent of the value of the field in all other bubbles.
This leads again to the formation of domains where the fields are
correlated, whereas no correlation exits between fields belonging to
different domains.  Then, after creation the bubble will expand at the
speed of light surrounded by a `sea' of false vacuum domains.  As
opposed to second--order phase transitions, here the nucleation
process is extremely inhomogeneous and $\la | \phi (\vec x) | \ra$ is
not a continuous function of time.

Let us turn now to the study of correlation lengths and their r\^ole
in the formation of topological defects.  One important feature in
determining the size of the domains where $\la | \phi (\vec x) | \ra$
is coherent is given by the spatial correlation of the field $\phi$.
Simple field theoretic considerations [see, \eg, Copeland, 1993]
for long wavelength fluctuations of $\phi$ lead to different functional
behaviors for the correlation function $G(r) \equiv \la
\phi(r_1)\phi(r_2) \ra$, where we noted $r = |r_1 - r_2|$.  What is
found depends radically on whether the wanted correlation is
computed between points in space separated by a distance $r$ much
smaller or much larger than a characteristic length $\xi^{-1} = m(T)
\simeq \sqrt{\lambda} ~ |\la\phi\ra |$, known as the {\sl correlation
length}.
We have
\be
\label{llave}
G(r) \simeq
\cases{
{T_c \over 4 \pi r} \exp (- {r\over \xi}) ~~~~~~~~  r >> \xi      \cr
~~~                                                               \cr
{T^2 \over 2 \pi^2}  ~~~~~~~~~~~~~~~~~~~                r << \xi ~.
}
\ee                                                                       

This tells us that domains of size $ \xi \sim m^{-1}$ arise where the
field $\phi$ is correlated.  On the other hand, well beyond $\xi$ no
correlations exist and thus points separated apart by $r >> \xi$ will
belong to domains with in principle arbitrarily different orientations
of the Higgs field.  This in turn leads, after the merging of these
domains in a cosmological setting, to the existence of defects, where
field configurations fail to match smoothly.

However, when $T \to T_c$ we have $m\to 0$ and so $\xi\to\infty$,
suggesting perhaps that for all points of space the field $\phi$
becomes correlated. This fact clearly violates causality.  The
existence of particle horizons in cosmological models (proportional to
the inverse of the Hubble parameter $H^{-1}$) constrains microphysical
interactions over distances beyond this causal domain.  Therefore we
get an upper bound to the correlation length as $\xi < H^{-1} \sim t$.

The general feature of the existence of uncorrelated domains has
become known as the Kibble mechanism [Kibble, 1976] and it seems to be
generic to most types of phase transitions.

\subsection{A survey of topological defects}
\label{sec-ASurv}

Different models for the Higgs field lead to the formation of a whole
variety of topological defects, with very different characteristics
and dimensions.  Some of the proposed theories have symmetry breaking
patterns leading to the formation of `domain walls' (mirror reflection
discrete symmetry): incredibly thin planar surfaces trapping enormous
concentrations of mass--energy which separate domains of conflicting
field orientations, similar to two--dimensional sheet--like structures
found in ferromagnets.  Within other theories, cosmological fields get
distributed in such a way that the old (symmetric) phase gets confined
into a finite region of space surrounded completely by the new
(non--symmetric) phase. This situation leads to the generation of
defects with linear geometry called `cosmic strings'.  Theoretical
reasons suggest these strings (vortex lines) do not have any loose
ends in order that the two phases not get mixed up.  This leaves
infinite strings and closed loops as the only possible alternatives
for these defects to manifest themselves in the early
universe\footnote{`Monopole' is another possible topological defect;
we defer its discussion to the next subsection.  Cosmic strings
bounded by monopoles is yet another possibility in GUT phase
transitions of the kind, \eg, ${\bf G}\to {\bf K}\times U(1)\to {\bf
K}$.  The first transition yields monopoles carrying a magnetic charge
of the $U(1)$ gauge field, while in the second transition the magnetic
field in squeezed into flux tubes connecting monopoles and
antimonopoles [Langacker \& Pi, 1980].}.

\begin{figure}[tbp]
  \begin{center}
    \leavevmode
    \epsfxsize = 15.5cm
    \epsffile{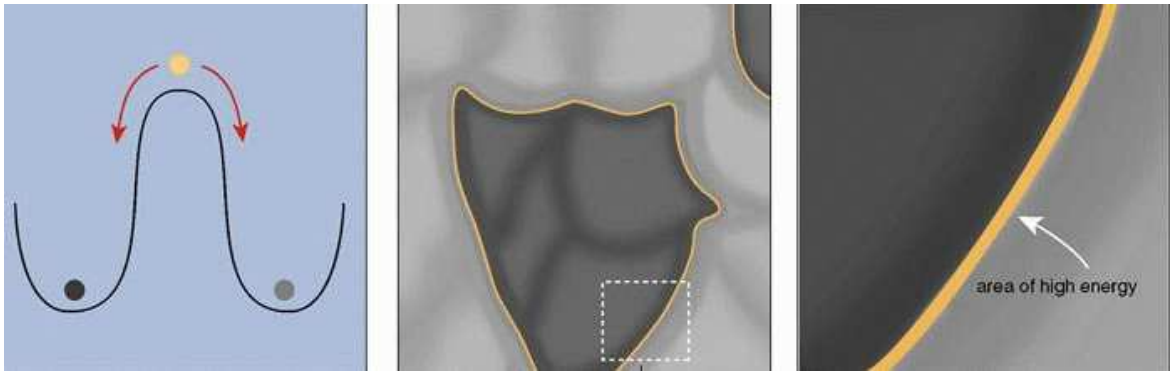}
%    \epsffile{ffffigg2s.ps}
%    \epsffile{ffffigg2.ps}
  \end{center}
\caption{{\sl In a simple model of symmetry breaking, the initial
symmetric ground state of the Higgs field (yellow dot) can fall into
the left- or right-hand valley of a double-well energy potential
(light and dark dots). In a cosmic phase transition, regions of the
new phase appear randomly and begin to grow and eventually merge as
the transition proceeds toward completion (middle). Regions in which
the symmetry has broken the same way can coalesce, but where regions
that have made opposite choices encounter each other, a topological
defect known as a domain wall forms (right). Across the wall, the
Higgs field has to go from one of the valleys to the other (in the
left panel), and must therefore traverse the energy peak. This creates
a narrow planar region of very high energy, in which the symmetry is
locally unbroken.}}
\label{fig-dwcolor}
\end{figure}               

With a bit more abstraction scientists have even conceived other
(semi) topological defects, called `textures'. These are
conceptually simple objects, yet, it is not so easy to imagine
them for they are just global field configurations living on
a three--sphere vacuum manifold (the minima of the effective
potential energy), whose non linear evolution perturbs spacetime.
Turok [1989] was the first to realize that many unified theories
predicted the existence of peculiar Higgs field configurations known
as (texture) knots, and that these could be of potential interest for
cosmology.  Several features make these defects interesting.  In
contrast to domain walls and cosmic strings, textures have no core and
thus the energy is more evenly distributed over space.  Secondly, they
are unstable to collapse and it is precisely this last feature which
makes these objects cosmologically relevant, for this instability
makes texture knots shrink to a microscopic size, unwind and radiate
away all their energy.  In so doing, they generate a gravitational
field that perturbs the surrounding matter in a way which can seed
structure formation.

%%-------------------------------------------------------------
\subsection{Conditions for their existence: topological criteria}
\label{subsec-condexist}      

Let us now explore the conditions for the existence of topological
defects. It is widely accepted that the final goal of particle physics
is to provide a unified gauge theory comprising strong, weak and
electromagnetic interactions (and some day also gravitation).  This
unified theory is to describe the physics at very high temperatures,
when the age of the universe was slightly bigger than the Planck time.
At this stage, the universe was in a state with the highest possible
symmetry, described by a symmetry group {\bf G}, and the Lagrangian
modeling the system of all possible particles and interactions present
should be invariant under the action of the elements of {\bf G}.

As we explained before, the form of the finite temperature effective
potential of the system is subject to variations during the cooling
down evolution of the universe.  This leads to a chain of phase
transitions whereby some of the symmetries present in the beginning
are not present anymore at lower temperatures.  The first of these
transitions may be described as {\bf G}$\to${\bf H}, where now {\bf H}
stands for the new (smaller) unbroken symmetry group ruling the
system.  This chain of symmetry breakdowns eventually ends up with
SU(3)$\times$SU(2)$\times$U(1), the symmetry group underlying the
`standard model' of particle physics.

A broken symmetry system (with a Mexican-hat potential for the Higgs
field) may have many different minima (with the same energy), all related
by the underlying symmetry.  Passing from one minimum to another is
included as one of the symmetries of the original group {\bf G}, and
the system will not change due to one such transformation.  If a
certain field configuration yields the lowest energy state of the
system, transformations of this configuration by the elements of the
symmetry group will also give the lowest energy state.  For example,
if a spherically symmetric system has a certain lowest energy value,
this value will not change if the system is rotated.

The system will try to minimize its energy and will spontaneously
choose one amongst the available minima.  Once this is done and the
phase transition achieved, the system is no longer ruled by {\bf G}
but by the symmetries of the smaller group {\bf H}.  So, if {\bf
G}$\to${\bf H} and the system is in one of the lowest energy states
(call it $S_1$), transformations of $S_1$ to $S_2$ by elements of {\bf
G} will leave the energy unchanged. However, transformations of $S_1$
by elements of {\bf H} will leave $S_1$ {\it itself} (and not just the
energy) unchanged.  The many distinct ground states of the system $S_1
, S_2 , \ldots $ are given by all transformations of {\bf G} that are
{\it not} related by elements in {\bf H}. This space of distinct
ground states is called the {\sl vacuum manifold} and denoted $\cal
M$.
%% thanx to Tanmay for this

\begin{quote}
$\cal M$ is the space of all elements of {\bf G} in which elements
related by transformations in {\bf H} have been identified.
Mathematicians call it the {\sl coset space} and denote it {\bf
G}$/${\bf H}. We then have ${\cal M}=$ {\bf G}$/${\bf H}.
\end{quote}

The importance of the study of the vacuum manifold lies in the fact
that it is precisely the {\sl topology} of ${\cal M}$ what determines the
type of defect that will arise.  Homotopy theory tells us how to map
${\cal M}$ into physical space in a non--trivial way, and what ensuing
defect will be produced.  For instance, the existence of non
contractible loops in ${\cal M}$ is the requisite for the formation of
cosmic strings.  In formal language this comes about whenever we have
the first homotopy group $\pi_1 ({\cal M}) \neq$ {\bf 1}, where {\bf
1} corresponds to the trivial group.  If the vacuum manifold is
disconnected we then have $\pi_0 ({\cal M}) \neq$ {\bf 1}, and domain
walls are predicted to form in the boundary of these regions where the
field $\phi$ is away from the minimum of the potential.  Analogously,
if $\pi_2 ({\cal M}) \neq$ {\bf 1} it follows that the vacuum manifold
contains non contractible two--spheres, and the ensuing defect is a
monopole.  Textures arise when ${\cal M}$ contains non contractible
three--spheres and in this case it is the third homotopy group, $\pi_3
({\cal M})$, the one that is non trivial. 
We summarize this in Table \ref{table-topo} .

\begin{table}[htbp]\begin{center}
%\phantom{.}
%\vspace{1cm}
\begin{tabular}{|c l l|}
\hline
$\pi_0 ({\cal M}) \neq${\bf 1} & ${\cal M}$ {\it disconnected}
     & {\sc Domain Walls}  \\
$\pi_1 ({\cal M}) \neq${\bf 1} & {\it non contractible loops} in
${\cal M}$
     & {\sc Cosmic Strings}  \\
$\pi_2 ({\cal M}) \neq${\bf 1} & {\it non contractible 2--spheres} in
${\cal M}$
     & {\sc Monopoles}   \\
$\pi_3 ({\cal M}) \neq${\bf 1} & {\it non contractible 3--spheres} in
${\cal M}$
     & {\sc Textures}  \\
\hline
\end{tabular}\end{center}
\caption{The topology of ${\cal M}$ determines the type of defect
that will arise.}
\label{table-topo}
\end{table}        

%%%%-----------------------------------------------------------
\section{Defects in the universe}
\label{sec-definuni}     

Generically topological defects will be produced if the conditions for
their existence are met. Then for example if the unbroken group {\bf
H} contains a disconnected part, like an explicit U(1) factor
(something that is quite common in many phase transition schemes
discussed in the literature), monopoles will be left as relics of the
transition. This is due to the fundamental theorem on the second
homotopy group of coset spaces [Mermin, 1979], which states that for a
simply--connected covering group {\bf G} we have\footnote{The
isomorfism between two groups is noted as $\cong$.  Note that by using
the theorem we therefore can reduce the computation of $\pi_2$ for a
coset space to the computation of $\pi_1$ for a group.  A word of
warning: the focus here is on the physics and the
mathematically--oriented reader should bear this in mind, especially
when we will become a bit sloppy with the notation.  In case this
happens, consult the book [Steenrod, 1951] for a clear exposition of
these matters.}  \be \pi_2({\bf G} / {\bf H}) \cong \pi_1({\bf H}_0)
~, \ee with ${\bf H}_0$ being the component of the unbroken group
connected to the identity.  Then we see that since monopoles are
associated with unshrinkable surfaces in {\bf G}$/${\bf H}, the
previous equation implies their existence if {\bf H} is
multiply--connected.  The reader may guess what the consequences are
for GUT phase transitions: in grand unified theories a semi--simple
gauge group {\bf G} is broken in several stages down to 
{\bf H} = SU(3)$\times$U(1). 
Since in this case $\pi_1({\bf H}) \cong {\cal Z}$,
the integers, we have $\pi_2 ({\bf G} / {\bf H}) \neq$ {\bf 1} and
therefore gauge monopole solutions exist [Preskill, 1979].

%%-------------------------------------------------------------
\subsection{Local and global monopoles and domain walls}
\label{sec-monoANDdo}
                        
Monopoles are yet another example of stable topological defects.
Their formation stems from the fact that the vacuum expectation value
of the symmetry breaking Higgs field has random orientations
($\la\phi^a\ra$ pointing in different directions in group space) on
scales greater than the horizon.  One expects therefore to have a
probability of order unity that a monopole configuration will result
after the phase transition (cf. the Kibble mechanism).  Thus, about
one monopole per Hubble volume should arise and we have for the number
density $n_{monop} \sim 1 / H^{-3} \sim T_c^6 / m_P^3$, where $T_c$ is
the critical temperature and $m_P$ is Planck mass, 
when the transition occurs.  We also know
the entropy density at this temperature, $s \sim T_c^3$, and so the
monopole to entropy ratio is $n_{monop} / s \simeq 100 (T_c / m_P)^3$.
In the absence of non--adiabatic processes after monopole creation
this constant ratio determines their present abundance.  For the
typical value $T_c\sim 10^{14}$ GeV we have $n_{monop} / s \sim
10^{-13}$. This estimate leads to a present $\Omega_{monop} h^2 \simeq
10^{11}$, for the superheavy monopoles $m_{monop}\simeq 10^{16}$ GeV
that are created\footnote{These are the actual figures for a gauge
SU(5) GUT second--order phase transition. Preskill [1979] has shown
that in this case monopole antimonopole annihilation is not effective
to reduce their abundance. Guth \& Weinberg [1983] did the case for a
first--order phase transition and drew qualitatively similar
conclusions regarding the excess of monopoles.}.  This value
contradicts standard cosmology and the presently most attractive way
out seems to be to allow for an early period of inflation: the massive
entropy production will hence lead to an exponential decrease of the
initial $n_{monop} / s$ ratio, yielding $\Omega_{monop}$ consistent
with observations.\footnote{The inflationary expansion reaches an end
in the so--called reheating process, when the enormous vacuum energy
driving inflation is transferred to coherent oscillations of the
inflaton field. These oscillations will in turn be damped by the
creation of light particles (\eg, via preheating) 
whose final fate is to thermalise and
reheat the universe.} In summary, the
broad--brush picture one has in mind is that of a mechanism that could
solve the monopole problem by `weeping' these unwanted relics out of
our sight, to scales much bigger than the one that will eventually
become our present horizon today.

Note that these arguments do not apply for global monopoles as
these (in the absence of gauge fields) possess long--range
forces that lead to a decrease of their number in comoving
coordinates. The large attractive force between global monopoles and
antimonopoles leads to a high annihilation probability and
hence monopole over--production does not take place.
Simulations performed by Bennett \&  Rhie [1990] showed
that global monopole evolution rapidly settles into a scale
invariant regime with only a few monopoles per horizon
volume at all times.

Given that global monopoles do not represent a danger for cosmology
one may proceed in studying their observable consequences. The
gravitational fields of global monopoles may lead to matter clustering
and CMB anisotropies. Given an average number of monopoles per horizon
of $\sim 4$, Bennett \& Rhie [1990] estimate a scale invariant
spectrum of fluctuations $( \delta\rho / \rho )_H \sim 30 G
\eta^2$ at horizon crossing\footnote{The spectrum of density
fluctuations on smaller scales has also been computed.  They normalize
the spectrum at $8 h^{-1}$ Mpc and agreement with observations lead
them to assume that galaxies are clustered more strongly than the
overall mass density, this implying a `biasing' of a few [see Bennett,
Rhie \& Weinberg, 1993 for details].}.  In a subsequent paper they
simulate the large--scale CMB anisotropies and, upon normalization
with {\sl COBE}--DMR, they get roughly $G \eta^2 \sim 6 \times
10^{-7}$ in agreement with a GUT energy scale $\eta$ [Bennett \&
Rhie, 1993]. However, as we will see in the CMB sections below,
current estimates for the angular power spectrum of global defects do
not match the most recent observations, their main problem being the
lack of power on the degree angular scale once the spectrum is
normalized to {\sl COBE} on large scales. 

Let us concentrate now on domain walls, and briefly try to show why
they are not welcome in any cosmological context (at least in the
simple version we here consider -- there is always room for more
complicated (and contrived) models).  If the symmetry breaking pattern
is appropriate at least one domain wall per horizon volume will be
formed.  The mass per unit surface of these two-dimensional objects is
given by $\sim \lambda^{1/2} \eta^3$, where $\lambda$ as usual
is the coupling constant in the symmetry breaking potential for the
Higgs field.  Domain walls are generally horizon--sized and therefore
their mass is given by $\sim \lambda^{1/2} \eta^3 H^{-2}$. This
implies a mass energy density roughly given by $\rho_{DW}\sim
\eta^3 t^{-1}$ and we may readily see now how the problem
arises: the critical density goes as $\rho_{crit} \sim t^{-2}$ which
implies $\Omega_{DW}(t) \sim (\eta / m_P)^2 \eta t$.
Taking a typical GUT value for $\eta$ we get $\Omega_{DW}(t\sim
10^{-35}{\rm sec}) \sim 1$ {\sl already} at the time of the phase
transition. It is not hard to imagine that today this will be at
variance with observations; in fact we get $\Omega_{DW}(t \sim
10^{18}{\rm sec}) \sim 10^{52}$. This indicates that models where
domain walls are produced are tightly constrained, and the general
feeling is that it is best to avoid them altogether [see Kolb \&
Turner, 1990 for further details; see also 
Dvali \etal, 1998, 
Pogosian \& Vachaspati, 2000 
\footnote{Animations of monopoles colliding with domain walls can be found in 
`LEP' page at {\tt http://theory.ic.ac.uk/\~{}LEP/figures.html}}
and Alexander \etal, 1999 for an alternative solution].

%%-------------------------------------------------------------
\subsection{Are defects inflated away?}
\label{sec-topoandinfla}
                              
It is important to realize the relevance that the Kibble's mechanism
has for cosmology; nearly every sensible grand unified theory (with
its own symmetry breaking pattern) predicts the existence of defects.
We know that an early era of inflation helps in getting rid of the
unwanted relics.  One could well wonder if the very same Higgs field
responsible for breaking the symmetry would not be the same one
responsible for driving an era of inflation, thereby diluting the
density of the relic defects.
This would get rid not only of (the unwanted) monopoles and
domain walls but also of any other (cosmologically appealing) defect.
Let us follow [Brandenberger, 1993] and sketch why this actually does
not occur. 
Take first the symmetry breaking potential of Eq. (\ref{VfiniT})
at zero temperature and add to it
a harmless $\phi$--independent term $3 m^4 / (2\lambda)$. This will
not affect the dynamics at all. Then we are led to
\be
\label{VfiniT2}
V(  \phi  ) =
{\lambda \over 4!} \left(
\phi^2 - {\eta }^2
\right)^2 ~,
\ee
with $\eta = ( 6 m^2 / \lambda )^{1/2}$
the symmetry breaking energy scale,
and where for the present heuristic digression we just took a real
Higgs field. Consider now the equation of motion for $\phi$,
\be                                             
\label{Higgapprox}
\ddot \phi \simeq - {\partial V\over\partial\phi }
= - {\lambda\over 3!} \phi^3 + m^2 \phi
\approx  m^2 \phi ~,
\ee
for $\phi << \eta$ very near the false vacuum of the effective
Mexican hat potential and
where, for simplicity, the expansion of the universe and
possible interactions of $\phi$ with other fields were neglected.
The typical time scale of the solution is $\tau\simeq m^{-1}$.
For an inflationary epoch to be effective we need
$\tau >> H^{-1}$, \ie, a sufficiently large number of e--folds
of slow--rolling solution. Note, however, that after some
e--folds of exponential expansion the curvature term in
the Friedmann equation becomes subdominant and we have
$H^2 \simeq 8\pi G ~V(0) / 3 \simeq (2\pi m^2 / 3 )(\eta /
m_P)^2$.
So, unless $\eta  > m_P$, which seems unlikely for a GUT phase
transition, we are led to $\tau << H^{-1}$ and therefore    
the amount of inflation is not enough for getting rid of
the defects generated during the transition by hiding them
well beyond our present horizon.  

Recently, there has been a large amount of work in getting defects,
particularly cosmic strings, after post-inflationary preheating.
Reaching the latest stages of the inflationary phase, the inflaton
field oscillates about the minimum of its potential. In doing so,
parametric resonance may transfer a huge amount of energy to other
fields leading to cosmologically interesting nonthermal phase
transitions.  Just like thermal fluctuations can restore broken
symmetries, here also, these large fluctuations may lead to the whole
process of defect formation again. Numerical simulations employing
potentials similar to that of Eq. (\ref{VfiniT2}) have shown that
strings indeed arise for values $\eta\sim 10^{16}$ GeV 
[Tkachev \etal , 1998, Kasuya \& Kawasaki, 1998]. Hence,
preheating after inflation helps in generating cosmic defects. 

%%-------------------------------------------------------------
\subsection{Cosmic strings}
\label{sec-cosmi}         

Cosmic strings are without any doubt the topological defect
most thoroughly studied, both in cosmology and solid--state
physics (vortices).
The canonical example, also describing flux tubes in superconductors,
is given by the Lagrangian
\be
\label{lagraCS}
{\cal L} = -{1\over 4} F_{\mu\nu} F^{\mu\nu} + {1\over 2}
|D_\mu\phi|^2 - {\lambda \over 4!} \left( |\phi |^2 - {\eta }^2
\right)^2 ~, \ee 
with $F_{\mu\nu} = \partial_{[\mu}A_{\nu ]}$, where
$A_{\nu}$ is the gauge field and the covariant derivative is $D_\mu =
\partial_\mu + i e A_{\mu}$, with $e$ the gauge coupling constant.
This Lagrangian is invariant under the action of the Abelian group
{\bf G} = U(1), and the spontaneous breakdown of the symmetry leads to
a vacuum manifold ${\cal M}$ that is a circle, $S^1$, \ie, the
potential is minimized for $\phi = \eta\exp (i\theta)$, with
arbitrary $0\leq\theta\leq 2\pi$.  Each possible value of $\theta$
corresponds to a particular `direction' in the field space.

\begin{figure}[t]
  \begin{center}
    \leavevmode
    \epsfxsize = 15.5cm
    \epsffile{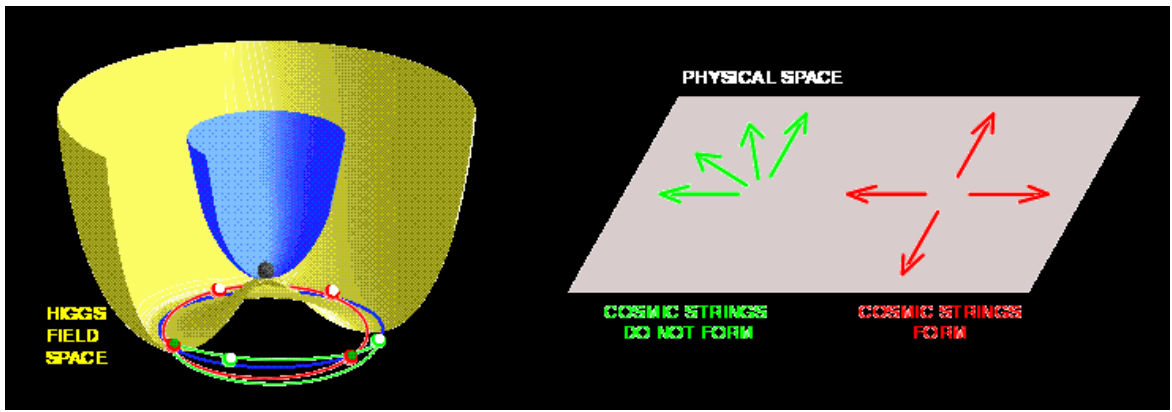}
%    \epsffile{recuadro_uk_as.ps}
%    \epsffile{recuadro_uk_a.eps}
%\vspace{-0.5cm}
  \end{center}
\caption{{\sl The complex scalar Higgs field evolves in a
temperature-dependent potential $V(\phi)$. At high temperatures
(violet surface) the vacuum expectation value of the field lies at the
bottom of $V$. For lower temperatures, the potential adopts the
``Mexican hat'' form (yellow surface) and the field spontaneously
chooses one amongst the new available (degenerate) lowest energy states
(the violet circle along the valley of the hat). This isolates a
single value/direction for the phase of the field, spontaneously
breaking the symmetry possessed by the system at high
energies. Different regions of the universe, with no causal
connection, will end up having arbitrarily different directions for
the field (arrows on the right). As separate regions of broken
symmetry merge, it is not always possible for the field orientations
to match. It may happen that a closed loop in physical space
intersects regions where the Higgs phase varies from 0 to 2$\pi$ (red
arrows, corresponding to the red dashed-line on the left panel). In
that situation, a cosmic string will pass somewhere inside the
loop. On the contrary, green arrows (and green dashed-line on the left
panel) show a situation where no string is formed after the phase
transition.}}
%\vspace{-0.5cm}
\label{fig-recuadro-a}
\end{figure}               

Now, as we have seen earlier, due to the overall cooling down of the
universe, there will be regions where the scalar field rolls down to
different vacuum states.  The choice of the vacuum is totally independent
for regions separated apart by one correlation length or more, thus
leading to the formation of domains of size $\xi\sim \eta^{-1}$.
When these domains coalesce they give rise to edges in the interface.
If we now draw a imaginary circle around one of these edges and the
angle $\theta$ varies by $2\pi$ then by contracting this loop we reach
a point where we cannot go any further without leaving the manifold
${\cal M}$. This is a small region where the variable $\theta$ is not
defined and, by continuity, the field should be $\phi = 0$.  In order
to minimize the spatial gradient energy these small regions line up
and form a line--like defect called cosmic string.

The width of the string is roughly $m_\phi^{-1} \sim (\sqrt{\lambda}
\eta)^{-1}$, $m_\phi$ being the Higgs mass. The string mass per
unit length, or tension, is $\mu \sim \eta^2$. This means that
for GUT cosmic strings, where $\eta\sim 10^{16}$ GeV, we have
$G\mu \sim 10^{-6}$.  We will see below that the dimensionless
combination $G\mu$, present in all signatures due to strings, is of
the right order of magnitude for rendering these defects
cosmologically interesting.

There is an important difference between global and gauge (or local)
cosmic strings: local strings have their energy confined mainly
in a thin core, due to the presence of gauge fields $A_\mu$ that
cancel the gradients of the field outside of it. Also these gauge
fields make it possible for the string to have a quantized
magnetic flux along the core.
On the other hand, if the string was
generated from the breakdown of a {\sl global} symmetry there are no
gauge fields, just Goldstone bosons, which, being massless, give
rise to long--range forces. No gauge fields can compensate
the gradients of $\phi$ this time and therefore there is an
infinite string mass per unit length.

Just to get a rough idea of the kind of models studied in the
literature, consider the case ${\bf G} = SO(10)$ that is broken to
${\bf H} = SU(5)\times {\cal Z}_2$.  For this pattern we have
$\pi_1({\cal M}) = {\cal Z}_2$, which is clearly non trivial and
therefore
cosmic strings are formed [Kibble \etal, 1982].\footnote{In the
        analysis one uses the
        fundamental theorem stating that, for a simply--connected Lie
        group {\bf G} breaking down to {\bf H}, we have
        $\pi_1({\bf G} / {\bf H}) \cong \pi_0({\bf H})$;
        see [Hilton, 1953].}                          

\begin{figure}[t]
  \begin{center}
    \leavevmode
    \epsfxsize = 15.5cm
    \epsffile{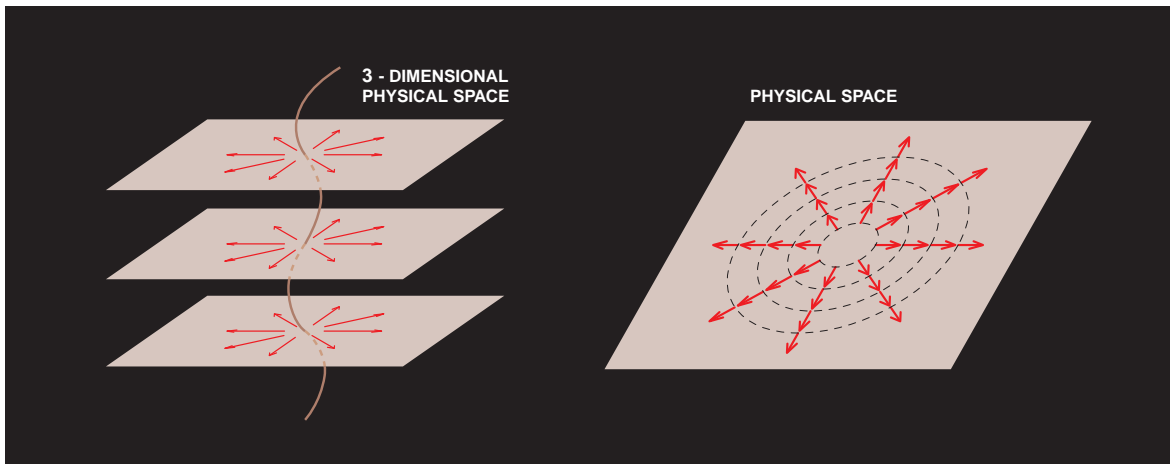}
  \end{center}
\caption{{\sl We can now extend the mechanism shown in the previous
figure to the full three-dimensional space. Regions of the various
planes that were traversed by strings can be superposed to show the
actual location of the cosmic string (left panel).  The figure on the
right panel shows why we are sure a string crosses the plane inside
the loop in physical space (the case with red arrows in the previous
figure). Continuity of the field imposes that if we gradually contract
this loop the direction of the field will be forced to wind
``faster''. In the limit in which the loop reduces to a point, the
phase is no longer defined and the vacuum expectation value of the
Higgs field has to vanish. This corresponds to the central tip of the
Mexican hat potential in the previous figure and is precisely the
locus of the false vacuum. Cosmic strings are just that, narrow,
extremely massive line-like regions in physical space where the Higgs
field adopts its high-energy false vacuum state.}}
\label{fig-recuadro-b}
\end{figure}               

%%-------------------------------------------------------------
\subsection{String loops and scaling}
\label{sec-loopheurist}      

We saw before the reasons why gauge monopoles and domain walls were a
bit of a problem for cosmology. Essentially, the problem was that
their energy density decreases more slowly than the critical density
with the expansion of the universe. This fact resulted in their
contribution to $\Omega_{\rm def}$ (the density in defects normalized
by the critical density) being largely in excess compared to 1, hence
in blatant conflict with modern observations. The question now arises
as to whether the same might happened with cosmic strings. Are strings
dominating the energy density of the universe? Fortunately, the answer
to this question is {\sl no}; strings evolve in such a way to make
their density $\rho_{\rm strings}\propto \eta^2 t^{-2}$. Hence, one
gets the same temporal behavior as for the critical density. The
result is that $\Omega_{\rm strings} \sim G\mu \sim (\eta/m_P)^2 \sim
10^{-6}$ for GUT strings, \ie, we get an interestingly small enough,
constant fraction of the critical density of the universe and strings
never upset standard observational cosmology.

Now, why this is so? The answer is simply the efficient way in which a
network of strings looses energy. The evolution of the string network
is highly nontrivial and loops are continuously chopped off from the
main infinite strings as the result of (self) intersections within 
the infinite--string network. Once they are produced, loops
oscillate due to their huge tension and slowly decay by emitting
gravitational radiation. Thus, energy is transferred from the cosmic 
string network to radiation.\footnote{High--resolution cosmic string 
simulations can be found in the Cambridge cosmology page at 
{\tt http://www.damtp.cam.ac.uk/user/gr/public/cs\_{}evol.html}} 

\begin{figure}[t]
  \begin{center}
    \leavevmode
    \epsfxsize = 15.5cm
    \epsffile{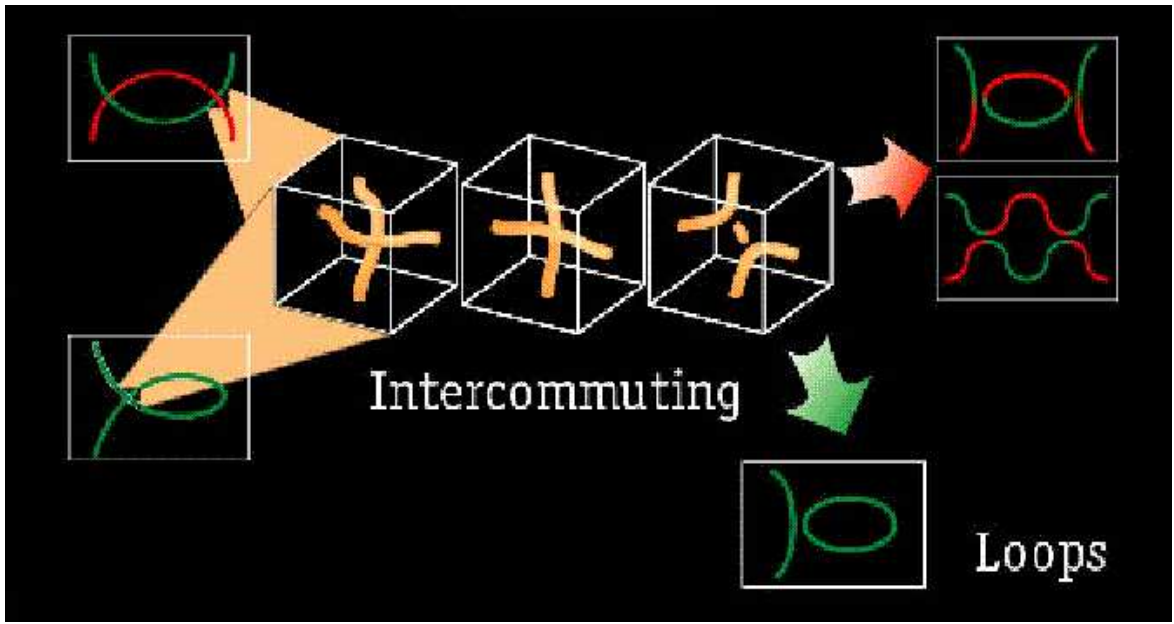}
%    \epsffile{Laroche2001s.ps}
%    \epsffile{Laroche2001.ps}
  \end{center}
\caption{{\sl Global string interactions leading to loop
formation. Whenever two string segments intersect, they reconnect or
intercommute (green and red strings -- upper part of the
figure). Analogously, if a string intersects itself, it can break off
a closed loop (green string -- bottom part of the figure). In both
cases, the interacting string segments first suffer a slight
deformation (due to the long--range forces present for global
strings), they subsequently fuse and finally exchange partners. A
ephemeral unstable amount of energy in the form of a small loop
remains in the middle where the energy is high enough to place the
Higgs field in the false vacuum. It then quickly collapses, radiating
away its energy. The situation is roughly the same for local strings,
as simulations have shown.}}
\label{fig-loopsform}
\end{figure}               

It turns out from simulations that most of the energy in the string
network (roughly a 80\%) is in the form of infinite strings. Soon after formation one
would expect long strings to have the form of random-walk with
characteristic step given by the correlation length $\xi$.  Also, the
typical distance between long string segments should also be of order
$\xi$. Monte Carlo simulations show that these strings are Brownian on
sufficiently large scales, which means that the length $\ell$ of a
string is related to the end-to-end distance ${\rm d}$ of two given points
along the string (with ${\rm d} \gg \xi$) in the form 
\be 
\ell = {\rm d}^2/\xi .
\ee 
What remains of the energy is given in the form of closed loops with
no preferred length scale (a scale invariant distribution) which
implies that the number density of loops having sizes between $R$ and
$R+ dR$ follows just from dimensional analysis
\be
d n_{\rm loops} \propto {dR\over R^4}
\ee
which is just another way of saying that $n_{\rm loops}\propto 1/R^3$,
loops behave like normal nonrelativistic matter.
The actual coefficient, as usual, comes from string simulations.

There are both analytical and numerical indications in favor of the
existence of a stable ``scaling solution'' for the
cosmic string network. After generation, the network quickly evolves
in a self similar manner with just a few infinite string segments 
per Hubble volume and Hubble time. 
A heuristic argument for the scaling solution due to Vilenkin [1985]
is as follows. 

If we take $\nu(t)$ to be the mean number of infinite string segments
per Hubble volume, then the energy density in infinite strings
$\rho_{\rm strings} = \rho_{\rm s}$ is
\be
\label{vilen1}
\rho_{\rm s}(t) = \nu(t) \eta^2 t^{-2} = \nu(t) \mu t^{-2} .  
\ee
Now, $\nu$ strings will typically have $\nu$ intersections, and so 
the number of loops $n_{\rm loops}(t) =  n_{\rm l}(t)$ produced per 
unit volume will be proportional to $\nu^2$. We find
\be
d n_{\rm l} \sim \nu^2 R^{-4} dR .
\ee
Hence, recalling now that the loop sizes grow with the expansion like
$R\propto t$ we have 
\be
\label{vilen2}
{d n_{\rm l}(t)\over dt} \sim p \nu^2 t^{-4} 
\ee
where $p$ is the probability of loop formation per intersection, a
quantity related to the intercommuting probability, both roughly of
order 1. 
We are now in a position to write an energy conservation equation for
strings plus loops in the expanding universe. Here it is
\be
\label{vilen3}
{d \rho_{\rm s} \over dt} + {3 \over 2 t} \, \rho_{\rm s} \sim 
 - m_{\rm l} {d n_{\rm l} \over dt } \sim 
 - \mu t     {d n_{\rm l} \over dt }
\ee
where $m_{\rm l} = \mu t$ is just the loop mass and where 
the second on the left hand side is the dilution term 
$3 H \rho_{\rm s}$ for an expanding radiation--dominated universe. 
The term on the right hand side amounts to the loss of energy from
the long string network by the generation of small closed loops. 
Plugging Eqs. (\ref{vilen1}) and (\ref{vilen2}) into (\ref{vilen3})
Vilenkin finds the following kinetic equation for $\nu(t)$
\be
{d\nu \over dt} - \, {\nu\over{2 t}} \sim - p {\nu^2 \over t}
\ee
with $p\sim 1$. Thus if $\nu \gg 1$ then ${d\nu / dt} < 0$ and $\nu$
tends to decrease in time, while if $\nu \ll 1$ then ${d\nu / dt} > 0$ and
$\nu$ increases. Hence, there will be a stable solution with 
$\nu \sim {\rm a ~few}$. 

%%-------------------------------------------------------------
\subsection{Global textures}
\label{sec-texuuu}      

Whenever a global non--Abelian symmetry is spontaneously and
completely broken (\eg\ at a grand unification scale), global defects
called textures are generated.  Theories where this global symmetry is
only partially broken do not lead to global textures, but instead to
global monopoles and non--topological textures.  As we already
mentioned global monopoles do not suffer the same constraints as their
gauge counterparts: essentially, having no associated gauge fields,
the long--range forces between pairs of monopoles lead to the
annihilation of their eventual excess and as a result monopoles scale
with the expansion.
On the other hand,
non--topological textures are a generalization that allows the broken
subgroup {\bf H} to contain non--Abelian factors. It is then possible
to have $\pi_3$ trivial as in, \eg, SO(5)$\to$SO(4) broken by a
vector, for which case we have ${\cal M} = S^4$, the four--sphere
[Turok, 1989].  Having explained this, let us concentrate in global
topological textures from now on.

Textures, unlike monopoles or cosmic strings, are not well localized
in space. This is due to the fact that the field remains in the vacuum
everywhere, in contrast to what happens for other defects, where the
field leaves the vacuum manifold precisely where the defect core is.
Since textures do not possess a core, all the energy of the field
configuration is in the form of field gradients.  This fact is what
makes them interesting objects {\sl only} when coming from global
theories: the presence of gauge fields $A_\mu$ could (by a suitable
reorientation) compensate the gradients of $\phi$ and yield $D_\mu\phi
= 0$, hence canceling out (gauging away) the energy of the
configuration\footnote{This does not imply, however, that the
classical dynamics of a gauge texture is trivial. The evolution of the
$\phi$--$A_\mu$ system will be determined by the competing tendencies
of the global field to unwind and of the gauge field to compensate the
$\phi$ gradients. The result depends on the characteristic size $L$ of
the texture: in the range $m_\phi^{-1} << L << m_A^{-1} \sim
(e\eta)^{-1}$ the behavior of the gauge texture resembles that
of the global texture, as it should, since in the limit $m_A$ very
small ($e\to 0$) the gauge texture turns into a global one [Turok \&
Zadrozny, 1990].}.

One feature endowed by textures that really makes these defects
peculiar is their being unstable to collapse.  The initial field
configuration is set at the phase transition, when $\phi$ develops a
nonzero vacuum expectation value.  $\phi$ lives in the vacuum manifold
${\cal M}$ and winds around ${\cal M}$ in a non--trivial way on scales
greater than the correlation length, $\xi \lsim t$.  The evolution is
determined by the nonlinear dynamics of $\phi$.  When the typical size
of the defect becomes of the order of the horizon, it collapses on
itself.  The collapse continues until eventually the size of the
defect becomes of the order of $\eta^{-1}$, and at that point
the energy in gradients is large enough to raise the field from its
vacuum state.  This makes the defect unwind, leaving behind a trivial
field configuration.  As a result $\xi$ grows to about the horizon
scale, and then keeps growing with it.  As still larger scales come
across the horizon, knots are constantly formed, since the field $\phi$
points in different directions on ${\cal M}$ in different Hubble
volumes.  This is the scaling regime for textures, and when it holds
simulations show that one should expect to find of order 0.04
unwinding collapses per horizon volume per Hubble time [Turok, 1989].
However, unwinding events are not the most frequent feature [Borrill
\etal, 1994], and when one considers random field configurations
without an unwinding event the number raises to about 1 collapse per
horizon volume per Hubble time. 

%%%%-----------------------------------------------------------
\subsection{Evolution of global textures}
\label{sec-texevol} 

We mentioned earlier that the breakdown of any non--Abelian global
symmetry led to the formation of textures. The simplest possible
example involves the breakdown of a global SU(2) by a complex
doublet $\phi^a$, where the latter may be expressed as a
four--component scalar field, \ie, $a=1\ldots 4$.  We may write the
Lagrangian of the theory much in the same way as it was done in
Eq. (\ref{lagraCS}), but now we drop the gauge fields (thus the
covariant derivatives become partial derivatives).  Let us take the
symmetry breaking potential as follows, $V( \phi ) = {\lambda \over 4}
\left( |\phi|^2 - {\eta }^2 \right)^2$. The situation in which a
global SU(2) in broken by a complex doublet with this potential $V$
is equivalent to the theory where SO(4) is broken by a
four--component vector to SO(3), by making $\phi^a$ take on a vacuum
expectation value.  We then have the vacuum manifold ${\cal M}$ given
by SO(4)/SO(3) = $S^3$, namely, a three--sphere with $\phi^a\phi_a =
\eta^2$.  As $\pi_3 (S^3) \not= {\bf 1}$ (in fact, $\pi_3 (S^3)
= {\cal Z}$) we see we will have non--trivial solutions of the field
$\phi^a$ and global textures will arise.

As usual, variation of the action with respect to the
field  $\phi^a$ yields the equation of motion
\be
\label{phieqn}
{\phi^b}'' + 2 {a' \over a} {\phi^b}' - \nabla^2 \phi^b = - a^2
{\partial V \over\partial\phi^b } ~, \ee where primes denote
derivatives with respect to conformal time and $\nabla$ is computed in
comoving coordinates.  When the symmetry in broken three of the
initially four degrees of freedom go into massless Goldstone bosons
associated with the three directions tangential to the vacuum
three--sphere. The `radial' massive mode that remains ($m_\phi \sim
\sqrt{\lambda}\eta$) will not be excited, provided we
concentrate on length scales much larger than $m_\phi^{-1}$.

To solve for the dynamics of the field $\phi^b$, two different
approaches have been implemented in the literature.  The first one
faces directly the full equation (\ref{phieqn}), trying to solve it
numerically.  The alternative to this exploits the fact that, at
temperatures smaller than $T_c$, the field is constrained to live in
the true vacuum.  By implementing this fact via a Lagrange
multiplier\footnote{In fact, in the action the coupling constant
$\lambda$ of the `Mexican hat' potential is interpreted as the
Lagrange multiplier.}  we get
\be
\label{phieqnsigma}
\nabla^\mu\nabla_\mu\phi^b =
- {\nabla^\mu\phi^c\nabla_\mu\phi_c \over \eta^2} \phi^b ~~ ; ~~
\phi^2 = \eta^2  ~,
\ee                               
with $\nabla^\mu$ the covariant derivative operator.
Eq. (\ref{phieqnsigma}) represents a non--linear sigma model
for the  interaction of the three massless modes
[Rajaraman, 1982].
This last approach  is only valid when probing length scales
larger than the inverse of the mass $m_\phi^{-1}$.
As we mentioned before, when this condition is not met the
gradients of the field are strong enough to make it leave
the vacuum manifold and unwind.                   

The approach (cf. Eqs. (\ref{phieqnsigma})) is suitable for analytic
inspection. In fact, an exact flat space solution was found assuming a
spherically symmetric ansatz. This solution represents the collapse
and subsequent conversion of a texture knot into massless Goldstone
bosons, and is known as the spherically symmetric self--similar (SSSS)
exact unwinding solution.  We will say no more here with regard to the
this solution, but just refer the interested reader to the original
articles [see, \eg, Turok \& Spergel, 1990; Notzold, 1991].
Simulations taking full account of the energy stored in gradients of
the field, and not just in the unwinding events, like in
Eq. (\ref{phieqn}), were performed, for example, in [Durrer \& Zhou,
1995]. \footnote{Simulations of the collapse of `exotic' textures
can be found at {\tt http://camelot.mssm.edu/\~{}ats/texture.html}} 

%%%%-----------------------------------------------------------
\section{Currents along strings}
\label{sec-currstrings}      

In the past few years it has become clear that topological defects,
and in particular strings, will be endowed with a considerably richer
structure than previously envisaged.  In generic grand unified models
the Higgs field, responsible for the existence of cosmic strings, will
have interactions with other fundamental fields. This should not
surprise us, for well understood low energy particle theories include
field interactions in order to account for the well measured masses of
light fermions, like the familiar electron, and for the masses of
gauge bosons $W$ and $Z$ discovered at CERN in the eighties.  Thus,
when one of these fundamental (electromagnetically charged) fields
present in the model condenses in the interior space of the string,
there will appear electric currents flowing along the string core.

Even though these strings are the most attractive ones, the fact of
them having electromagnetic properties is not actually fundamental for
understanding the dynamics of circular string loops. In fact, while in
the uncharged and non current-carrying case symmetry arguments do not
allow us to distinguish the existence of rigid rotations around the
loop axis, the very existence of a small current breaks this symmetry,
marking a definite direction, which allows the whole loop
configuration to rotate.  This can also be viewed as the existence of
spinning particle--like solutions trapped inside the core.  The
stationary loop solutions where the string tension gets balanced by
the angular momentum of the charges is what Davis and Shellard [1988]
dubbed {\it vortons}.

Vorton configurations do not radiate classically.  Because they have
loop shapes, implying periodic boundary conditions on the charged
fields, it is not surprising that these configurations are
quantized. At large distances these vortons look like point masses
with quantized electric charge (actually they can have more than a
hundred times the electron charge) and angular momentum.  They are
very much like particles, hence their name.  They are however very
peculiar, for their characteristic size is of order of their charge
number (around a hundred) times their thickness, which is essentially
some fourteen orders of magnitude smaller than the classical electron
radius.  Also, their mass is often of the order of the energies of
grand unification, and hence vortons would be some twenty orders of
magnitude heavier than the electron.

But why should strings become conducting in the first place?  The
physics inside the core of the string differs somewhat from outside of
it.  In particular the existence of interactions among the Higgs field
forming the string and other fundamental fields, like that of charged
fermions, would make the latter loose their masses inside the
core. Then, only small energies would be required to produce pairs of
trapped fermions and, being effectively massless inside the string
core, they would propagate at the speed of light.
These zero energy fermionic states, also called zero modes, endow the
string with currents and in the case of closed loops they provide the
mechanical angular momentum support necessary for stabilizing the
contracting loop against collapse.

%%-------------------------------------------------------------
\subsection{Goto--Nambu Strings}
\label{subsec-gnstrings}   

Our aim now is to introduce extra fields into the problem. The simple
Lagrangian we saw in previous sections was a good approximation for
ideal structureless strings, known under the name of Goto--Nambu
strings [Goto, 1971; Nambu, 1970].  Additional fields coupled with the
string--forming Higgs field often lead to interesting effects in the
form of generalized currents flowing along the string core.

But before taking into full consideration the internal structure of
strings we will start by setting the scene with the simple Abelian
Higgs model (which describes scalar electrodynamics) in order to fix
the notation etc. This is a prototype of gauge field theory with
spontaneous symmetry breaking G = U(1) $\to$ \{1\}.  The Lagrangian reads
[Higgs, 1964]
\be 
\label{higgs964}   
{\cal L}_{_{\rm H}}
           = -{1\over 2}[{D}^\mu \Phi][{D}_\mu\Phi]^*
           - {1\over 4} (F^{(\phi)}_{\mu \nu})^2
           - {\lambda_{\phi}\over 8}(\vert\Phi\vert^2 - \eta^2)^2 , 
\ee
with gauge covariant derivative 
$D_\mu = \partial_\mu + i q A^{(\phi)}_\mu$,
antisymmetric tensor 
$F^{(\phi)}_{\mu \nu}= \nabla_{\mu}A^{(\phi)}_{\nu}-
\nabla_{\nu}A^{(\phi)}_{\mu}$ for the gauge vector field
$A^{(\phi)}_{\nu}$, and complex scalar field 
$\Phi=\vert\Phi\vert e^{i\alpha}$ with gauge coupling $q$. 

The first solutions for this theory were found by 
Nielsen \& Olesen [1973]. A couple of relevant properties are
noteworthy: 
\begin{itemize}
\item{}  
the mass per unit length for the string is $\mu = U \sim \eta^2$. 
For GUT local strings this gives $\mu \sim 10^{22}{\rm g}/{\rm cm}$, 
while one finds  
$\mu \sim \eta^2\ln(r/m_{\rm s}^{-1}) \to \infty$ if strings are
global, due to the absence of compensating gauge fields. This
divergence is in general not an issue, because global strings only
in few instances are isolated; in a string network, a natural cutoff is the
distance to the neighboring string.

\item{} 
There are essentially two characteristic mass scales (or inverse
length scales) in the problem: $m_{\rm s}\sim
\lambda_{\phi}^{1/2}\eta$ and $m_{\rm v}\sim q\eta$, corresponding to
the inverse of the Compton wavelengths of the scalar (Higgs) and vector
($A^{(\phi)}_{\nu}$) particles, respectively.

\item{}    
There exists a sort of screening of the energy, called `Higgs
screening', implying a finite energy configuration, thanks to the way in
which the vector field behaves far from the string core: 
$A_\theta\to (1 / qr) d\alpha / d\theta \,\, ,  {\rm ~for~} r\to\infty$.

After a closed path around the vortex one has $\Phi(2\pi)=\Phi(0)$,
which implies that the winding phase $\alpha$ should be an integer
times the cylindrical angle $\theta$, namely $\alpha=n\theta$. 
This integer $n$ is dubbed the `winding number'. In turn, from this fact
it follows that there exists a tube of quantized  `magnetic' flux,
given by   
\be
\Phi_{_{\rm B}}= \oint \vec A . \vec {d\ell}
= {1\over q}\int_0^{2\pi}{d\alpha\over d\theta} d\theta
= {2\pi n\over q}
\ee

\end{itemize}

In the string there is a sort of competing effect between the fields: 
the gauge field acts in a repulsive manner; the flux doesn't like to
be confined to the core and $B$ lines repel each other. On the other
hand, the scalar field behaves in an attractive way; it tries to
minimize the area where $V(\Phi)\not=0$, that is, where the field
departs from the true vacuum. 

\begin{figure}[t]
%\vspace{-2cm}
\begin{center}
\leavevmode
{\hbox %
{\epsfxsize = 8cm \epsffile{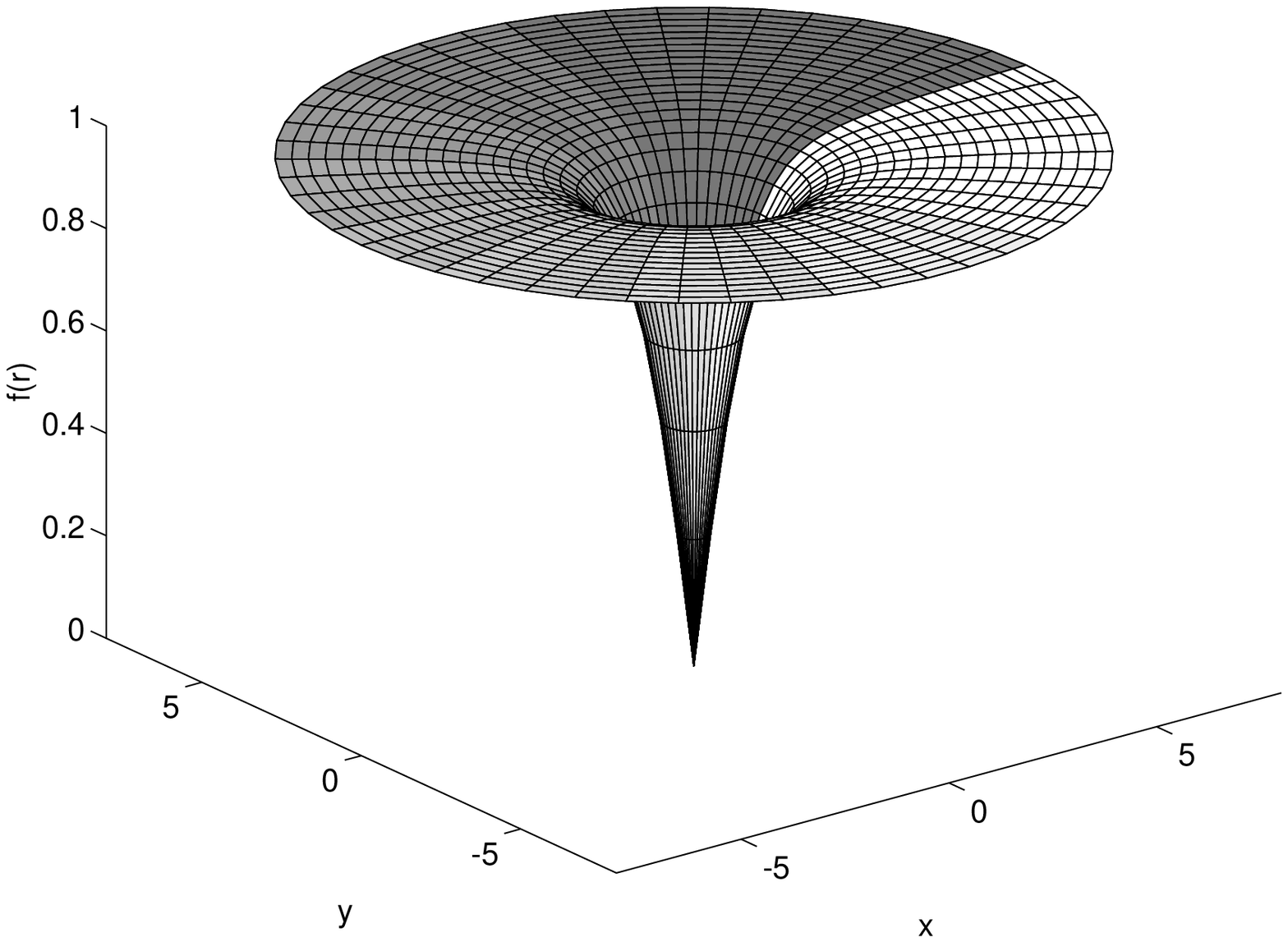} }
{\epsfxsize = 8cm \epsffile{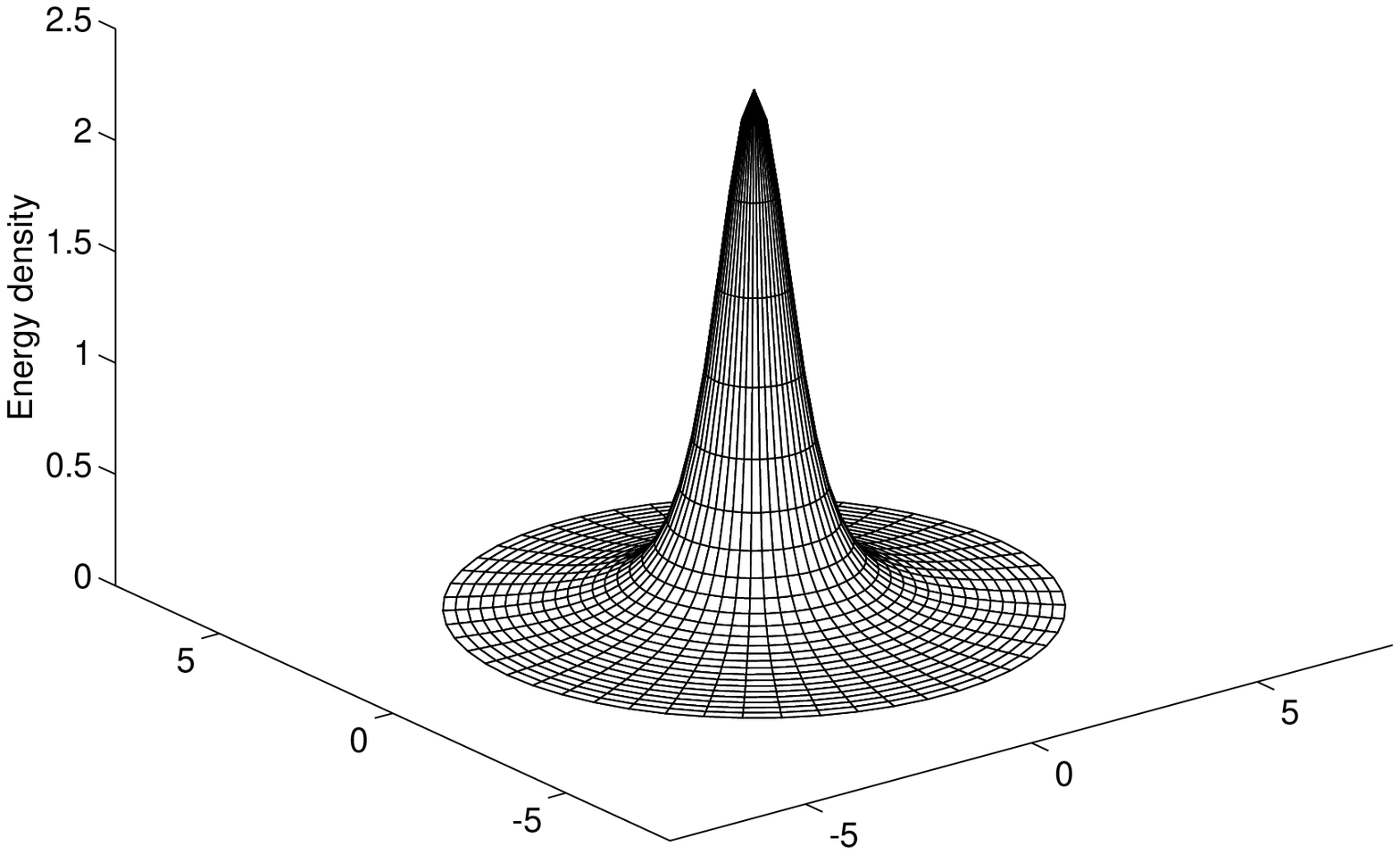} } }
\end{center}
%\vspace{-3cm}  
\caption{{\sl 
Higgs field and energy profiles for Goto--Nambu cosmic strings.
The left panel shows the amplitude of the Higgs field around the
string. The field vanishes at the origin (the false vacuum) and
attains its asymptotic value (normalized to unity in the figure) far
away from the origin. The phase of the scalar field (changing from 0
to $2\pi$) is shown by the shading of the surface. In the right panel
we show the energy density of the configuration. The maximum value is
reached at the origin, exactly where the Higgs is placed in the false
vacuum. [Hindmarsh \& Kibble, 1995].}}
\label{fig-HinKib}\end{figure}   

Finally, we can mention a few condensed--matter `cousins' of
Goto--Nambu strings: flux tubes in superconductors [Abrikosov, 1957]
for the nonrelativistic version of gauge strings ($\Phi$ corresponds
to the Cooper pair wave function). Also, vortices in superfluids, for the
nonrelativistic version of global strings ($\Phi$ corresponds to the
Bose condensate wave function).
Moreover, the only two relevant scales of the problem we mentioned
above are the Higgs mass $m_{\rm s}$ and the gauge vector mass $m_{\rm
v}$. Their inverse give an idea of the characteristic scales on which
the fields acquire their asymptotic solutions far away from the string
`location'. In fact, the relevant core widths of the string are given
by $m_{\rm s}^{-1}$ and $m_{\rm v}^{-1}$. It is the comparison of
these scales that draws the dividing line between two qualitatively
different types of solutions. If we define the parameter 
$\beta = ({m_{\rm s} / m_{\rm v}})^2$, superconductivity theory
says that $\beta <1$ corresponds to Type I behavior while
          $\beta >1$ corresponds to Type II. For us, $\beta <1$
implies that the characteristic scale for the vector field is smaller
than that for the Higgs field and so magnetic field $B$ flux lines are
well confined in the core; eventually, an $n$--vortex string with high
winding number $n$ stays stable. On the contrary, $\beta >1$ says that
the characteristic scale for the vector field exceeds that for the
scalar field and thus $B$ flux lines are not confined; 
the $n$--vortex string will eventually split into $n$ vortices of flux
$2\pi/q$. In summary:
\be
\beta = ({m_{\rm s}\over m_{\rm v}})^2
\cases {<1 {\it ~n\!\!-\!\!vortex ~stable~ (B ~flux ~lines
           ~confined ~in ~core) {\rm ~~- Type ~I}} \cr
        >1 {\it ~Unstable: ~splitting ~into ~n ~vortices ~of 
           ~flux~} 2\pi/q {\rm ~- Type ~II}}
\ee
                          
%%-------------------------------------------------------------
\subsection{Witten strings}
\label{subsec-wittstrings}   

The first model giving rise to scalar superconductivity in strings was
proposed by Witten [1985]. His is a toy Abelian U(1)$\times$U(1)
model, in which two complex scalar fields, together with their
associated gauge vector fields, interact through a term in the
potential. In a way analogous to the structureless strings, one of the
U(1) gauge groups is broken to produce standard strings.  
The other U(1) factor is the responsible for the current-carrying
capabilities of the defect. 

So, we now add a new set of terms, corresponding to a new complex
scalar field $\Sigma$, to the Lagrangian of Eq. (\ref{higgs964}).
This new scalar field will be coupled to the also new vector field
$A^{(\sigma)}_{\mu}$ (eventually the photon field), with coupling
constant $e$ ($e^2 \sim1/137$). The extra Lagrangian for the current is
\be
{\cal L}_{\rm current}
           = -{1\over 2}[{D}^\mu \Sigma][{D}_\mu\Sigma]^*
           - {1\over 4} (F^{(\sigma)}_{\mu \nu})^2
           - V_{_{\Phi,\Sigma}} 
\ee
with the additional interaction potential 
\be
\label{extrapot}
V_{{\Phi,\Sigma}}=f(\vert\Phi\vert^2 - v^2)|\Sigma|^2+
{\lambda_{\sigma}\over 4}|\Sigma|^4
\ee
and where, as usual, 
$D_\mu\Sigma = (\partial_\mu + i e A^{(\sigma)}_\mu)\Sigma$
and 
$F^{(\sigma)}_{\mu \nu}=
\nabla_{\mu}A^{(\sigma)}_{\nu} - \nabla_{\nu}A^{(\sigma)}_{\mu}$.
Remark that the complete potential term of the full theory 
under consideration now is the
sum of Eq. (\ref{extrapot}) and the potential term of
Eq. (\ref{higgs964}). The first thing one does, then, is to try and
find the minimum of this {\it full} potential $V(\Phi,\Sigma)$.
It turns out that, provided the parameters are chosen as
$\eta^2>v^2$ and 
$f^2v^4<{1\over 8}\lambda_{\phi}\lambda_{\sigma}\eta^4$,
one gets the minimum of the potential for
$\vert\Phi\vert=\eta$ and $|\Sigma|=0$.
In particular we have 
$V(|\Phi|=\eta,|\Sigma|=0) < V(|\Phi|=0,|\Sigma|\not=0)$ and 
the group U(1) associated with $A^{(\sigma)}_{\mu}$ remains unbroken.
In the case of electromagnetism, this tells us that outside of the
core, where the Higgs field takes on its true vacuum value
$|\Phi|=\eta$, electromagnetism remains a symmetry of the theory, in
agreement with the standard model. 
Hence, there exists a solution where $(\Phi,A^{(\phi)}_{\mu})$ result in
the Nielsen--Olesen vortex and where the new fields 
$(\Sigma,A^{(\sigma)}_{\mu})$ vanish.  

This is ok for the exterior region of the string, where the Higgs
field attains its true vacuum. However, inside the core we have 
$|\Phi|=0$ and the full potential reduces to
\be
V_{{\Phi=0}} = {\lambda_{\phi}\over 8} \eta^4 - f v^2 |\Sigma|^2 +
{\lambda_{\sigma}\over 4} |\Sigma|^4
\ee
Here, a vanishing $\Sigma$ is {\sl not} the value that minimizes the
potential inside the string core. On the contrary, within the string
the value $|\Sigma|=\sqrt{2f/\lambda_{\sigma}}\, v \not= 0$ is favored.
Thus, a certain nonvanishing amplitude for this new field exists in
the center of the string and slowly decreases towards the exterior, as
it should to match the solution we wrote in the previous paragraph.  
In sum, the conditions in the core favor the formation of a
$\Sigma$-condensate.  
In a way analogous to what we saw for the Nielsen--Olesen vortex, now 
the new gauge group U(1), associated with $A^{(\sigma)}_{\mu}$, {\it
is} broken. Then, it was $\Sigma=|\Sigma|e^{i\varphi}$ and now the
phase $\varphi(t,z)$ is an additional internal degree of freedom of
the theory: the Goldstone boson carrying U(1) charge (eventually,
electric charge) up and down the string. 

Let us now concentrate on the currents and field profiles. 
For the new local group U(1), the current can be computed as
\be
{\cal J}^{\mu}=
{\delta {\cal L}_{\rm current} \over \delta A^{(\sigma)}_{\mu}} =
{ i \over 2} e \Sigma^*
\stackrel {\leftrightarrow}{\partial^\mu}
\Sigma - e^2 A^{(\sigma) \mu} |\Sigma|^2
\ee
Given the form for the `current carrier' field $\Sigma$ 
we get 
\be
{\cal J}^{\mu}= e J^{\mu}  \quad {\rm with} \quad 
J^{\mu}=-|\Sigma|^2 (\partial^\mu\varphi + e A^{(\sigma) \mu})
\ee
{}From the classical Euler--Lagrange equation for $\Phi$, $J^{\mu}$ is conserved and
well-defined even in the global or neutral case (\ie, when the coupling $e=0$).

Now, let us recall the symmetry of the problem under consideration. 
The string is taken along the vertical $z$--axis and we are studying a
stationary flow of current. Hence, the current $J^{\mu}$ cannot depend 
on internal coordinates $a=t,z$ (by `internal' one generally means
internal to the worldsheet of the string).

\begin{figure}[t]
  \begin{center}
    \leavevmode
    \epsfxsize = 10cm
    \epsffile{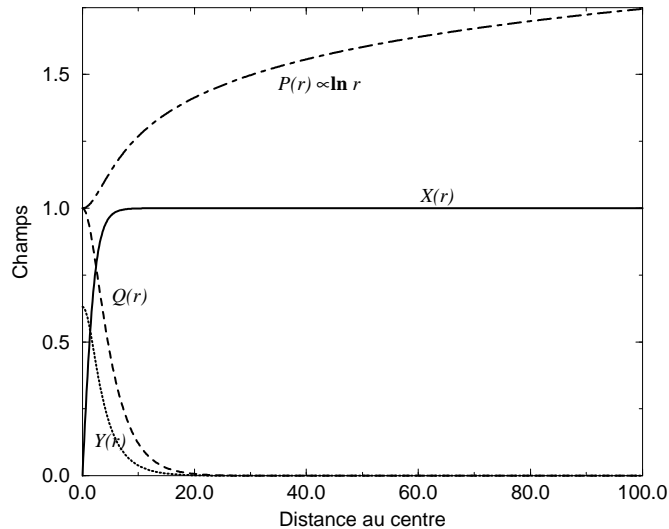}
  \end{center}
\caption{{\sl 
Profiles for the different fields around a conducting cosmic strings
[Peter, 1992]. The figure shows the Higgs field (noted with the
rescaled function X(r)), exactly as in the left panel of 
Figure \ref{fig-HinKib}. The profile Q(r) is essentially (the
$\theta$--component of) the gauge vector field $A^{(\phi)}_{\mu}$,
whose gradient helps in canceling the otherwise divergent energy
density of the (global) string and concentrates the
energy of the configuration inside a narrow core, as in the right
panel of Figure \ref{fig-HinKib}. 
The profile Y(r) is a rescaled function for the amplitude of the
current--carrier field $\Sigma$. Its form shows clearly the existence
of a boson condensate in the core of the string, signaling the flow of
a current along the string. Finally, P(r) is essentially the
electromagnetic field $A_z^{(\sigma)}$ with its standard logarithmic
divergence.}}
\label{fig-champs}
\end{figure}               

Conventionally, one takes the phase varying linearly with time and
position along the string $\varphi=\omega t - k z$ and solves the full
set of  Euler--Lagrange equations, as in Peter [1992]. In so doing,
one can write, along the core, $J^{a}=-|\Sigma|^2 P^a$ and, in turn, 
$P_a(r) = P_a(0) P(r)$ for each one of the internal coordinates, this
way separating the value at the origin of the configuration from a
common (for both coordinates) $r$--dependent solution $P$ with the
condition $P(0) = 1$. In this way, one can define the 
parameter $w$ (do not confuse with $\omega$) such that 
$w=P_z^2(0)-P_t^2(0)$ or, equivalently, $P^aP_a=wP^2$. 
Then the current satisfies $J^{a}J_{a}=|\Sigma|^4 w P^2$.

The parameter $w$ is important because from its sign one can know in
which one of a set of qualitatively different regimes we are working.
Actually, $w$ leads to the following classification [Carter, 1997]
\be
w \cases {
>0 {\it ~magnetic~ regime~} \,\,\,
\exists {\it ~reference ~frame ~where~} J^{a} {\it ~is ~pure ~spatial}
\cr
<0 {\it ~electric~ regime~} \,\,\,\,\,\,\,\,
\exists {\it ~reference ~frame ~where~} J^{a} {\it ~is ~mainly
                                ~charge ~density}
\cr                                                             
=0 {\it ~null}
}
\ee
{}From the solution of the field equations one gets the standard
logarithmic behavior for 
$P_z=\partial_z\varphi + e A_z^{(\sigma)} \propto \ln(r)$ 
far from the (long) string. This is the expected logarithmic
divergence of the electromagnetic potential around an infinite
current--carrier wire with `dc' current $I$ that gives rise to 
a magnetic field  $B^{(\sigma)} \propto 1/r$ (see Figure \ref{fig-champs}).

%%-------------------------------------------------------------
\subsection{Superconducting strings !}
\label{subsec-supstrings}        

One of the most amazing things of the strings we are now treating is
the fact that, provided some general conditions (\eg, the appropriate
relation between the free parameters of the model) are satisfied,
these objects can turn into superconductors. 
So, under the conditions that the $e A^{\mu}$ term dominates in the
expression for the current ${\cal J}^{z}$, we can write
\be
{\cal J}^{z}= - e^2 |\Sigma|^2 A^{z}
%\hspace*{0.2in}
%{\rm London}^2 {\rm ~equation ~(1935)}
\ee
which is no other than the London equation [London \& London,
1935]. From it, recalling the Faraday's law of the set of Maxwell
equations, we can take derivatives on both sides to get
\be
\partial_t {\cal J}^{z}= e^2 |\Sigma|^2 E^{ z} .
\label{persicurr}
\ee
Then, the current grows up linearly in time with an amplitude
proportional to the electric field. 
This behavior is exactly the one
we would expect for a superconductor [Tinkham, 1995].  In particular,
the equation signals the existence of persistent currents. To see it,
just compare with the corresponding equation for a wire of finite
conductivity ${\cal J}^{z}= \sigma E^{z}$. 
One clearly sees in this equation that when
the applied electric field is turned off, after a certain
characteristic time, the current stops. On the contrary,
in Eq. (\ref{persicurr}), when the electric field vanishes, the
current does not stop but stays constant, \ie, it persists flowing
along the string.

At sufficiently low temperatures certain materials undergo a phase
transition to a new (superconducting) phase, characterized notably by
the absence of resistance to the passage of currents. Unlike in these
theories, no critical temperature is invoked in here, except for the
temperature at which the condensate forms inside the string, the
details of the phase transition being of secondary importance.
Moreover, no gap in the excitation spectrum is present, unlike in the
solid--state case where the amount of energy required to excite the
system is of the order of that to form a Cooper pair, and hence the
existence of the gap.  

The very same considerations of the above paragraphs 
are valid for fermion (massless) zero
modes along the string [Witten, 1985]. In fact, a generic prediction
of these models is the existence of a maximum current above which the
current--carrying ability of the string saturates. In his pioneering
paper, Witten pointed out that for a fermion of charge $q$ and mass
in vacuum $m$, its Fermi momentum along the string should be below its
mass (in natural units). If this were not the case, \ie, if the
momenta of the fermions exceeded this maximum value, then it would be
energetically favorable for the particle to jump out of the core of the
string [Gangui \etal, 1999]. 
This implies that the current saturates and reaches a maximum
value 
\be
{\cal J}_{\rm max} \sim {q m c^2 \over 2\pi \hbar}
\ee
If we take electrons as the charge carriers, then one gets 
currents of size ${\cal J}_{\rm max}\sim$ tens of amp\`eres,
interesting but nothing exceptional (standard superconducting
materials at low temperature reach thousands of amp\`eres and more). 
On the other hand, if we focus in
the early universe and consider that the current is carried by GUT
superheavy fermions, whose normal mass would be around $10^{16}$ GeV, 
then currents more like ${\cal J}_{\rm max}\sim 10^{20}$A are
predicted. Needless to say, these currents are enormous, even by
astrophysical standards!

Und Meissner..? It has long been known that superconductors exclude
static magnetic fields from their interior. This is an effect called 
the Meissner effect, known since the 1930s and that was later
explained by the BCS (or Bardeen-Cooper-Schrieffer) theory in 1957.
One can well wonder what the situation is in our present case, \ie, do
current--carrying cosmic strings show this kind of behavior?

To answer this question, let us write Amp\`ere's law (in the
Coulomb, or radiation, gauge $\bar\nabla\cdot\bar A=0$)
\be
\nabla^2 A^{z}=-4\pi {\cal J}^{z}
\ee
Also, let us rewrite the London equation
\be
{\cal J}^{z}= - e^2 |\Sigma|^2 A^{ z}
\ee
Putting these two equations together we find 
\be
\nabla^2 A^{ z}=\lambda^{-2}A^{z}
\ee
where we wrote the electromagnetic penetration
depth $\lambda\sim(e|\Sigma(0)|)^{-1}$.

Roughly, for Cartesian coordinates, if we take $\hat x$ perpendicular
to the surface, we have $A^{z}\propto e^{-x/\lambda}$, which is
nothing but the expected exponential decrease of the vector potential 
inside the core [Meissner, 1933]. 
[to be more precise, in the string case we expect 
$\nabla^2 P_a = e^2 |\Sigma|^2 P_a$, with
$P_a=\partial_a\varphi+e A_{a}$].

For a lump of standard metal a penetration depth of roughly 
$\lambda\sim 10^{-5}$cm is ok. In the string case, however, 
\be
\lambda\sim e^{-1} |\Sigma(0)|^{-1}\sim e^{-1}
v^{-1}
\ee
which is roughly the Compton wavelength of $A_\mu$.  Now, recall that
we had $v^{-1}>\eta^{-1}$, and that $\eta^{-1}$ was the characteristic
(Compton) size of the string core.  Hence we finally get that
$\lambda$ can be bigger than the size of the string -- unlike what
happens with standard condensed--matter superconductors,
electromagnetic fields {\it can} penetrate the string core!

%%-------------------------------------------------------------
\subsection{Macroscopic string description}
\label{subsec-macrostrings}     

Let us recapitulate briefly the microphysics setting before we see its
connection with the macroscopic string description we will develop
below. We consider a Witten--type bosonic superconductivity model in
which the fundamental Lagrangian is invariant under the action of a
U(1)$\times$U(1) symmetry group. The first U(1) is spontaneously
broken through the usual Higgs mechanism in which the Higgs field
$\Phi$ acquires a non--vanishing vacuum expectation value.  Hence, at
an energy scale $m_{\rm s}\sim \lambda_{\phi}^{1/2}\eta$ (we will call
$m_{\rm s}= m$ hereafter) we are left with a network of ordinary
cosmic strings with tension and energy per unit length $T \sim U \sim
m^2$, as dictated by the Kibble mechanism.  

The Higgs field is coupled not only with its associated gauge vector
but also with a second charged scalar boson $\Sigma$, the {\it current
carrier} field, which in turn obeys a quartic potential.  A second
phase transition breaks the second U(1) gauge (or global, in the case
of neutral currents) group and, at an energy scale $\sim m_*$, the
generation of a current--carrying condensate in the vortex makes the
tension no longer constant, but dependent on the magnitude of the
current, with the general feature that $T \leq m^2 \leq U$, breaking
therefore the degeneracy of the Nambu--Goto strings (more below).  
The fact that
$|\Sigma|\neq 0$ in the string results in that either electromagnetism
(in the case that the associated gauge vector $A^{(\sigma)}_{\mu}$ is
the electromagnetic potential) or the global U(1) is spontaneously
broken in the core, with the resulting Goldstone bosons carrying
charge up and down the string.  

%%-------------------------------------------------------------
\subsubsection{Macroscopic quantities}

So, let us define the relevant macroscopic quantities needed to find
the string equation of state. For that, we have to first express the
energy momentum tensor as follows
\be
T^\mu_\nu = -2 g^{\mu\alpha}{\delta{\cal L}\over \delta g^{\alpha\nu}}
+ \delta^\mu_\nu {\cal L} .
\ee
One then calculates the macroscopic quantities internal to the string
worldsheet (recall `internal' means coordinates $t,z$)
\be
\bar T^{ab}=2\pi\int r dr T^{ab} \quad \quad 
\bar J^{a}=2\pi\int r dr J^{a} \quad \quad {\rm for} \quad a,b = t,z
\ee
The macroscopic charge density/current intensity is defined as 
\be
C=2\pi\int r dr \sqrt{|J^{a}J_{a}|} 
  =2\pi\sqrt{|w|}\int r dr |\Sigma|^2 P 
\ee      
Now, the state parameter is $\nu\equiv {\rm sgn}(w)\sqrt{|w|}$.  For
vanishing coupling $e$ we have $w\sim k^2-\omega^2$ and $\nu$ yields
the energy of the carrier (in the case $w<0$) or its momentum ($w>0$).

We get the energy per unit length $U$ and the tension of the string
$T$ by diagonalizing $\bar T^{ab}$ 

\be
U=\bar T^{tt} \quad \quad T=-\bar T^{zz}
\ee

\begin{figure}[t]
\begin{center}
\leavevmode
{\hbox %
{\epsfxsize = 8cm \epsffile{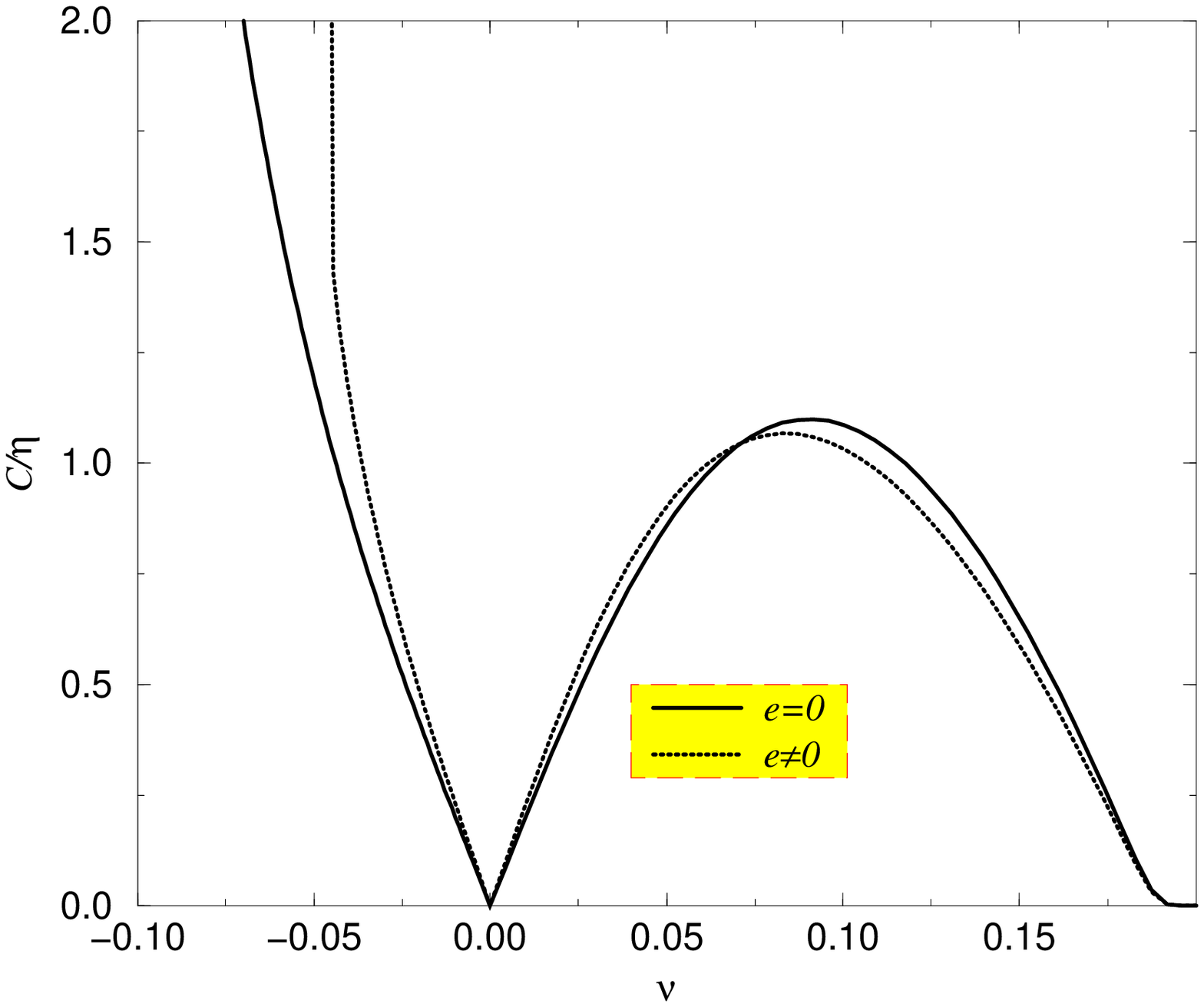} }
{\epsfxsize = 8cm \epsffile{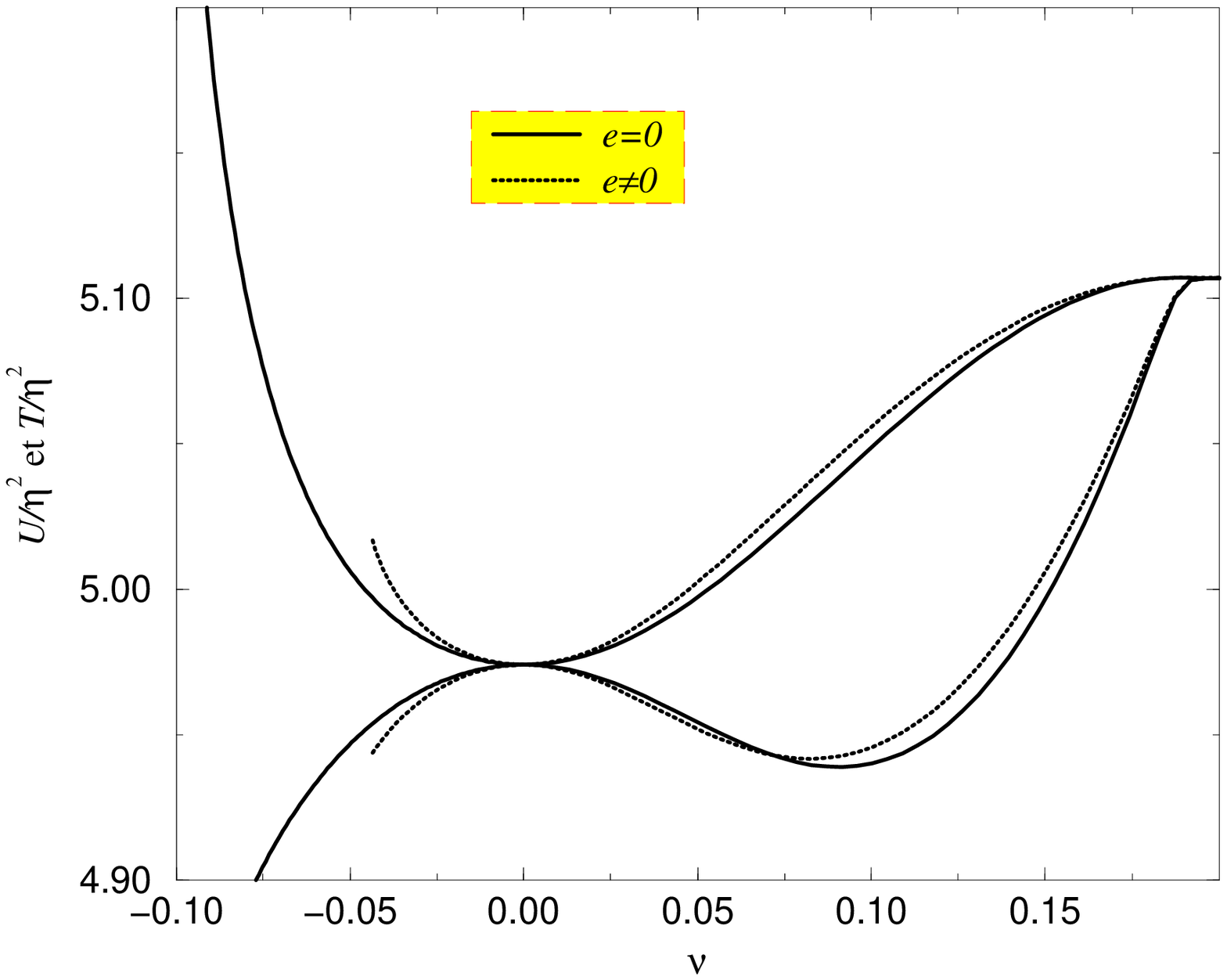} } }
\end{center}
%\vspace{-3cm}  
\caption{{\sl 
Variation of the relevant macroscopic quantities with the state parameter.  
In the left panel we show the variation of the amplitude of the
macroscopic (integrated) charge density (for $w<0$) and current
intensity (for $w>0$) along the string core versus the state
parameter, as defined by $\nu\equiv {\rm sgn}(w)\sqrt{|w|}$.
In the right panel one can see the corresponding variations of the
integrated energy per unit length (upper set of curves) and 
tension (lower set of curves) for the string. 
Both the neutral ($e=0$) and the charged cases are shown with, in the
latter case, a rather exaggerated value of the coupling, in order to
distinguish the curves in each set [Peter, 1992].}}
\label{fig-eqnstateppp}\end{figure}   

As shown in Figure (\ref{fig-eqnstateppp}) the general string dynamics
in the neutral case does not get much modified when the
electromagnetic $e$-coupling is included.
Nevertheless, a couple of main features are worth to note:
\begin{itemize}
\item{}
In the magnetic regime there is {\it saturation}. In this situation ($w>0$) 
the current intensity $C$ reaches a maximum value and, at the same
time, $T$ passes through a minimum.
\item{} 
In the electric regime there is a {\it phase frequency threshold}. In
this case ($w<0$) the charge density of the conducting string diverges 
$C\to\infty$ and the tension tends to vanish $T\to 0^+$. An analytic 
treatment shows that $C\propto(w+m_\sigma^2)^{-1}$, with
$m_\sigma^2 = 2 f (\eta^2-v^2)$. Note that this threshold changes with
the coupling, when $e$ is very large.
\item{} 
We always find $T>0$ in $w>0$ case. Hence, there is no place for {\it
springs}, a conjecture first announced by Peter [1993]. 
Note that $T$ diminishes just a few percent, and then the current saturates. 
If this were not the case, $c^2_{\rm T}=T/U$ would be negative and
this would imply instabilities [Carter, 1989]. Hence, there would be
no static equilibrium configurations.
\end{itemize} 

%%-------------------------------------------------------------
\subsubsection{Macroscopic description}        

Now, let us focus on the macroscopic string description. 
For a local U(1) we have
\be
{\cal J}^{\mu}=
{\delta {\cal L}_{\rm current} \over \delta A^{(\sigma)}_{\mu}} =
+ e J^{\mu}
\ee
[to stick to usual notation in the literature, we are now changing
$e\to -e$ in our expressions of previous sections]. 
In this equation we have the conserved Noether current
\be
J_{\mu}=|\Sigma|^2 (\partial_\mu\varphi - e A_\mu^{(\sigma)})
\ee
Now, recall that $A_\mu^{(\sigma)}$ varies little inside the core, as
the penetration depth was bigger than the string core radius. 
We can then integrate to find the macroscopic current
\be
I_a=2{\tilde {\cal K}} (\partial_a\varphi - e A_a^{(\sigma)})
= 2{\tilde {\cal K}} \varphi_{|a}
\hspace*{0.2in}
{\rm with~~~} {\tilde {\cal K}}={1\over 2}\int dx dy |\Sigma|^2
\label{currmicro}
\ee
which is well--defined even for electromagnetic coupling $e\to 0$.

The macroscopic dynamics is describable in terms of a Lagrangian
function ${\cal L}(w)$ depending only on the internal degrees of
freedom of the string. Now it is $\varphi$'s gradient that 
characterizes local state of string through 
\be
w={\kappa}_0 \gamma^{ab}\varphi_{|a}\varphi_{|b}
{\rm ~~with~~}
\gamma_{ab}=g_{\mu\nu}x^\mu_{,a}x^\nu_{,b}
\ee
where $\gamma_{ab}$ is the induced metric on the worldsheet.
The latter is given in terms of the background
spacetime metric $g_{\mu\nu}$ with respect to the 4--dimensional
background coordinates $x^\mu$ of the worldsheet. 
We use a comma to denote simple partial differentiation with respect to
the worldsheet coordinates $\xi^a$ and using Latin indices for the
worldsheet coordinates $\xi^{_1}=\sigma$ (spacelike),
$\xi^{_0}=\tau$ (timelike). As we saw above, the gauge covariant derivative
$\varphi_{|a}$ is expressible in the presence of a background
electromagnetic field with Maxwellian gauge covector 
$A_\mu^{(\sigma)}$ ($A_\mu$ hereafter) by
$\varphi_{|a}=\varphi_{,a}\! -\! eA_\mu x^\mu_{\, ,a}$.  
So, now a key r\^ole is played by the squared of
the gradient of $\varphi$ in characterizing the local state of the
string through $w$.

The dynamics of the system is determined by the Lagrangian
${\cal L}(w)$. Note there is no explicit appearance 
of $\varphi$ in ${\cal L}$. From it we get the conserved particle
current vector $z_a$, such that
\be
z^a_{\,\,;a}=0 {\rm ~~~with~~~}
z_a=-{\partial{\cal L}\over \partial(\varphi^{|a})}
\label{currmacro}
\ee

Let's define $-d{\cal L} / dw  = {1\over 2} {\cal K}^{-1}$.
Matching Eqns. (\ref{currmacro}) and (\ref{currmicro}), {\sl viz.} 
$z_a {\it (macro)} \stackrel{.}{=} I_a {\it (micro)}$
we find
\be
-{\partial{\cal L}\over \partial(\varphi^{|a})}=
-{d{\cal L}\over dw } {\partial w\over \partial(\varphi^{|a})}=
{1\over 2} {\cal K}^{-1} 2 {\kappa}_0 \varphi_{|a}=
{{\kappa}_0 \over {\cal K}}\varphi_{|a} \stackrel{.}{=}
2 {\tilde {\cal K}}\varphi_{|a}
\ee
which allows us to see the interpretation of the quantity
${\cal K}^{-1}$. 
In fact, we have ${\cal K}^{-1}\propto {\tilde {\cal K}}\propto $
{\it amplitude of $\Sigma$-condensate}.
When $w\to 0$ (null) we have ${\cal K}\to 1$.
(with ${\kappa}_0$ the zero current limit of ${\tilde {\cal K}}$).

%%-------------------------------------------------------------
\subsection{The dual formalism}   
\label{subsec-dualf}  

The usual procedure for treating a specific cosmic string dynamical
problem consists in writing and varying an action which is assumed to
be the integral over the worldsheet of a Lagrangian function depending
on the internal degrees of freedom of the worldsheet. In particular,
for the structureless string, this is taken to be the Goto--Nambu
action, \ie\ the integral over the surface of the constant string
tension. In more general cases, various functions have been suggested
that supposedly apply to various microscopic field
configurations. They share the feature that the description is
achieved by means of a scalar function $\varphi$, identified with the
phase of a physical field trapped on the string, whose squared
gradient, namely the state parameter $w$, has values which completely
determine the dynamics through a Lagrangian function ${\cal
L}(w)$. This description has the pleasant feature that it is easily
understandable, given the clear physical meaning of
$\varphi$. However, as we shall see, there are instances for which it
is not so easily implemented and for which an alternative, equally
valid, dual formalism is better adapted [Carter, 1989].

%%-------------------------------------------------------------
\subsubsection{Macroscopic equation of state}          

But first, let us concentrate on the macroscopic equation of state.  At this
point, it is clear that conducting strings have a considerably richer
structure than Goto--Nambu strings. In particular, Witten strings have
and internal structure with its own equation of state $U=U(T)$. This,
in turn, allows us to compute the characteristic perturbations speeds
[Carter, 1989] :
\begin{itemize}  
\item{}  
A transverse (wiggle) speed $c^2_{\rm T}=T/U$ for  
{\it extrinsic} perturbations of the worldsheet.
\item{}    
A longitudinal (`woggle') speed $c^2_{\rm L}=-dT/dU$ for 
{\it sound-type} perturbations within the worldsheet.
\end{itemize}  
Of course, these characteristic speeds are not defined for a
structureless Goto--Nambu string, but are fully meaningful for any
other model. 
Numerical results for Witten strings by Peter [1992] yield $c_{\rm
L}<c_{\rm T}$, \ie\ the regime is {\it supersonic}. 

We will now explore the different ans\"atze proposed in the literature
over the years. Clearly, the simplest case is that one without any
currents, namely the Goto--Nambu action. In the present formalism it
is expressed by the action
\be
S_{\rm GN}=-m^2\int \sqrt{-\gamma}d\sigma^2
\ee
which is proportional to string worldsheet area. The corresponding
Lagrangian is given simply by ${\cal L}_{\rm GN}=-m^2$ and its
equation of state results $U=T=m^2$. 

The first thing that comes to the mind when trying to extend this
simple action to the case including currents is of course to add a
small (linear) term proportional to the state parameter $w$, which
itself includes the relevant information on the currents. Hence, a 
first try would be 
${\cal L}_{\rm linear}=-m^2-{w\over 2}$. It turns out that this simple
model is also self--dual 
(with $\Lambda_{\rm linear}=-m^2-{\chi\over 2}$, to be precised below) 
and the equation of state resulting is (for both electric and magnetic regimes)
$U+T=2m^2$. However, it follows that $c_{\rm T}<c_{\rm L}=1$, \ie, the
model is {\it subsonic} and this goes at odds with the numerical
results for Witten strings.

2nd try: keeping with minimal modifications {\sl autour} the
Goto--Nambu solution, another, Kaluza--Klein inspired, model was
proposed: ${\cal L}_{\rm KK}=-m\sqrt{m^2+w}$. This model is also
self--dual and the resulting equation of state is $UT=m^4$. Moreover,
in the limit of small currents it reproduces the linear model of the
last paragraph. However, this time both characteristic perturbation
speeds are equal and smaller than unity, $c_{\rm T}=c_{\rm L}<1$, \ie\
the model is {\it transonic} and this fact disqualifies it for
modeling Witten strings.

At this point, one may think that there is an additional relevant
parameter in the theory, the scale associated with the
current--carrier mass, which we shall note $m_*$ ($=m_\sigma$). It is
only by introducing this extra mass scale that the precise numerical
solutions for Witten strings can be recovered. Two models were
proposed, the first one with

\be
{\cal L}_{\rm rational}=-m^2-{w\over 2}(1+{w\over m^2_*})^{-1}
\ee
for which we get the amplitude of the $\Sigma$--condensate 
${\cal K}^{-1}= (1+{w\over m^2_*})^{-2}$
(recall that it was ${\cal K}^{-1}\propto \int dx dy |\Sigma|^2$
and $C \propto \sqrt{|w|} \int dx dy |\Sigma|^2$).
This ansatz fits well the $w\to -m^2_\sigma$ divergence in 
the macroscopic charge density $C$ [see Figure
(\ref{fig-eqnstateppp})] and it is the best choice for spacelike currents.

The second model is given by 
\be
\label{lagrlog}
{\cal L}_{\rm log}=-m^2-{m^2_*\over 2}\ln(1+{w\over m^2_*})
\ee
and we get ${\cal K}^{-1}= (1+{w\over m^2_*})^{-1}$. This one is the
best for timelike currents and is OK for spacelike currents as well
[Carter \& Peter, 1995].

These two two--scale models we will employ below to study the dynamics
of conducting string loops and the influence of electromagnetic
self--corrections on this dynamics at first order between the current
and the self--generated electromagnetic field. But before that, let
us introduce the formal framework we need for the job.

%%-------------------------------------------------------------
\subsubsection{The dual formalism}        

Here we will derive in parallel expressions for the currents and state
parameters in two representations, which are dual to each other.  This
will not be specific to superconducting vacuum vortex defects, but is
generally valid to the wider category of elastic string models
[Carter, 1989]. In this formalism one works with a two--dimensional
worldsheet supported master function $\Lambda(\chi)$ considered as the
dual of ${\cal L}(w)$, these functions depending respectively on the
squared magnitude of the gauge covariant derivative of the scalar
potentials $\psi$ and $\varphi$ as given by
\begin{equation}
\chi ={\tilde\kappa}_{_0}\gamma^{ab}\psi_{|a }\psi_{|b} \ \
\longleftrightarrow \ \
w =\kappa_{_0}\gamma^{ab}\varphi_{|a }\varphi_{|b} \ ,
\label{state-parameters}
\end{equation}
where $\kappa_{_0}$ and ${\tilde\kappa}_{_0}$ are adjustable,
respectively positive and negative, dimensionless normalization
constants that, as we will see below, are related to each other.
The arrow in the previous equation stands to mean an exact
correspondence between quantities appropriate to each dual
representation.   

In Eq.~(\ref{state-parameters}) the scalar potentials $\psi$ and
$\varphi$ are such that their gradients are orthogonal to each other,
namely
\begin{equation}
\gamma^{ab}\varphi_{|a }\psi_{|b} = 0 \ ,
\label{ortho}
\end{equation}
implying that if one of the gradients, say $\varphi_{|a }$ is
timelike, then the other one, say $\psi_{|a}$, will be spacelike,
which explains the different signs of the dimensionless constants
$\kappa_{_0}$ and ${\tilde\kappa}_{_0}$.

Whether or not background electromagnetic and gravitational fields are
present, the dynamics of the system can be described in the two
equivalent dual representations which are
governed by the master function $\Lambda$ and the Lagrangian scalar
${\cal L}$, that are functions only of the state parameters $\chi$ and
$w$, respectively.  The corresponding conserved current vectors, $n^a$
and $z^a$, in the worldsheet, will be given according to the
Noetherian prescription
\begin{equation}
n^a=- {\partial {\Lambda}\over\partial \psi_{|a} }\ \
\longleftrightarrow \ \
z^a=- {\partial {\cal L}\over\partial \varphi_{|a} }\ .
\end{equation}
This implies
\begin{equation}
{\cal K}_{\Lambda} n^a= {\tilde\kappa}_{_0} \psi^{|a}\ \
\longleftrightarrow \ \
{\cal K} z^a=  \kappa_{_0} \varphi^{|a}\ ,
\label{zcur-}
\end{equation}
where we use the induced metric for internal index raising, and where
${\cal K}$ and ${\cal K}_{\Lambda}$ can be written as
\begin{equation}
{\cal K}_{\Lambda}^{-1} = - 2 { d{\Lambda}\over d\chi }\ \
\longleftrightarrow \ \
{\cal K}^{-1} = -2 { d{\cal L} \over dw} .
\label{calk-}
\end{equation}    
As it will turn out, the equivalence of the two mutually dual
descriptions is ensured provided the relation
\begin{equation} {\cal K}_{\Lambda} = -{\cal K}^{-1},\label{KK}
\end{equation}
holds. This means one can define ${\cal K}$ in two alternative ways,
depending on whether it is seen it as a function of $\Lambda$ or of
${\cal L}$. We shall therefore no longer use the function ${\cal
K}_{\Lambda}$ in what follows.

Based on Eq.~(\ref{ortho}) that expresses the orthogonality of the
scalar potentials we can conveniently write the relation between
$\psi$ and $\varphi$ as follows
\begin{equation}
\varphi_{|a } = {\cal K}
{\sqrt{-{\tilde\kappa}_{_0}}\over \sqrt{\kappa_{_0}}} \epsilon_{ab}
\psi^{|b} \ ,
\label{ansatzzz}
\end{equation}
where $\epsilon$ is the antisymmetric surface measure tensor (whose
square is the induced metric, $\epsilon_{ab}\epsilon^b{_ c}
=\gamma_{ac}$). From this and using Eq. (\ref{state-parameters}) we
easily get the relation between the state variables,
\begin{equation} w={\cal K}^2\chi .\label{wK} \end{equation}

Both the master function $\Lambda$ and the Lagrangian ${\cal L}$ are
related by a Legendre type transformation that gives
\begin{equation}
\Lambda={\cal L}+{\cal K}\chi\ .
\label{Lamb-} \end{equation}                
\begin{table}[t]
%\begin{table}[h]
\label{tatable}
\begin{center}
%\phantom{.}
%\vspace{1cm}
\begin{tabular}{*{5}{c}}
\multicolumn{5}{c}{}\\
\multicolumn{5}{c}{\large  Equations of state for both regimes}\\
\multicolumn{5}{c}{}\\
\hline
regime & $U$ & $T$ & $\chi$ and $w$ & current\\[0.5ex]
\hline
electric & $-\Lambda$  & $-{\cal L}$ & $< 0$ & timelike\\[0.5ex]
magnetic & $-{\cal L}$ & $-\Lambda$ & $> 0$ & spacelike\\[0.5ex]
%\hline
\end{tabular}
%\vspace{-3cm}
\caption{Values of the energy per unit length $U$ and tension $T$
depending on the timelike or spacelike character of the current,
expressed as the negative values of either $\Lambda$ or ${\cal L}$.}
%\vspace{-4cm}
\end{center}
\end{table}
The functions ${\cal L}$ and $\Lambda$ can be seen [Carter, 1997] to
provide values for the energy per unit length $U$ and the tension $T$
of the string depending on the signs of the state parameters $\chi$
and $w$. (Originally, analytic forms for these functions ${\cal L}$
and $\Lambda$ were derived as best fits to the eigenvalues of the
stress--energy tensor in microscopic field theories). The necessary
identifications are summarized in Table 1.2.
%\ref{tatable}.

This way of identifying the energy per unit length and tension with
the Lagrangian and master functions also provides the constraints on
the validity of these descriptions: the range of
variation of either $w$ or $\chi$ follows from the requirement of
local stability, which is equivalent to the demand that the squared
speeds $c_{_{\rm E}}^{\, 2}=T/U$ and $c_{_{\rm L}}^{\,2} =-dT/dU$ of
extrinsic and longitudinal (sound type) perturbations be
positive. This is thus characterized by the unique relation
\begin{equation}
{{\cal L} \over\Lambda}>0>{d{\cal L} \over d\Lambda}\ ,
\label{stabc-}
\end{equation}
which should be equally valid in both the electric and magnetic
ranges.
Having defined the internal quantities, we now turn to the actual
dynamics of the worldsheet and prove explicitly the equivalence
between the two descriptions.            

%%-------------------------------------------------------------
\subsubsection{Equivalence between ${\cal L}$ and $\Lambda$}

The dynamical equations for the string model can be
obtained either from the master function $\Lambda$ or from the
Lagrangian ${\cal L}$ in the usual way, by applying the variation
principle to the surface action integrals
\begin{equation}
{\cal S}_\Lambda = \int d\sigma\,d\tau\,\sqrt{-\gamma}\, \Lambda(\chi) , 
\label{action} 
\end{equation}
and
\begin{equation}
{\cal S} _{\cal L} = \int d\sigma\,d\tau\,\sqrt{-\gamma}\, {\cal L}(w) ,
\label{action2} 
\end{equation} 
(where $\gamma\equiv \det
\{\gamma_{ab}\}$) in which the independent variables are either the
scalar potential $\psi$ or the phase field $\varphi$ on the worldsheet
and the position of the worldsheet itself, as specified by the
functions $x^\mu\{\sigma,\tau\}$.

Independently of the detailed form of the complete system, one knows
in advance, as a consequence of the local or global $U(1)$ phase
invariance group, that the corresponding Noether currents will be
conserved, namely
\begin{equation}
\big(\sqrt{-\gamma}\, n^a\big)_{,a}=0\ \
\longleftrightarrow \ \
\big(\sqrt{-\gamma}\, z^a\big)_{,a}=0\ .
\end{equation}
For a closed string loop, this implies (by Green's theorem) the
conservation of the corresponding flux integrals
\begin{equation}
N=\oint d\xi^a \epsilon_{ab} n^b\ \
\longleftrightarrow \ \
Z=\oint d\xi^a \epsilon_{ab} z^b\ ,
\label{zin}
\end{equation}
meaning that for any circuit round the loop one will obtain the same
value for the integer numbers $N$ and $Z$, respectively.  $Z$ is
interpretable as the integral value of the number of carrier particles
in the loop, so that in the charge coupled case, the total electric
charge of the loop will be $Q=Ze$. Moreover, the angular momentum 
of the closed loop turns out to be simply $J = Z N$. 

The loop is also characterized by a second independent integer
number $N$ whose conservation is trivially obvious.
Thus we have the topologically conserved numbers defined by
\begin{eqnarray}
2\pi Z=\oint d\psi & = & \oint d\xi^a \psi_{|a} = \oint d\xi^a
\psi_{,a}
\nonumber \\
& \longleftrightarrow & \nonumber \\
2\pi N = \oint d\varphi & = & \oint d\xi^a \varphi_{|a} =
\oint d\xi^a \varphi_{,a} \ ,
\label{win}
\end{eqnarray}
where it is clear that $N$, being related to the phase of a physical
microscopic field, has the meaning of what is usually referred to as
the winding number of the string loop.  The last equalities in
Eqs.~(\ref{win}) follow just from explicitly writing the covariant
derivative $_{|a}$ and noting that the circulation integral
multiplying $A_\mu$ vanishes.  Note however that, although $Z$ and $N$
have a clearly defined meaning in terms of underlying microscopic
quantities, because of Eqs.~(\ref{zin}) and (\ref{win}), the roles of
the dynamically and topologically conserved integer numbers are
interchanged depending on whether we derive our equations from
$\Lambda$ or from its dual ${\cal L}$. 

As usual, the stress momentum energy density distributions $\hat
T^{\mu\nu}_\Lambda$ and $\hat T^{\mu\nu}_{\cal L}$ on the background
spacetime are derivable from the action by varying the actions with
respect to the background metric, according to the specifications
\begin{equation}
\hat T^{\mu\nu}_\Lambda \equiv {2\over\sqrt{-g}}{\delta{\cal
S}_\Lambda
\over \delta g_{\mu\nu}} \equiv  {2\over\sqrt{-g}}
{\partial(\sqrt{-g}\,\Lambda )\over\partial g_{\mu\nu}}, \label{tmunu}
\end{equation}
and
\begin{equation}
\hat T^{\mu\nu}_{\cal L} \equiv {2\over\sqrt{-g}}{\delta{\cal S}_{\cal
L}
\over \delta g_{\mu\nu}} \equiv  {2\over\sqrt{-g}}
{\partial(\sqrt{-g}\,{\cal L})\over\partial g_{\mu\nu}} .
\end{equation}
This leads to expressions of the standard form, \ie\ expressible as an
integral over the string itself 
\begin{equation}
\sqrt{-g}\, \hat T^{\mu\nu}=\int d\sigma\,d\tau\,\sqrt{-\gamma}\,
\delta^{(4)} \left[x^\rho - x^\rho \{\sigma,\tau \}\right]\,
\overline T{^{\mu\nu}} \end{equation}
in which the {\it surface} stress energy momentum tensors on the
worldsheet (from which the surface energy density $U$ and the string
tension $T$ are obtainable as the negatives of its eigenvalues) can be
seen to be given by                      
\begin{equation}
\overline T^{\mu\nu}_\Lambda = {\Lambda}\eta^{\mu\nu} +{\cal K}^{-1}
\omega^\mu \omega^\nu \ \
\longleftrightarrow \ \
\overline T^{\mu\nu}_{\cal L} = {\cal L}\eta^{\mu\nu} +{\cal K}
c^\mu c^\nu \ , \label{stress}
\end{equation}
where the (first) fundamental tensor of the worldsheet is given by
\begin{equation}
\eta^{\mu\nu}= \gamma^{ab} x^\mu_{,a} x^\nu_{,b} 
\end{equation}                  
and the
corresponding rescaled currents $\omega^\mu$ and $c^\mu$ 
are obtained by setting
\begin{equation}
n^\mu=\sqrt{-{\tilde\kappa}_{_0}}\,\omega^\mu \ \
\longleftrightarrow \ \
z^\mu=\sqrt{\kappa_{_0}}\, c^\mu \ .
\label{scur-}\end{equation}            

Plugging Eqs.~(\ref{scur-}) into Eqs.~(\ref{stress}), and using
Eqs.~(\ref{KK}), (\ref{wK}) and (\ref{Lamb-}), we find that the two
stress--energy tensors coincide:
\begin{equation} \overline T^{\mu\nu}_{\cal L} = \overline
T^{\mu\nu}_\Lambda \equiv \overline T^{\mu\nu}.\end{equation} This is
indeed what we were looking for since the dynamical equations for the
case at hand, namely
\begin{equation} \eta ^\rho _\mu \nabla _\rho \overline T^{\mu\nu}
=0,\end{equation} which hold for the uncoupled case, are then strictly
equivalent whether we start with the action $S_\Lambda$ or with
$S_{\cal L}$.

%%-------------------------------------------------------------
\subsubsection{Inclusion of Electromagnetic Corrections}

Implementing electromagnetic corrections [Carter, 1997b], even at the
first order, is not an easy task as can already be seen by the much
simpler case of a charged particle for which a mass renormalization is
required even before going on calculating anything in effect related
to electromagnetic field. The same applies in the current--carrying
string case, and the required renormalization now concerns the master
function $\Lambda$. However, provided this renormalization is
adequately performed, inclusion of electromagnetic corrections, at
first order in the coupling between the current and the
self--generated electromagnetic field, then becomes a very simple
matter of shifting the equation of state, everything else being left
unchanged. Let us see how this works explicitly.

Defining $K{_{\mu\nu}}^\rho \equiv \eta^\tau_\mu \eta^\sigma_\nu \nabla
_\tau \eta^\rho_\sigma$ the second fundamental tensor of the
worldsheet, the equations of motion of a charge coupled string read
\begin{equation} 
\overline T^{\mu\nu} K{_{\mu\nu}}^\rho = \perp
^{\rho\mu}F_{\mu\nu} j^\nu,
\label{eq-mot}\end{equation} where $\perp^{\rho\mu}$ is
the tensor of orthogonal projection to the worldsheet ($\perp^\rho_\mu
= g^\rho_\mu - \eta^\rho_\mu$), $F_{\mu\nu}=2\nabla_{[\mu} A_{\nu ]}$
is the external electromagnetic tensor and $j^\mu$ stands for the
electromagnetic current flowing along the string, namely in our case
\begin{equation}
j^\mu = r e z^\mu \equiv q c^\mu ,\end{equation} with $r$ the
effective charge of the current carrier in unit of the electron charge
$e$ (working here in units where $e^2 \simeq 1/137$).  

Before going on, let us explain a bit the last equations.  The above
Eq. (\ref{eq-mot}) is no other than an extrinsic equation of motion
that governs the evolution of the string worldsheet in the presence of
an external field. In fact we readily recognize the external force
density acting on the worldsheet $f_\rho = F_{\rho\mu} \, j^\mu$, just
a Lorentz--type force with $j^\mu$ the corresponding surface current.

Let us also give a simple example where the above seemingly
complicated equation of motion proves to be something very well known
to all of us. In fact, the above is the two--dimensional analogue of
Newton's second law. For a point particle of mass $m$ the Lagrangian
is ${\cal L}=-m$, which implies that its stress energy momentum tensor
is given by $\bar T^{\mu\nu} = m u^\mu u^\nu$ (with $u^\mu u_\mu =
-1$, for the unit tangent vector $u^\mu$ of the particle's
worldline). Then, the first fundamental tensor is $\eta^{\mu\nu} = - u^\mu
u^\nu$. From this it follows that the second fundamental tensor can
be constructed, giving $K_{\mu\nu}^{\,\,\,\,\,\rho}=u_\mu u_\nu \dot
u^\rho$.  Hence, the extrinsic equation of motion yields $m \dot
u^\rho = \perp^\rho_\mu f^\mu$, \ie, the external to the 
worldline force $\perp^\rho_\mu f^\mu$ being equal to
the mass times the acceleration [Carter, 1997b].

As we mentioned, we are now interested in Eq. (\ref{eq-mot}) which is
the natural generalization to two dimensions of Newton's second
law. But now we want to include self interactions.  
The self interaction electromagnetic field on the worldsheet itself
can be evaluated [Witten, 1985] and one finds
\begin{equation} A^\mu\Big|_{_{\rm string}} = \lambda j^\mu = \lambda q
c^\mu,
\label{selfA}\end{equation}
with
\begin{equation} \lambda = 2 \ln (m_\sigma \bar\Delta ),
\label{lambda}\end{equation}
where $\bar\Delta$ is an infrared cutoff scale to compensate for the
asymptotically logarithmic behavior of the electromagnetic potential
and $m_\sigma$ the ultraviolet cutoff corresponding to the effectively
finite thickness of the charge condensate, \ie, the Compton wavelength
of the current-carrier $m_\sigma^{-1}$. In the practical situation of
a closed loop, $\bar\Delta$ should at most be taken as the total length of
the loop.

\begin{figure}[t]\begin{center}\leavevmode\epsfxsize = 3.3in
\epsffile{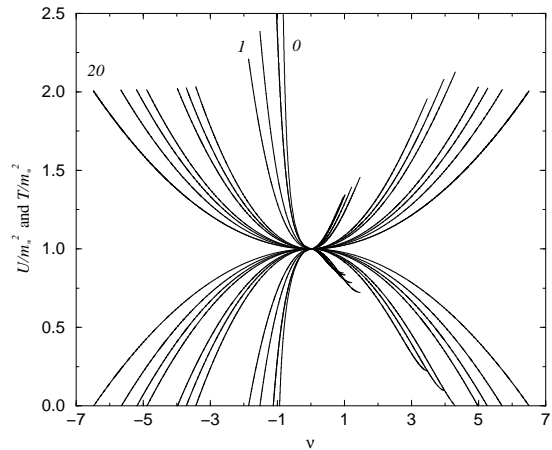} \end{center} \caption{{\sl
Variation of the equation of state with the electromagnetic
self--correction $\lambda q^2$. It relates the energy per unit length
$U$ (upper set of curves) and the tension $T$ (lower set of curves),
both in units of $m_*^2$, the current--carrier mass, and is plotted
against $\nu$, which is the (sign preserving) square root of the state
parameter $w$. Values used for this correction are in the 
set $[0, 0.1, 0.5, 1, 2, 5, 7, 8, 9, 10, 20]$, and the figure is calculated for
$\alpha=(m/m_*)^2=1$. Increasing the value of $\lambda q^2$ enlarges the
corresponding curve in such a way that for very large values (in this
particular example, it is for for $\lambda q^2 \geq 7$), the tension
on the magnetic side becomes negative before saturation is reached
[Gangui, Peter \& Boehm, 1998].}}
\label{fig-celine1}\end{figure}              

The contribution of the self field of Eq. (\ref{selfA}) in the
equations of motion~(\ref{eq-mot}) was calculated by Carter [1997b]
and the result is interpretable as a renormalization of the stress
energy tensor. That is, the result including electromagnetic
corrections is recovered if, in Eq.~(\ref{tmunu}), one uses
\begin{equation} \Lambda \to
\Lambda + {1\over 2} \lambda q^2 \chi \label{start}
\end{equation}
instead of $\Lambda$.  So, electromagnetic corrections are simply
taken into account in the dual formalism employing the master function
$\Lambda(\chi)$ unlike the case if we used ${\cal L}(w)$. In fact, it
is not always possible to invert the above relation to get an
appropriate replacement for the Lagrangian.  That the correction
enters through a simple modification of $\Lambda(\chi)$ and not of
${\cal L}(w)$ is understandable if one remembers that $\chi$ is the
amplitude of the current, so that a perturbation in the
electromagnetic field acts on the current linearly, so that an
expansion in the electromagnetic field and current yields, to first
order in $q$, $\Lambda \to \Lambda + {1\over 2}j_\mu A^\mu$, which
transforms easily into Eq.~(\ref{start}).

One example of the implementation of the above formalism is the study
of circular conducting cosmic string loops [Carter, Peter \& Gangui,
1997]. In fact, the mechanics of strings developed above allows a
complete study of the conditions under which loops endowed with
angular momentum will present an effective centrifugal potential
barrier. Under certain conditions, this barrier will prevent the loop 
collapse and, if saturation is avoided, one would expect that loops
will eventually radiate away their excess energy and settle down into 
a vorton type equilibrium state. 

If this were the whole story then we would of course be in a big
problem, for these vortons, as stable objects, would not decay and
would most probably be too abundant to be compatible with the standard
cosmology.  It may however be possible that in realistic models of
particle physics the currents could not survive subsequent phase
transitions so that vortons could dissipate.  Another way of getting
rid of (at least some of) the excess of abundance of these objects is
to take account of the electromagnetic self interactions in the
macroscopic state of the conducting string: as we said above, the
electromagnetic field in the vicinity of the string will interact with
the very same string current that generated it, with the resulting
effect of modifying its macroscopic equation of state (see Figure
\ref{fig-celine1}). These modifications make a departure of the
resulting vorton distribution from that expected otherwise,
diminishing their relic abundance.

%%-------------------------------------------------------------
\subsection{The Future of the Loops}
\label{subsec-futloops}      

Loops are formed through string interactions. Their shape is arbitrary
and they will (like their progenitors) move relativistically and emit
gravitational radiation. This will make the loops shrink while the
currents (the rotation of the current carriers), initially weak, will
begin to affect the dynamics at some point.  Also, the string tension
will try to minimize the bending, leading to a final state of a
circular and rotating ring.

\begin{figure}[t]\begin{center}\leavevmode\epsfxsize = 3.3in 
\epsffile{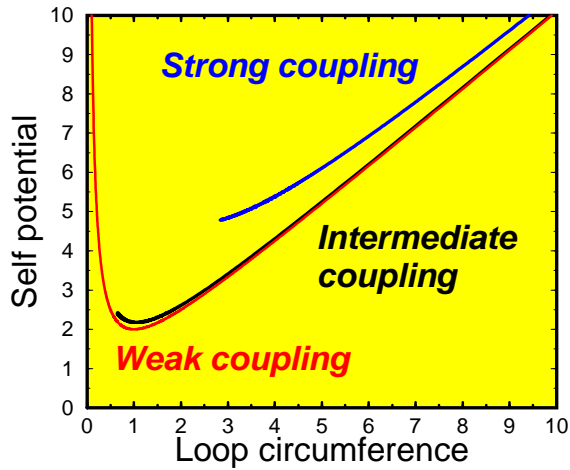} \end{center} \caption{{\sl 
Variations of the self potential $\Upsilon$ with the ring's
circumference $\ell=2\pi r$ and the electromagnetic self coupling 
($\lambda q^2$ in the text). 
The red curve stands for various values of
$\lambda q^2 < 0.1$ for which they are indistinguishable, and in the
``safe'' zone vorton--forming case; the minimum value of $\Upsilon$ is then $M$,
the vorton mass. $\Upsilon$ for $\lambda q^2 =1$ is represented as the
black line, where it is clear that we now are in a zone where the
potential has a minimum (new value for $M$) but now terminates at
some point. Finally the blue curve represents
the potential for $\lambda q^2 = 10$, an unrealistically large value,
and this time the curve terminates even before reaching a minimum:
this is a situation in which all loops with such parameters will
eventually decay [Gangui, Peter \& Boehm, 1998].}}
\label{fig-potloops}\end{figure}                    

Once a string loop has reached the state of a ring, it still has to be
decided whether it'll become a vorton (an equilibrium configuration)
or not.  In [Carter, Peter \& Gangui, 1997] and [Gangui, Peter \&
Boehm, 1998] we studied the dynamics of circular rings, including the
possibility that the current be charged, so that even the
contributions of the electric and magnetic fields surrounding and
generated by the string were considered.  This dynamics was
describable in terms of a very limited number of variables, namely the
ring total mass $M$, its rotation velocity, and the number of charges
it carried. Given this, it was found that a typical loop of radius $r$
lives in a potential $\Upsilon(r)$ whose functional form depends on
the Lagrangian ${\cal L}$ of Eq. (\ref{lagrlog}) and looks like the 
one shown in Figure \ref{fig-potloops} 
\be
M^2 \dot r^2 = M^2 - \Upsilon^2 .
\ee
{}From this the force it exerts onto itself can be derived.  What is
represented there is the force strength, in arbitrary units of energy,
exerted on the loop by itself as a function of its circumference.

The loop evolution follows that of the potential: it first goes down
(therefore shrinking) until it reaches the valley in the bottom of
which the force vanishes, then its inertia makes it climb up again on
the opposite direction where the force now tends to stop its
shrinking (centrifugal barrier). 
At this point, two possibilities arise, depending on the
initial mass available. Either this mass is not too big, less than the
value of the energy where the potential ends (see the black curve), or
it exceeds this energy: in the former case, the loop will bounce back
and eventually oscillate around the equilibrium position at the bottom
of the valley (in order to stabilize itself there, the loop will loose
some energy in the form of radiation); in the latter case, it will
shrink so much that its size will eventually approach the limit (its
Compton length) where quantum effects will disintegrate away
the ring into a burst of particles.
Note the divergence for very large values of the radius $r$. This is
nothing but the evidence that an infinite amount of energy is needed
to enlarge infinitely the loop, a sort of confinement effect.  In the
Figure \ref{fig-potloops} we also see how the magnitude of the
electromagnetic corrections, when strings are coupled with electric
and magnetic fields, tends to reduce the number of surviving vortons:
a stable configuration (red line) for a weak coupling may become
unstable and collapse if its initial mass is too big for intermediate
couplings (black line), or it will do so regardless of its mass in the
strong coupling case (blue line).

%%%%-----------------------------------------------------------
\section{Structure formation from defects}
\label{sec-lls}      

%%-------------------------------------------------------------
\subsection{Cosmic strings}
\label{sec-llsstrings}       

In this section we will provide just a quick description of the
remarkable cosmological features of cosmic strings.
Many of the proposed observational tests for the existence of cosmic
strings are based on their gravitational interactions.  In fact, the
gravitational field around a straight static string is very unusual
[Vilenkin, 1981].
As is well known, the Newtonian limit of Einstein field equations with
source term given by $T^\mu_\nu = {\rm diag}(\rho, -p_1, -p_2, -p_3)$
in terms of the Newtonian potential $\Phi$ is given by
$\nabla^2\Phi = 4\pi G (\rho + p_1 + p_2 + p_3)$, just a statement of
the well known fact that pressure terms also contribute to the
`gravitational mass'. For an infinite string in the $z$--direction one
has $p_3 = -\rho$, \ie, strings possess a large relativistic
tension (negative pressure). Moreover, averaging on the string core results in vanishing
pressures for the $x$ and $y$ directions yielding $\nabla^2\Phi = 0$
for the Poisson equation. This indicates that space is flat outside of
an infinite straight cosmic string and therefore test particles in its
vicinity should not feel any gravitational attraction. 

In fact, a full general relativistic analysis confirms this and test
particles in the space around the string feel no Newtonian attraction;
however there exists something unusual, a sort of wedge missing from
the space surrounding the string and called the `deficit angle',
usually noted $\Delta$, that makes the topology of space around the
string that of a cone.
To see this, consider the metric of a source with energy--momentum
tensor [Vilenkin 1981, Gott 1985]
\be T_\mu ^\nu = \delta (x) \delta (y) {\rm diag}(\mu ,0,0,T) \ . \ee
In the case with $T= \mu$ (a rather simple equation of state) 
this is the effective energy--momentum tensor of an
unperturbed string with string tension $\mu$ as seen from distances
much larger than the thickness of the string (a Goto--Nambu string). 
However, real strings develop small--scale structure and are therefore
not well described by the Goto--Nambu action. When perturbations are
taken into account $T$ and $\mu$ are no longer equal and can only be
interpreted as effective quantities for an observer who cannot resolve
the perturbations along its length. And in this case we are left without
an effective equation of state. Carter [1990] has proposed that these
`noisy' strings should be such that both its speeds of propagation of
perturbations coincide. Namely, 
the transverse (wiggle) speed $c_{\rm T}=(T/\mu)^{1/2}$ for
extrinsic perturbations should be equal to 
the longitudinal (woggle) speed $c_{\rm L}=(-dT/d\mu)^{1/2}$ for
sound--type perturbations. 
This requirement yields the new equation of state    
\be \mu T = \mu_0^2 \ee 
and, when this is satisfied, it 
describes the energy-momentum tensor of a wiggly string as seen by an
observer who cannot resolve the wiggles or other irregularities along
the string [Carter 1990, Vilenkin 1990].

\begin{figure}[t]\begin{center}\leavevmode
{\hbox %
{\epsfxsize = 7cm \epsffile{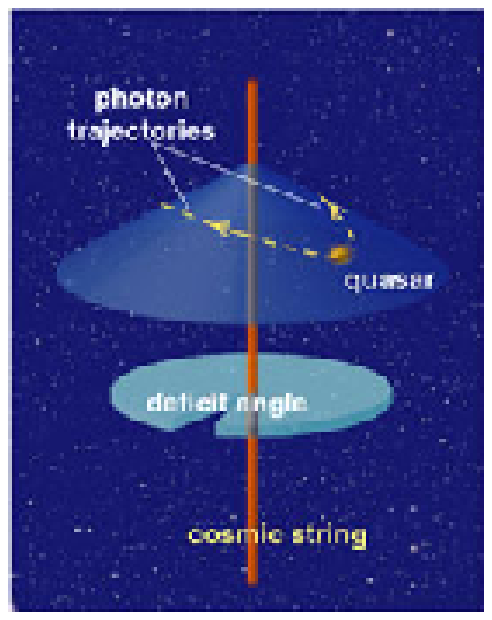} }
{\epsfxsize = 5.6cm \epsffile{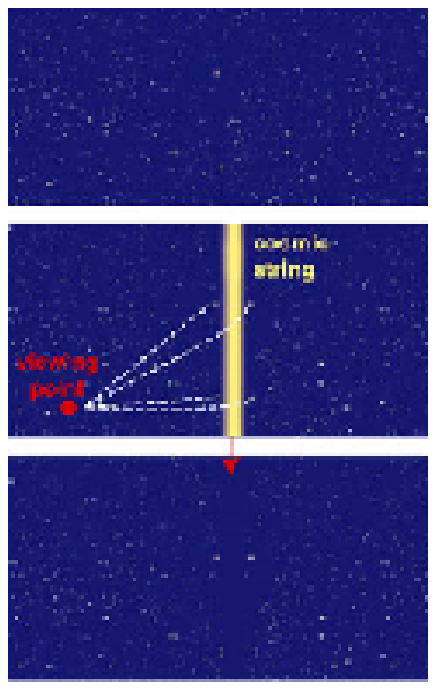}}}
%{\epsfxsize = 7cm \epsffile{fig5a.ps} }
%{\epsfxsize = 5.6cm \epsffile{fig5b.ps}}}
\end{center}
\caption{{\sl Cosmic strings affect surrounding spacetime by removing
a small angular wedge, creating a conelike geometry (left). Space
remains flat everywhere, but a circular path around the string
encompasses slightly less than 360 degrees. The deficit angle is tiny,
about $10^{-5}$ radian. To an observer, the presence of a cosmic
string would be betrayed by its effect on the trajectory of passing
light rays, which are deflected by an amount equal to the deficit
angle. The resultant gravitational lensing reveals itself in the
doubling of images of objects behind the string (right panel).}}
\label{fig-fig5ab}\end{figure}   
%FIGURE

The gravitational field around the cosmic string [neglecting terms of
order $(G\mu)^2$] is found by solving the linearized Einstein
equations with the above $T_\mu ^\nu$. One gets
\be
h_{00} = h_{33} = 4G(\mu -T) \ln(r/r_0 ) ,
\label{h00}
\ee
\be
h_{11} = h_{22} = 4G(\mu +T) \ln(r/r_0 ) ,
\ee
where $h_{\mu \nu} = g_{\mu \nu} - \eta _{\mu \nu}$ is the metric
perturbation, the radial distance from the string is 
$r = (x^2 + y^2 )^{1/2}$, and $r_0$ is a constant of
integration.                                                    

For an ideal, straight, unperturbed string, the tension and mass per
unit length are $T = \mu = \mu_0$ and one gets
\be
h_{00} = h_{33} = 0, \ \ \
h_{11} = h_{22} = 8G\mu_0 \ln(r/r_0 ) .
\ee
By a coordinate transformation one can bring this metric to a locally flat form
\be
ds^2 = dt^2 - dz^2 - dr^2 - (1-8 G\mu_0 ) r^2 d\phi ^2 ,
\ee
which describes a conical and flat (Euclidean) space with a wedge of
angular size $\Delta = 8\pi G \mu_0$ (the deficit angle) removed from
the plane and with the two faces of the wedge identified.

%%-------------------------------------------------------------
\subsubsection{Wakes and gravitational lensing}

We saw above that test particles\footnote{If one takes into account
the own gravitational field of the particle living in the spacetime
around a cosmic string, then the situation changes. In fact, the
presence of the conical `singularity' introduced by the string
distorts the particle's own gravitational field and results in the
existence of a weak attractive force proportional to $G^2\mu m^2/r^2$,
where $m$ is the particle's mass [Linet, 1986].} at rest in the
spacetime of the straight string experience no gravitational force,
but if the string moves the situation radically changes. Two particles
initially at rest while the string is far away, will suddenly begin
moving towards each other after the string has passed between
them. Their head--on velocities will be proportional to $\Delta$ or,
more precisely, the particles will get a boost $v = 4\pi G\mu_0 v_s
\gamma$ in the direction of the surface swept out by the string.
Here, $\gamma = (1-v_s^2)^{-1/2}$ is the Lorentz factor and $v_s$ the
velocity of the moving string.  Hence, the moving string will built up
a {\sl wake} of particles behind it that may eventually form the
`seed' for accreting more matter into sheet--like structures [Silk \&
Vilenkin 1984].

\begin{figure}[t]\begin{center}\leavevmode \epsfxsize = 5cm
\epsffile{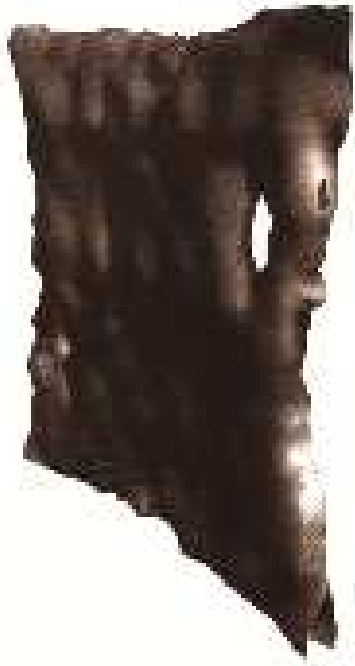} \end{center} \caption{{\sl By deflecting
the trajectory of ordinary matter, strings offer an interesting means
of forming large-scale structure. A string sweeping through a
distribution of interstellar dust will draw particles together in its
wake, giving them lateral velocities of a few kilometers per
second. The trail of the moving string will become a planar region of
high-density matter, which, after gravitational collapse, could turn
into thin, sheetlike distributions of galaxies [Image courtesy of
Pedro Avelino and Paul Shellard].}}
\label{fig-wake}\end{figure} 
%FIGURE

Also, the peculiar topology around the string makes it act as a
cylindric gravitational lens that may produce double images of distant
light sources, \eg, quasars.  The angle between the two images
produced by a typical GUT string would be $\propto G\mu$ and of order
of a few seconds of arc, independent of the impact parameter and with
no relative magnification between the images [see Cowie \& Hu, 1987,
for a recent observational attempt].

The situation gets even more interesting when we allow the string to
have small--scale structure, which we called wiggles above, as in fact
simulations indicate. Wiggles not only modify the string's effective
mass per unit length, $\mu$, but also built up a Newtonian
attractive term in the velocity boost inflicted on nearby test
particles. To see this, let us consider the formation of a wake behind
a moving wiggly string. Assuming the string moves along the $x$--axis,
we can describe the situation in the rest frame of the string. In this
frame, it is the particles that move, and these flow past the string
with a velocity $v_s$ in the opposite direction. Using conformally
Minkowskian coordinates we can express the relevant components of the
metric as
\be
ds^2 = (1+h_{00})[dt^2 - (dx^2 + dy^2)] ,
\ee
where the missing wedge is reproduced by identifying the half-lines
$y=\pm 4 \pi G \mu x$, $x \ge 0$.  
The linearized geodesic equations in this metric can be written as
\be
2 \ddot x = - ( 1- { \dot x }^2 - {\dot y} ^2 ) \partial _x h_{00} ,
\ee
\be
2 \ddot y = - ( 1- { \dot x }^2 - {\dot y} ^2 ) \partial _y h_{00} ,
\ee
where over--dots denote derivatives with respect to $t$.   
Working to first order in $G\mu$, the second of these equations can be
integrated over the unperturbed trajectory $x = v_s t$, $y = y_0$. 
Transforming back to the frame in which the string has a velocity
$v_s$ yields the result for the velocity impulse in the $y$--direction after the
string has passed [Vachaspati \& Vilenkin, 1991; Vollick, 1992]
\be
v = - {{2\pi G (\mu -T)} \over {v_s \gamma}}
             - 4 \pi G \mu v_s \gamma
\ee

The second term is the velocity impulse due to the conical deficit
angle we saw above. This term will dominate for large string
velocities, case in which big planar wakes are predicted.  In this
case, the string wiggles will produce inhomogeneities in the wake and
may easy the fragmentation of the structure. The `top--down' scenario
of structure formation thus follows naturally in a universe with
fast-moving strings.  On the contrary, for small velocities, it is the
first term that dominates over the deflection of particles. The origin
of this term can be easily understood 
[Vilenkin \& Shellard, 2000]. From Eqn. (\ref{h00}), the gravitational force on a
non--relativistic particle of mass $m$ is $F  \sim m G(\mu - T) /r$. A
particle with an impact parameter $r$ is exposed to this force for a
time $\Delta t \sim r/v_s$ and the resulting velocity is $v \sim
(F/m) \Delta t \sim G(\mu - T) / v_s$.

%%-------------------------------------------------------------
\subsection{Textures}
\label{sec-llstextures}       

During the radiation era, and when the correlation length is already
growing with the Hubble radius, the texture field has energy density
$\rho_{texture}\sim (\nabla\phi)^2 \sim \eta^2 / H^{-2}$, and remains
a fixed fraction of the total density $\rho_{c} \sim t^{-2}$ yielding
$\Omega_{texture} \sim G \eta^2$. This is the scaling behavior for
textures and thus we do not need to worry about textures dominating
the universe.

But as we already mentioned, textures are unstable to collapse, and
this collapse generates perturbations in the metric of spacetime that
eventually lead to large scale structure formation. These
perturbations in turn will affect the photon geodesics leading to CMB
anisotropies, the clearest possible signature to probe the existence
of these exotic objects being the appearance of hot and cold {\sl
spots} in the microwave maps.  Due to their scaling behavior, the
density fluctuations induced by textures on any scale at horizon
crossing are given by $(\delta\rho / \rho )_H \sim G \eta^2$.  CMB
temperature anisotropies will be of the same amplitude.
Numerically--simulated maps, with patterns smoothed over $10^\circ$
angular scales, by Bennett \& Rhie [1993] yield, upon normalization to
the {\sl COBE}--DMR data, a dimensionless value $G \eta^2 \sim
10^{-6}$, in good agreement with a GUT phase transition energy
scale. It is fair to say, however, that the texture scenario is having
problems in matching current data on smaller scales [see, \eg, Durrer,
2000].

%%%%-----------------------------------------------------------
\section{CMB signatures from defects}
\label{sec-cmbdefects}      

\begin{figure}[t]
%\vspace{-2cm}
\begin{center}
\leavevmode
{\hbox %
%{\epsfxsize = 7cm \epsffile{joao2_9702131.ps} }
{\epsfxsize = 7cm \epsffile{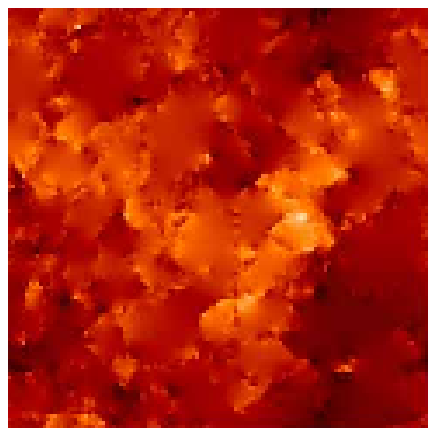} }
{\epsfxsize = 7cm \epsffile{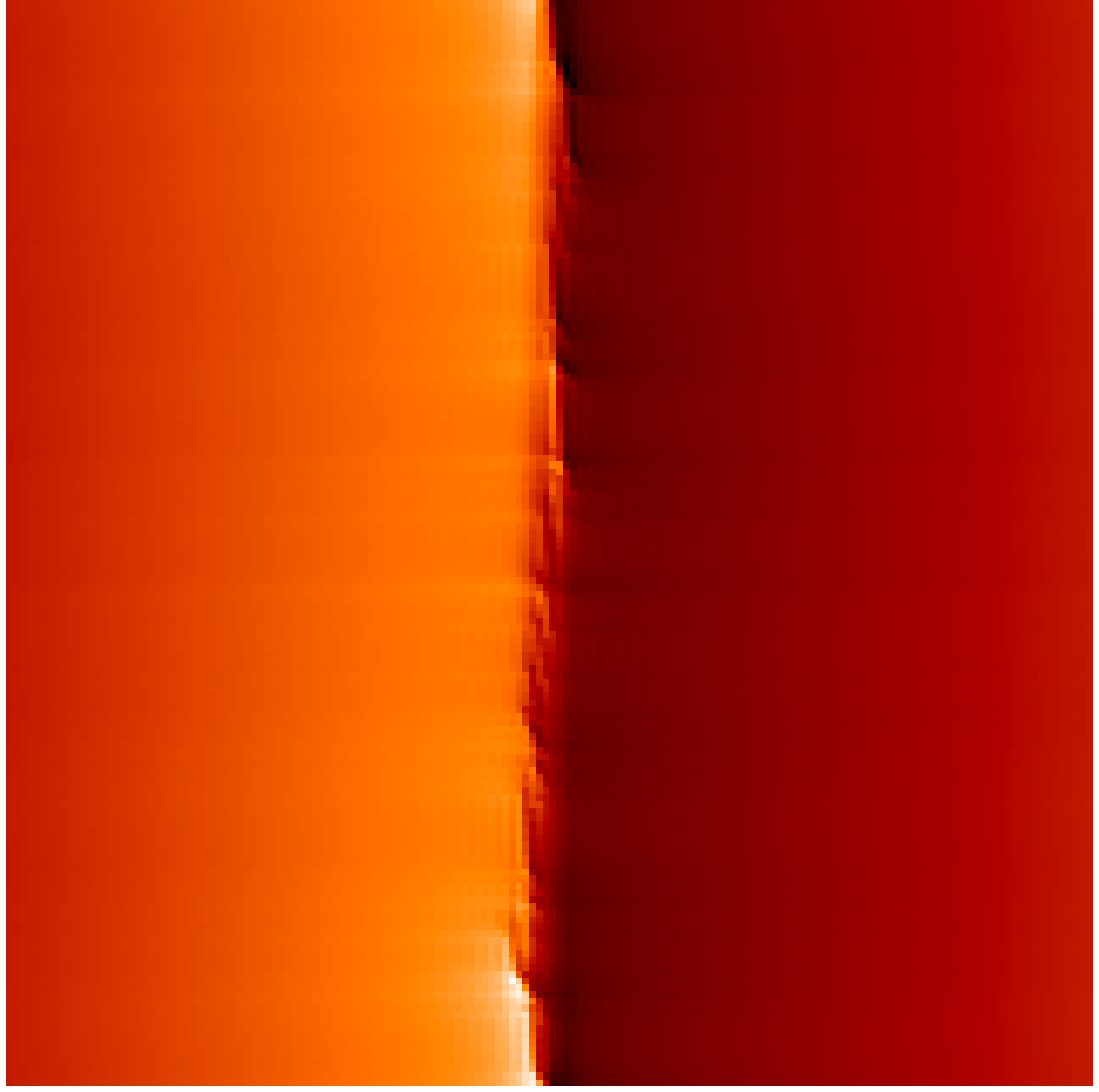} } }
\end{center}
%\vspace{-3cm}  
\caption{{\sl The Kaiser-Stebbins effect for cosmic strings. 
A string network evolves into a self-similar scaling regime,
perturbing matter and radiation during its evolution. The effect on
the CMB after recombination leads to distinct steplike
discontinuities on small angular scales that were first studied 
by Kaiser \& Stebbins [1984]. The left panel shows a simulated patch
of the sky that fits in one of the pixels of the COBE experiment.
Hence, higher resolution observatories are needed in order to detect
strings. The right panel shows a patch on the CMB sky of order 20'
across. However, recent studies indicate that this clean tell-tale signal gets
obscured at subdegree angular scales due to the temperature
fluctuations generated before recombination.
[Magueijo \& Ferreira 1997].}}
\label{fig-KSs}\end{figure}

If cosmic defects have really formed in the early universe and some of
them are still within our present horizon today, the anisotropies in
the CMB they produce would have a characteristic signature.  Strings,
for example, would imprint the background radiation in a very
particular way due to the Doppler shift that the background radiation
suffers when a string intersects the line of sight.  The conical
topology of space around the string will produce a differential redshift of
photons passing on different sides of it, resulting in step--like
discontinuities in the effective CMB temperature, given by 
$\D \approx 8 \pi G \mu v_s
\gamma$ with, as before, $\gamma = (1-v_s^2)^{-1/2}$ the Lorentz
factor and $v_s$ the velocity of the moving string.  This `stringy'
signature was first studied by Kaiser \& Stebbins [1984] and Gott
[1985] (see Figure \ref{fig-KSs}).

Anisotropies of the CMB are directly related to the origin of
structure in the universe. Galaxies and clusters of galaxies
eventually formed by gravitational instability from primordial density
fluctuations, and these same fluctuations left their imprint on the
CMB. Recent balloon [de Bernardis, \etal, 2000; Hanany, \etal, 2000]
and ground-based interferometer [Halverson, \etal, 2001] experiments
have produced reliable estimates of the power spectrum of the CMB
temperature anisotropies.  While they helped eliminate certain
candidate theories for the primary source of cosmic perturbations, the
power spectrum data is still compatible with the theoretical estimates
of a relatively large variety of models, such as $\Lambda$CDM,
quintessence models or some hybrid models including cosmic defects.

There are two main classes of models of structure formation
--\textit{passive} and \textit{active} models. In passive models,
density inhomogeneities are set as initial conditions at some early
time, and while they subsequently evolve as described by
Einstein--Boltzmann equations, no additional perturbations are
seeded. On the other hand, in active models the sources of density
perturbations are time--dependent.

All specific realizations of passive models are based on the idea of
inflation. In simplest inflationary models it is assumed that there
exists a weakly coupled scalar field $\phi$, called the inflaton,
which ``drives'' the (quasi) exponential expansion of the
universe. The quantum fluctuations of $\phi$ are stretched by the
expansion to scales beyond the horizon, thus ``freezing'' their
amplitude.  Inflation is followed by a period of thermalization,
during which standard forms of matter and energy are formed. Because
of the spatial variations of $\phi$ introduced by quantum
fluctuations, thermalization occurs at slightly different times in
different parts of the universe. Such fluctuations in the
thermalization time give rise to density fluctuations. Because of
their quantum nature and because of the fact that initial
perturbations are assumed to be in the vacuum state and hence well
described by a Gaussian distribution, perturbations produced during
inflation are expected to follow Gaussian statistics to a high degree
[Gangui, Lucchin, Matarrese \& Mollerach, 1994], or either be products
of Gaussian random variables. This is a fairly general prediction that
will be tested shortly with MAP and more thoroughly in the future with
Planck.\footnote{Useful CMB resources can be found at 
{\tt http://www.mpa-garching.mpg.de/\~{}banday/CMB.html}}

Active models of structure formation are motivated by cosmic
topological defects with the most promising candidates being cosmic
strings. As we saw in previous sections, it is widely believed that
the universe underwent a series of phase transitions as it cooled down
due to the expansion. If our ideas about grand unification are
correct, then some cosmic defects should have formed during phase
transitions in the early universe. Once formed, cosmic strings could
survive long enough to seed density perturbations.  Defect models
possess the attractive feature that they have no parameter freedom, as
all the necessary information is in principle contained in the
underlying particle physics model. Generically, perturbations produced
by active models are not expected to be Gaussian distributed
[Gangui, Pogosian \& Winitzki, 2001a].

%%-------------------------------------------------------------
\subsection{CMB power spectrum from strings}
\label{sec-powerspectrum}      

The narrow main peak and the presence of the second and the third
peaks in the CMB angular power spectrum, as measured by BOOMERANG,
MAXIMA and DASI [de Bernardis, \etal, 2000; Hanany, \etal, 2000;
Halverson, \etal, 2001], is an evidence for coherent oscillations of
the photon--baryon fluid at the beginning of the decoupling epoch
[see, \eg, Gangui, 2001]. While such coherence is a property of all passive
model, realistic cosmic string models produce highly incoherent
perturbations that result in a much broader main peak. This excludes
cosmic strings as the primary source of density fluctuations unless
new physics is postulated, \textit{e.g.}~models with a varying speed of light
[Avelino \& Martins, 2000]. In addition to purely active or passive models, it
has been recently suggested that perturbations could be seeded by some
combination of the two mechanisms. For example, cosmic strings could
have formed just before the end of inflation and partially contributed
to seeding density fluctuations. It has been shown [Contaldi, \etal,
1999; Battye \& Weller, 2000; Bouchet, \etal, 2001] that such hybrid
models can be rather successful in fitting the CMB power spectrum
data.

Calculating CMB anisotropies sourced by topological defects is a
rather difficult task. In inflationary scenario the entire information
about the seeds is contained in the initial conditions for the
perturbations in the metric.  In the case of cosmic defects,
perturbations are continuously seeded over the period of time from the
phase transition that had produced them until today. The exact
determination of the resulting anisotropy requires, in principle, the
knowledge of the energy--momentum tensor [or, if only two point
functions are being calculated, the unequal time correlators, Pen,
Seljak, \& Turok, 1997] of the defect network and the products of its
decay at all times. This information is simply not available! Instead,
a number of clever simplifications, based on the expected properties
of the defect networks ({\it e.g.} scaling), are used to calculate the
source. The latest data from BOOMERANG and MAXIMA experiments clearly
disagree with the predictions of these simple models of defects
[Durrer, Gangui \& Sakellariadou, 1996].

The shape of the CMB angular power spectrum is determined by three
main factors: the geometry of the universe, coherence and causality.
The curvature of the universe directly affects the paths of light rays
coming to us from the surface of last scattering. In a closed
universe, because of the lensing effect induced by the positive
curvature, the same physical distances between points on the sky would
correspond to larger angular scales. As a result, the peak structure
in the CMB angular power spectrum would shift to larger angular scales
or, equivalently, to smaller values of the multipoles $\ell$'s.

\begin{figure}[t]
\begin{center}\leavevmode \epsfxsize = 7cm
\epsffile{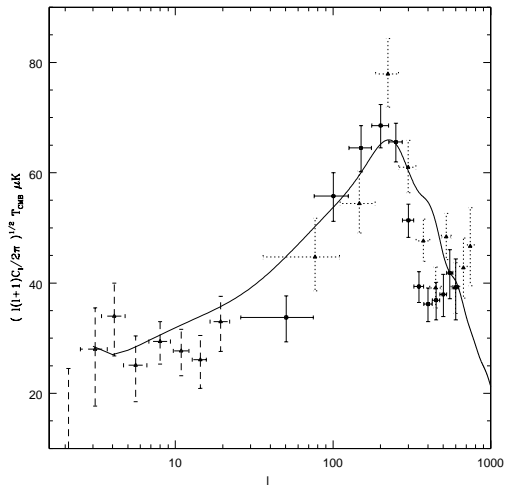} \end{center} \caption{{\sl 
The CMB power spectrum produced by the wiggly string model of
[Pogosian \& Vachaspati, 1999] in a closed universe with $\Omega_{\rm
total}=1.3$, $\Omega_{\rm baryon}=0.05$, $\Omega_{\rm CDM}=0.35$,
$\Omega_{\Lambda}=0.9$, and $H_0 = 65 {\rm ~km ~s}^{-1} {\rm
Mpc}^{-1}$ [Pogosian, 2000].}}
\label{omega}
\end{figure}

The prediction of the cosmic string model of [Pogosian \& Vachaspati, 1999] for
$\Omega_{\rm total}=1.3$ is shown in Figure \ref{omega}.
As can be seen, the main peak in the angular power spectrum can be
matched by choosing a reasonable value for $\Omega_{\rm
total}$. However, even with the main peak in the right place the
agreement with the data is far from satisfactory. The peak is
significantly wider than that in the data and there is no sign of a
rise in power at $l\approx 600$ as the actual data seems to
suggest [Hanany, \etal, 2000]. The sharpness and the height of the main peak
in the angular spectrum can be enhanced by including the effects of
gravitational radiation [Contaldi, Hindmarsh \& Magueijo, 1999]
and wiggles [Pogosian \& Vachaspati, 1999]. 
More precise high--resolution numerical simulations of
string networks in realistic cosmologies with a large contribution
from $\Omega_{\Lambda}$ are needed to determine the exact amount of
small--scale structure on the strings and the nature of the products of
their decay. It is, however, unlikely that including these effects
alone would result in a sufficiently narrow main peak and some
presence of a second peak.  

This brings us to the issues of causality and coherence and how the
random nature of the string networks comes into the calculation of the
anisotropy spectrum.
Both experimental and theoretical results for the CMB power spectra
involve calculations of averages. When estimating the correlations of
the observed temperature anisotropies, it is usual to compute the
average over all available patches on the sky. When calculating the
predictions of their models, theorists find the average over the
{\em ensemble} of possible outcomes provided by the model.

In inflationary models, as in all passive models, only the initial
conditions for the perturbations are random.  The subsequent evolution
is the same for all members of the ensemble. For wavelengths higher
than the Hubble radius, the linear evolution equations for the Fourier
components of such perturbations have a growing and a decaying
solution. The modes corresponding to smaller wavelengths have only
oscillating solutions. As a consequence, prior to entering the
horizon, each mode undergoes a period of phase ``squeezing'' which
leaves it in a highly coherent state by the time it starts to
oscillate. Coherence here means that all members of the ensemble,
corresponding to the same Fourier mode, have the same temporal
phase. So even though there is randomness involved, as one has to draw
random amplitudes for the oscillations of a given mode, the time
behavior of different members of the ensemble is highly
correlated. The total spectrum is the ensemble--averaged superposition
of all Fourier modes, and the predicted coherence results in an
interference pattern seen in the angular power spectrum as the
well--known acoustic peaks.

\begin{figure}[t]
\begin{center}\leavevmode \epsfxsize = 7cm
\epsffile{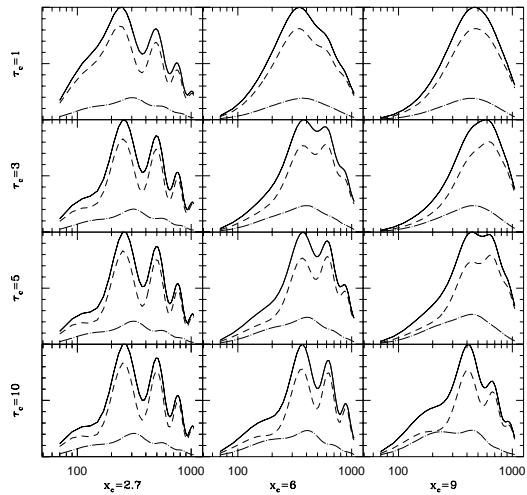} \end{center} \caption{{\sl 
The predictions of the toy model of Magueijo, \etal\ [1996]
for different values of parameters $x_c$, the
coherence length, and $\tau_c$, the coherence time. $x_c\propto \eta /
\lambda_c(\eta)$, where $\eta$ is the conformal time and
$\lambda_c(\eta)$ is the correlation length of the network at time
$\eta$. One can obtain oscillations in the CMB power spectrum by
fixing either one of the parameters and varying the other.}}
\label{magetal}\end{figure}

In contrast, the evolution of the string network is highly
non-linear. Cosmic strings are expected to move at relativistic
speeds, self--intersect and reconnect in a chaotic fashion. The
consequence of this behavior is that the unequal time correlators of
the string energy--momentum vanish for time differences larger than a
certain coherence time ($\tau_c$ in Figure \ref{magetal}). Members of
the ensemble corresponding to a given mode of perturbations will have
random temporal phases with the ``dice'' thrown on average once in
each coherence time. The coherence time of a realistic string network
is rather short. As a result, the interference pattern in the angular
power spectrum is completely washed out.

Causality manifests itself, first of all, through the initial
conditions for the string sources, the perturbations in the metric and
the densities of different particle species. If one assumes that the
defects are formed by a causal mechanism in an otherwise smooth
universe then the correct initial condition are obtained by setting
the components of the stress--energy pseudo--tensor $\tau_{\mu \nu}$
to zero [Veeraraghavan \& Stebbins, 1990; Pen, Spergel \& Turok,
1994].  These are the same as the isocurvature initial conditions [Hu,
Spergel \& White, 1997]. A generic prediction of isocurvature models
(assuming perfect coherence) is that the first acoustic peak is almost
completely hidden. The main peak is then the second acoustic peak and
in flat geometries it appears at $\ell\approx 300 - 400$. This is due to
the fact that after entering the horizon a given Fourier mode of the
source perturbation requires time to induce perturbations in the
photon density.
Causality also implies that no superhorizon correlations in the
string energy density are allowed. The correlation length of a
``realistic'' string network is normally between 0.1 and 0.4 of the
horizon size.

An interesting study was performed by Magueijo, Albrecht, Ferreira \&
Coulson [1996], where they constructed a toy model of defects
with two parameters: the coherence length and the coherence time. The
coherence length was taken to be the scale at which the energy density
power spectrum of the strings turns from a power law decay for large
values of $k$ into a white noise at low $k$. This is essentially the
scale corresponding to the correlation length of the string
network. The coherence time was defined in the sense described in the
beginning of this section, in particular, as the time difference
needed for the unequal time correlators to vanish.
Their study showed (see Figure \ref{magetal}) that by accepting any
value for one of the parameters and varying the other (within the
constraints imposed by causality) one could reproduce the oscillations
in the CMB power spectrum. Unfortunately for cosmic strings, at least
as we know them today, they fall into the parameter range
corresponding to the upper right corner in Figure \ref{magetal}.             

In order to get a better fit to present--day observations, cosmic
strings must either be more coherent or they have to be stretched over
larger distances, which is another way of making them more
coherent. To understand this imagine that there was just one long
straight string stretching across the universe and moving with some
given velocity. The evolution of this string would be linear and the
induced perturbations in the photon density would be coherent. By
increasing the correlation length of the string network we would move
closer to this limiting case of just one long straight string and so
the coherence would be enhanced.

The question of whether or not defects can produce a pattern of the
CMB power spectrum similar to, and including the acoustic peaks of,
that produced by the adiabatic inflationary models was repeatedly
addressed in the literature [Contaldi, Hindmarsh \& Magueijo 1999;
Magueijo, \etal\ 1996; Liddle, 1995; Turok, 1996; Avelino \& Martins,
2000]. In particular, it was shown [Magueijo, \etal\ 1996; Turok,
1996] that one can construct a causal model of active seeds which for
certain values of parameters can reproduce the oscillations in the CMB
spectrum. The main problem today is that current realistic models of
cosmic strings fall out of the parameter range that is needed to fit
the observations.  At the moment, only the (non-minimal) models with
either a varying speed of light or hybrid contribution of
strings+inflation are the only ones involving topological defects that
to some extent can match the observations. One possible way to
distinguish their predictions from those of inflationary models would
be by computing key non--Gaussian statistical quantities, such as the
CMB bispectrum.

%%-------------------------------------------------------------
\subsection{CMB bispectrum from active models}
\label{sec-}      

Different cosmological models differ in their predictions for the
statistical distribution of the anisotropies beyond the power
spectrum.  Future MAP and Planck satellite missions will provide
high-precision data allowing definite estimates of non-Gaussian
signals in the CMB. It is therefore important to know precisely which
are the predictions of all candidate models for the statistical
quantities that will be extracted from the new data and identify their
specific signatures.

Of the available non-Gaussian statistics, the CMB bispectrum, or the
three-point function of Fourier components of the temperature
anisotropy, has been perhaps the one best studied in the literature
[Gangui \& Martin, 2000a]. There are a few cases where the bispectrum
may be deduced analytically from the underlying model.  The bispectrum
can be estimated from simulated CMB sky maps; however, computing a
large number of full-sky maps resulting from defects is a much more
demanding task.  Recently, a precise numerical code to compute it, not
using CMB maps and similar to the \textsc{CMBFAST} 
code\footnote{{\tt http://physics.nyu.edu/matiasz/CMBFAST/cmbfast.html}} for the power
spectrum, was developed in [Gangui, Pogosian \& Winitzki 2001b]. What
follows below is an account of this work.
% sorry guys, no t to do much editing..

In a few words, given a suitable model, one can generate a statistical
\emph{ensemble} of realizations of defect matter perturbations. We
used a modified Boltzmann code based on \textsc{CMBFAST} to compute
the effect of these perturbations on the CMB and found the bispectrum
estimator for a given realization of sources. We then performed
statistical averaging over the ensemble of realizations to compute the
expected CMB bispectrum. (The CMB power spectrum was also obtained as
a byproduct.)  As a first application, we then computed the expected
CMB bispectrum from a model of simulated string networks first
introduced by Albrecht \etal\ [1997] and further developed in
[Pogosian \& Vachaspati, 1999] and in [Gangui, Pogosian \& Winitzki
2001].

We assume that, given a model of active perturbations, such as a
string
simulation, we can calculate the energy-momentum tensor $T_{\mu \nu
}({\mathbf
x},\tau )$ for a particular realization of the sources in a finite
spatial
volume $V_{0}$.
Here, ${\bf x}$ is a 3-dimensional
coordinate and $\tau $ is the cosmic time.
Many simulations are run to obtain an ensemble of random
realizations of sources with statistical properties appropriate for
the
given model. The spatial Fourier decomposition
of $T_{\mu \nu }$ can be written as
\begin{equation} \label{fouriersum}
T_{\mu \nu }({{\mathbf x}},\tau )=\sum _{{\mathbf k}}\Theta_{\mu \nu
}({{\mathbf k}},\tau )e^{i{{\mathbf k}}{{\mathbf x}}}\, \, \, ,
\end{equation}
where ${{\mathbf k}}$ are discrete. If $V_{0}$ is sufficiently large
we can approximate the summation by the integral \begin{eqnarray}
\sum _{{\mathbf k}}\Theta_{\mu \nu }({{\mathbf k}},\tau )e^{i{{\mathbf
k}}{{\mathbf x}}}\approx \frac{V_{0}}{(2\pi )^{3}}\int d^{3}{{\mathbf
k}}\Theta_{\mu \nu }({{\mathbf k}},\tau )e^{i{{\mathbf k}}{{\mathbf
x}}}\, \,
\, ,\label{sumtoint}
\end{eqnarray}
and the corresponding inverse Fourier transform will be
\begin{equation}
\label{inversefourier}
\Theta_{\mu \nu }({{\mathbf k}},\tau )=\frac{1}{V_{0}}\int
_{V_{0}}d^{3}{{\mathbf x}}\,T_{\mu \nu }({{\mathbf x}},\tau
)e^{-i{{\mathbf
k}}{{\mathbf x}}}\, \, \, .
\end{equation}
Of course, the final results, such as the CMB power spectrum or
bispectrum,
do not depend on the choice of $V_{0}$. To ensure this independence,
we shall keep $V_{0}$ in all expressions where it appears below.                          

It is conventional to expand the temperature fluctuations over the
basis of
spherical harmonics, \begin{equation}
{\Delta T/T}({\hat{{\mathbf n}}})=\sum
_{lm}a_{lm}Y_{lm}({\hat{{\mathbf n}}}),\end{equation}
where $\hat{{\mathbf n}}$ is a
unit vector. The coefficients $a_{lm}$ can be decomposed into Fourier
modes, \begin{equation} \label{eq:alm-def}
a_{lm}=\frac{V_{0}}{(2\pi )^{3}}\left( -i\right) ^{l}4\pi \int
d^{3}{{\mathbf k}}\, \Delta _{l}\left( {{\mathbf k}}\right)
Y^{*}_{lm}({\hat{{\mathbf k}}}).
\end{equation}
Given the sources $\Theta_{\mu \nu }({{\mathbf k}},\tau )$, the
quantities
$\Delta _{l}({{\mathbf k}})$ are found by solving linearized
Einstein-Boltzmann equations and integrating along the line of sight,
using
a code similar to CMBFAST [Seljak \& Zaldarriaga, 1996]. 
This standard procedure
can be written
symbolically as the action of a linear operator ${\hat{B}}_{l}^{\mu
\nu
}(k)$ on the source energy-momentum tensor, $\Delta _{l}({{\mathbf
k}})={\hat{B}}_{l}^{\mu \nu }(k)\Theta_{\mu \nu }({{\mathbf k}},\tau
)$,
so the third moment of $\Delta _{l}({{\mathbf k}})$ is linearly
related to the
three-point correlator of $\Theta_{\mu \nu }({{\mathbf k}},\tau
)$. Below
we consider the quantities $\Delta _{l}({{\mathbf k}})$, corresponding
to a
set of realizations of active sources, as given. The numerical
procedure for computing $\Delta _{l}({{\mathbf k}})$ 
was developed in  [Albrecht \etal\, 1997] and in [Pogosian \& Vachaspati, 1999].    

The third moment of $a_{lm}$, namely $\left\langle
a_{l_{1}m_{1}}a_{l_{2}m_{2}}a_{l_{3}m_{3}}\right\rangle $, can be
expressed
as 
\begin{eqnarray} 
%&  & 
\left( -i\right) ^{l_{1}+l_{2}+l_{3}}\left( 4\pi
\right) ^{3}\! \! \frac{V_{0}^{3}}{(2\pi )^{9}}\! \! \int \! \!
d^{3}{{\mathbf k}}_{1}d^{3}{{\mathbf k}}_{2}d^{3}{{\mathbf
k}}_{3}Y^{*}_{l_{1}m_{1}}\! ({\hat{{\mathbf k}}}_{1})
%\nonumber \\ &
%\times
%& 
Y^{*}_{l_{2}m_{2}}\! ({\hat{{\mathbf k}}}_{2})Y^{*}_{l_{3}m_{3}}\!
({\hat{{\mathbf k}}}_{3})\left\langle \Delta _{l_{1}}\! \! \left(
{{\mathbf
k}}_{1}\right) \Delta _{l_{2}}\! \! \left( {{\mathbf k}}_{2}\right)
\Delta
_{l_{3}}\! \! \left( {{\mathbf k}}_{3}\right) \right\rangle
.\label{eq:3alm-1}
\end{eqnarray}

A straightforward numerical evaluation of Eq.~(\ref{eq:3alm-1}) from
given sources $\Delta _{l}\left( {{\mathbf k}}\right) $ is
prohibitively difficult, because it involves too many integrations of
oscillating functions.  However, we shall be able to reduce the
computation to integrations over scalars [a similar method was
employed in Komatsu \& Spergel, 2001 and in Wang \& Kamionkowski, 2000].
Due to homogeneity, the 3-point function vanishes unless the triangle
constraint is satisfied,\begin{equation} \label{eq:triangle} {{\mathbf
k}}_{1}+{{\mathbf k}}_{2}+{{\mathbf k}}_{3}=0.
\end{equation}
We may write \begin{eqnarray}
\left\langle \Delta _{l_{1}}\left( {{\mathbf k}}_{1}\right)
\Delta
_{l_{2}}\left( {{\mathbf k}}_{2}\right) \Delta _{l_{3}}\left(
{{\mathbf
k}}_{3}\right) \right\rangle  =  \delta ^{(3)}\left(
{{\mathbf k}}_{1}+{{\mathbf k}}_{2}+{{\mathbf k}}_{3}\right)
P_{l_{1}l_{2}l_{3}}\left( {{\mathbf k}}_{1},{{\mathbf
k}}_{2},{{\mathbf
k}}_{3}\right) ,\label{p3lvector}
\end{eqnarray}                                     
where the three-point function $P_{l_{1}l_{2}l_{3}}\left( {{\mathbf
k}}_{1},{{\mathbf k}}_{2},{{\mathbf k}}_{3}\right) $ is defined only
for values of ${\mathbf k}_{i}$ that satisfy
Eq.~(\ref{eq:triangle}). Given the scalar values $k_{1}$, $k_{2}$,
$k_{3}$, there is a unique (up to an overall rotation) triplet of
directions ${\hat{{\mathbf k}}}_{i}$ for which the RHS of
Eq.~(\ref{p3lvector}) does not vanish. The quantity
$P_{l_{1}l_{2}l_{3}}\left( {{\mathbf k}}_{1},{{\mathbf
k}}_{2},{{\mathbf k}}_{3}\right) $ is invariant under an overall
rotation of all three vectors ${{\mathbf k}}_{i}$ and therefore may be
equivalently represented by a function of \emph{scalar} values
$k_{1}$, $k_{2}$, $k_{3}$, while preserving all angular
information. Hence, we can rewrite Eq.~(\ref{p3lvector}) as
\begin{eqnarray} \left\langle \Delta _{l_{1}}\! \! \left(
{{\mathbf k}}_{1}\right) \Delta _{l_{2}}\! \! \left( {{\mathbf
k}}_{2}\right) \Delta _{l_{3}}\! \! \left( {{\mathbf k}}_{3}\right)
\right\rangle =  \delta ^{(3)}\left( {{\mathbf
k}}_{1}+{{\mathbf k}}_{2}+{{\mathbf k}}_{3}\right)
P_{l_{1}l_{2}l_{3}}(k_{1},k_{2},k_{3}).\label{p3lscalar}
\end{eqnarray}
Then, using the simulation volume $V_{0}$ explicitly, we have
\begin{equation}
\label{p3l0}
P_{l_{1}l_{2}l_{3}}\! \left( k_{1},k_{2},k_{3}\right) \! =\!
\frac{(2\pi
)^{3}}{V_{0}}\left\langle \Delta _{l_{1}}\! \! \left( {{\mathbf
k}}_{1}\right) \Delta _{l_{2}}\! \! \left( {{\mathbf k}}_{2}\right)
\Delta
_{l_{3}}\! \! \left( {{\mathbf k}}_{3}\right) \right\rangle .
\end{equation}
Given an arbitrary direction $\hat{{\mathbf k}}_{1}$ and the
magnitudes
$k_{1}$, $k_{2}$ and $k_{3}$, the directions $\hat{{\mathbf k}}_{2}$
and $\hat{{\mathbf k}}_{3}$ are specified up to overall rotations by
the
triangle constraint. Therefore, both sides of Eq.~(\ref{p3l0}) are
functions of scalar $k_{i}$ only. The expression on the RHS of Eq.
(\ref{p3l0}) is evaluated numerically by averaging over different
realizations of the sources \textit{and} over permissible directions
$\hat{{\mathbf k}}_{i}$; below we shall give more details of the
procedure.      

Substituting Eqs.~(\ref{p3lscalar}) and (\ref{p3l0}) into
(\ref{eq:3alm-1}), Fourier transforming the Dirac delta and using the
Rayleigh identity, we can perform all angular integrations
analytically and obtain a compact form for the third moment,
\begin{equation}
\label{eq:3alm-res}
\left\langle a_{l_{1}m_{1}}a_{l_{2}m_{2}}a_{l_{3}m_{3}}\right\rangle
={\mathcal{H}}_{l_{1}l_{2}l_{3}}^{m_{1}m_{2}m_{3}}\int r^{2}dr\,
b_{l_{1}l_{2}l_{3}}(r),
\end{equation}
where, denoting the Wigner $3j$-symbol by
$\left( ^{\, \, l_{1}\, \; l_{2}\, \; l_{3}}_{m_{1}m_{2}m_{3}}\right)
$, we
have
\begin{eqnarray} {\mathcal{H}}_{l_{1}l_{2}l_{3}}^{m_{1}m_{2}m_{3}}
\equiv   \sqrt{\frac{\left( 2l_{1}+1\right) \left( 2l_{2}+1\right)
\left(
2l_{3}+1\right) }{4\pi }} \left(
\begin{array}{ccc}
l_{1} & l_{2} & l_{3}\\
0 & 0 & 0
\end{array}\right) \left( \begin{array}{ccc}
l_{1} & l_{2} & l_{3}\\
m_{1} & m_{2} & m_{3}
\end{array}\right) \, ,\label{eq:hlll}
\end{eqnarray}
and where we have defined the auxiliary quantities
$b_{l_{1}l_{2}l_{3}}$
using spherical Bessel functions $j_{l}$, \begin{eqnarray}
b_{l_{1}l_{2}l_{3}}(r) & \equiv  & \frac{8}{\pi
^{3}}\frac{V_{0}^{3}}{(2\pi
)^{3}}\int k_{1}^{2}dk_{1}\, k_{2}^{2}dk_{2}\, k_{3}^{2}dk_{3}\,
\nonumber
\\ & \times  &
j_{l_{1}}(k_{1}r)j_{l_{2}}(k_{2}r)j_{l_{3}}(k_{3}r)P_{l_{1}l_{2}l_{3}}(k_{1},k_{2},k_{3}).
\label{defineb}
\end{eqnarray}
The volume factor $V_{0}^{3}$ contained in this expression is correct:
as
shown in the next section, each term $\Delta _{l}$ includes a factor
$V_{0}^{-2/3}$, while the average quantity
$P_{l_{1}l_{2}l_{3}}(k_{1},k_{2},k_{3})\propto V_{0}^{-3}$
{[}cf.~Eq.~(\ref{p3l0}){]}, so that the arbitrary volume $V_{0}$ of
the
simulation cancels.
                                                                            
Our proposed numerical procedure therefore consists of computing the
RHS of
Eq.~(\ref{eq:3alm-res}) by evaluating the necessary integrals. For
fixed
$\left\{ l_{1}l_{2}l_{3}\right\} $, computation of the quantities
$b_{l_{1}l_{2}l_{3}}(r)$ is a triple integral over scalar $k_{i}$
defined
by Eq.~(\ref{defineb}); it is followed by a fourth scalar integral
over $r$
{[}Eq.~(\ref{eq:3alm-res}){]}. We also need to average over many
realizations of sources to obtain $P_{l_{1}l_{2}l_{3}}\! \left(
k_{1},k_{2},k_{3}\right) $. It was not feasible for us to precompute
the
values $P_{l_{1}l_{2}l_{3}}\! \left( k_{1},k_{2},k_{3}\right) $ on a
grid
before integration because of the large volume of data: for each set
$\left\{ l_{1}l_{2}l_{3}\right\} $ the grid must contain $\sim 10^{3}$
points for each $k_{i}$. Instead, we precompute $\Delta _{l}\! \!
\left(
{{\mathbf k}}\right) $ from one realization of sources and evaluate
the RHS
of Eq.~(\ref{p3l0}) on that data as an \emph{estimator} of
$P_{l_{1}l_{2}l_{3}}\! \left( k_{1},k_{2},k_{3}\right) $, averaging
over
allowed directions of $\hat{{\mathbf k}}_{i}$. The result is used for
integration in Eq.~(\ref{defineb}).

Because of isotropy and since the
allowed sets of directions $\hat{{\mathbf k}}_{i}$ are planar, it is
enough
to restrict the numerical calculation to directions $\hat{{\mathbf
k}}_{i}$
within a fixed two-dimensional plane. This significantly reduces the
amount
of computations and data storage, since $\Delta _{l}\! \left(
{{\mathbf
k}}\right) $ only needs to be stored on a two-dimensional grid of
${\mathbf
k}$.

In estimating $P_{l_{1}l_{2}l_{3}}\! \left( k_{1},k_{2},k_{3}\right) $
from Eq.~(\ref{p3l0}), averaging over directions of $\hat{{\mathbf
k}}_{i}$ plays a similar role to ensemble averaging over source
realizations.  Therefore if the number of directions is large enough
(we used 720 for cosmic strings), only a moderate number of different
source realizations is needed. The main numerical difficulty is the
highly oscillating nature of the function
$b_{l_{1}l_{2}l_{3}}(r)$. The calculation of the bispectrum for cosmic
strings presented in the next Section requires about 20 days of a
single-CPU workstation time per realization.

We note that this method is specific for the bispectrum and cannot be
applied to compute higher-order correlations. The reason is that
higher-order correlations involve configurations of vectors ${\mathbf
k}_{i}$ that are not described by scalar values $k_{i}$ and not
restricted to a plane. For instance, a computation of a 4-point
function would involve integration of highly oscillating functions
over four vectors ${\mathbf k}_{i}$ which is computationally
infeasible.

{}From Eq.~(\ref{eq:3alm-res}) we derive the CMB angular bispectrum
${\mathcal{C}}_{l_{1}l_{2}l_{3}}$, defined
as [Gangui \& Martin, 2000b]
\begin{eqnarray}
\bigl \langle a_{l_{1}m_{1}}a_{l_{2}m_{2}}a_{l_{3}m_{3}}\bigr \rangle
=\left( \begin{array}{ccc} l_{1} & l_{2} & l_{3}\\
m_{1} & m_{2} & m_{3}
\end{array}\right) {\mathcal{C}}_{l_{1}l_{2}l_{3}}\, .
\end{eqnarray}
The presence of the 3$ j$-symbol guarantees that the third moment
vanishes
unless $m_{1}+m_{2}+m_{3}=0$ and the $l_{i}$ indices satisfy the
triangle
rule $|l_{i}-l_{j}|\leq l_{k}\leq l_{i}+l_{j}$. Invariance under
spatial
inversions of the three-point correlation function implies the
additional
`selection rule' $l_{1}+l_{2}+l_{3}=\mbox {even}$, in order for the
third
moment not to vanish. Finally, from this last relation and
using standard properties of the 3$ j$-symbols, it follows that the
angular bispectrum ${\mathcal{C}}_{l_{1}l_{2}l_{3}}$ is left unchanged
under any arbitrary permutation of the indices $l_{i}$.
                                                                           
In what follows we will restrict our calculations to the angular bispectrum
$C_{l_{1}l_{2}l_{3}}$ in the `diagonal' case,
\textit{i.e.}~$l_{1}=l_{2}=l_{3}=l$.
This is a representative case and, in fact, the one most frequently
considered in the literature. Plots of the power spectrum are usually
done
in terms of $l(l+1)C_{l}$ which, apart from constant factors, is the
contribution to the mean squared anisotropy of temperature
fluctuations per
unit logarithmic interval of $l$. In full analogy with this, the
relevant
quantity to work with in the case of the bispectrum is
\begin{eqnarray} G_{lll} = l(2l+1)^{3/2}\left( \begin{array}{ccc}
l & l & l\\
0 & 0 & 0
\end{array}\right) C_{lll}\, .\label{eq:QtP}
\end{eqnarray}
For large values of the multipole index $l$, $G_{lll}\propto
l^{3/2}C_{lll}$.  Note also what happens with the 3$ j$-symbols
appearing in the definition of the coefficients
${\mathcal{H}}_{l_{1}l_{2}l_{3}}^{m_{1}m_{2}m_{3}}$: the symbol
$\left( ^{\, \, l_{1}\, \; l_{2}\, \; l_{3}}_{m_{1}m_{2}m_{3}}\right)
$ is absent from the definition of $C_{l_{1}l_{2}l_{3}}$, while in
Eq.~(\ref{eq:QtP}) the symbol $\left( ^{\, l\; l\; l}_{0\: 0\:
0}\right) $ is squared. Hence, there are no remnant oscillations due
to the alternating sign of $\left( ^{\, l\; l\; l}_{0\: 0\: 0}\right)$.

However, even more important than the value of $C_{lll}$ itself is the
relation between the bispectrum and the cosmic variance associated
with it.  In fact, it is their comparison that tells us about the
observability `in principle' of the non-Gaussian signal. The cosmic
variance constitutes a theoretical uncertainty for all observable
quantities and comes about due to the fact of having just one
realization of the stochastic process, in our case, the CMB sky
[Scaramella \& Vittorio, 1991].

The way to proceed is to employ an estimator
$\hat{C}_{l_{1}l_{2}l_{3}}$ for the bispectrum and compute the
variance from it. By choosing an unbiased estimator we ensure it
satisfies $C_{l_{1}l_{2}l_{3}}=\langle
\hat{C}_{l_{1}l_{2}l_{3}}\rangle $. However, this condition does not
isolate a unique estimator. The proper way to select the {\it best
unbiased} estimator is to compute the variances of all candidates and
choose the one with the smallest value.  The estimator with this
property was computed in [Gangui \& Martin, 2000b] and is
\begin{equation} \label{eq:clll-best}
\hat{C}_{l_{1}l_{2}l_{3}}=\! \! \! \sum _{m_{1},m_{2},m_{3}}\! \!
\left(
\begin{array}{ccc} l_{1} & l_{2} & l_{3}\\
m_{1} & m_{2} & m_{3}
\end{array}\right) a_{l_{1}m_{1}}a_{l_{2}m_{2}}a_{l_{3}m_{3}}.
\end{equation}
The variance of this estimator, assuming a mildly non-Gaussian
distribution, can be expressed in terms of the angular power spectrum
$C_{l}$ as follows
\begin{equation}
\label{eq:sigma}
 \sigma
^{2}_{\hat{C}_{l_{1}l_{2}l_{3}}}\! \! \! \!
=C_{l_{1}}C_{l_{2}}C_{l_{3}}\!
\left( 1\! +\! \delta _{l_{1}l_{2}}\! \! +\! \delta _{l_{2}l_{3}}\! \!
+\!
\delta _{l_{3}l_{1}}\! \! +\! 2\delta _{l_{1}l_{2}}\delta
_{l_{2}l_{3}}\right) .
\end{equation}
The theoretical signal-to-noise ratio for the bispectrum is then given
by
\begin{equation}
(S/N)_{l_{1}l_{2}l_{3}} =
|C_{l_{1}l_{2}l_{3}}/\sigma_{\hat{C}_{l_{1}l_{2}l_{3}}}|.
\end{equation}
In turn, for the diagonal case $l_{1}=l_{2}=l_{3}=l$ we have
\begin{equation}
(S/N)_{l} = |C_{lll}/\sigma _{\hat{C}_{lll}}|.
\end{equation}
                                                 
Incorporating all the specifics of the particular experiment, such as
sky coverage, angular resolution, etc., will allow us to give an
estimate of the particular non-Gaussian signature associated with a
given active source and, if observable, indicate the appropriate range
of multipole $l$'s where it is best to look for it.
                                                             
%%-------------------------------------------------------------
\subsection{CMB bispectrum from strings}
\label{sec-bispectrum}      

To calculate the sources of perturbations we have used an updated
version of the cosmic string model first introduced by Albrecht \etal\
[1997] and further developed in [Pogosian \& Vachaspati, 1999], where
the wiggly nature of strings was taken into account. In these previous
works the model was tailored to the computation of the two-point
statistics (matter and CMB power spectra). When dealing with
higher-order statistics, such as the bispectrum, a different strategy
needs to be employed.

In the model, the string network is represented by a collection of
uncorrelated straight string segments produced at some early epoch and
moving with random uncorrelated velocities. At every subsequent epoch,
a certain fraction of the number of segments decays in a way that
maintains network scaling. The length of each segment at any time is
taken to be equal to the correlation length of the network. This and
the root mean square velocity of segments are computed from the
velocity-dependent one-scale model of Martins \& Shellard [1996].
The positions of segments are drawn from a uniform distribution in
space, and their orientations are chosen from a uniform distribution
on a two-sphere.

The total energy of the string network in a volume $V$ at any time is
$E=N\mu L$, where $N$ is the total number of string segments at that
time, $\mu $ is the mass per unit length, and $L$ is the length of one
segment. If $L$ is the correlation length of the string network then,
according to the one-scale model, the energy density is $\rho
={E/V}={\mu /L^{2}}$, where $V=V_{0}a^{3}$, the expansion factor $a$
is normalized so that $a=1$ today, and $V_{0}$ is a constant
simulation volume. It follows that $N=V/L^{3}=V_{0}/\ell^{3}$, where
$\ell=L/a$ is the comoving correlation length.  In the scaling regime
$\ell$ is approximately proportional to the conformal time $\tau $ and
so the number of strings $N(\tau )$ within the simulation volume
$V_{0}$ falls as $\tau ^{-3}$.
                                                     
To calculate the CMB anisotropy one
needs to evolve the string network over at least four orders of
magnitude
in cosmic expansion. Hence, one would have to start with $N\gsim
10^{12}$
string segments in order to have one segment left at the present time.
Keeping track of such a huge number of segments is numerically
infeasible.
A way around this difficulty was suggested in
Ref.\cite{ABR97}, where the idea was to consolidate all string
segments
that decay at the same epoch. The number of segments that decay by the
(discretized) conformal time $\tau _{i}$ is \begin{equation}
\label{eq:nd}
N_{d}(\tau _{i})=V_{0}\left( n(\tau _{i-1})-n(\tau _{i})\right) ,
\end{equation}
where $n(\tau )=[\ell(\tau )]^{-3}$ is the number density of strings
at time
$\tau $. The energy-momentum tensor in Fourier space, $\Theta^{i}_{\mu
\nu }$,
of these $N_{d}(\tau _{i})$ segments is a sum \begin{equation}
\label{emtsum}
\Theta^{i}_{\mu \nu }=\sum _{m=1}^{N_{d}(\tau _{i})}\Theta^{im}_{\mu
\nu }\,
\, \, , \end{equation}
where $\Theta^{im}_{\mu \nu }$ is the Fourier transform of the
energy-momentum
of the $m$-th segment. If segments are uncorrelated, then
\begin{equation}
\label{eq:theta2}
\langle\Theta^{im}_{\mu\nu}\Theta^{im'}_{\sigma\rho}\rangle =
\delta_{m m'}
\langle\Theta^{im}_{\mu\nu}\Theta^{im}_{\sigma\rho}\rangle
\end{equation}
and
\begin{equation}
\langle\Theta^{im}_{\mu\nu}\Theta^{im'}_{\sigma\rho}
\Theta^{im''}_{\gamma\delta}\rangle =
\delta_{m m'}\delta_{m m''}
\langle\Theta^{im}_{\mu\nu}\Theta^{im}_{\sigma\rho}
\Theta^{im}_{\gamma\delta}\rangle .
\end{equation}                      
Here the angular brackets $\langle \ldots \rangle $ denote the
ensemble average, which in our case means averaging over many
realizations
of the string network. If we are calculating power spectra, then the
relevant quantities are the two-point functions of $\Theta^{i}_{\mu
\nu }$,
namely
\begin{eqnarray} \langle \Theta^{i}_{\mu \nu }\Theta^{i}_{\sigma \rho
}\rangle =\langle \sum _{m=1}^{N_{d}(\tau _{i})}\sum
_{m'=1}^{N_{d}(\tau
_{i})}\Theta^{im}_{\mu \nu }\Theta^{im'}_{\sigma \rho }\rangle
.\label{thefix1} \end{eqnarray}
Eq.~(\ref{eq:theta2}) allows us to write \begin{eqnarray}
\langle \Theta^{i}_{\mu \nu }\Theta^{i}_{\sigma \rho }\rangle =\sum
_{m=1}^{N_{d}(\tau _{i})}\langle \Theta^{im}_{\mu \nu
}\Theta^{im}_{\sigma
\rho }\rangle =N_{d}(\tau _{i})\langle \Theta^{i1}_{\mu \nu
}\Theta^{i1}_{\sigma \rho }\rangle ,\label{thefix2}
\end{eqnarray}
where $\Theta^{i1}_{\mu \nu }$ is of the energy-momentum
of one of the segments that decay by the time $\tau _{i}$. The last
step in
Eq.~(\ref{thefix2}) is possible because the segments are statistically
equivalent. Thus, if we only want to reproduce the correct power
spectra in
the limit of a large number of realizations, we can replace the sum in
Eq.~(\ref{emtsum}) by \begin{equation} \label{thefix3}
\Theta^{i}_{\mu \nu }=\sqrt{N_{d}(\tau _{i})}\Theta^{i1}_{\mu \nu }.
\end{equation}
The total energy-momentum tensor of the network in Fourier space is a
sum over
the consolidated segments:
\begin{equation}
\label{emtsum1}
\Theta_{\mu \nu }=\sum _{i=1}^{K}\Theta^{i}_{\mu \nu }=\sum
_{i=1}^{K}\sqrt{N_{d}(\tau _{i})}\Theta^{i1}_{\mu \nu }\, .
\end{equation}
So, instead of summing over $\sum _{i=1}^{K}N_{d}(\tau _{i})\gsim
10^{12}$
segments we now sum over only $K$ segments, making $K$ a parameter.

For the three-point functions we extend the above procedure. Instead
of
Eqs.~(\ref{thefix1}) and (\ref{thefix2}) we now write
\begin{eqnarray}
\langle \Theta^{i}_{\mu \nu }\Theta^{i}_{\sigma \rho
}\Theta^{i}_{\gamma
\delta }\rangle \! = \! \langle \sum _{m=1}^{N_{d}(\tau _{i})}\sum
_{m'=1}^{N_{d}(\tau _{i})}\sum _{m''=1}^{N_{d}(\tau
_{i})}\Theta^{im}_{\mu \nu
}\Theta^{im'}_{\sigma \rho }\Theta^{im''}_{\gamma \delta }\rangle  
=\!\!\! \sum _{m=1}^{N_{d}(\tau _{i})}\langle \Theta^{im}_{\mu
\nu
}\Theta^{im}_{\sigma \rho }\Theta^{im}_{\gamma \delta }\rangle
= N_{d}(\tau
_{i})\langle \Theta^{i1}_{\mu \nu }\Theta^{i1}_{\sigma \rho
}\Theta^{i1}_{\gamma \delta }\rangle \, 
\label{newfix} \end{eqnarray}
Therefore, for the purpose of calculation of three-point functions,
the sum
in Eq.~(\ref{emtsum}) should now be replaced by \begin{equation}
\label{newfix1}
\Theta^{i}_{\mu \nu }=[N_{d}(\tau _{i})]^{1/3}\Theta^{i1}_{\mu \nu }\,
.
\end{equation}

Both expressions in Eqs.~(\ref{thefix3}) and (\ref{newfix1}), depend
on the simulation volume, $V_{0}$, contained in the definition of
$N_{d}(\tau _{i})$ given in Eq.~(\ref{eq:nd}). This is to be expected
and is consistent with our calculations, since this volume cancels in
expressions for observable quantities.

Note also that the simulation model in its present form does not allow
computation of CMB sky maps. This is because the method of finding the
two- and three-point functions as we described involves
{}``consolidated{}'' quantities $\Theta^{i}_{\mu \nu }$ which do not
correspond to the energy-momentum tensor of a real string
network. These quantities are auxiliary and specially prepared to give
the correct two- or three-point functions after ensemble averaging.

\begin{figure}[t]\begin{center}\leavevmode \epsfxsize = 9cm
\epsffile{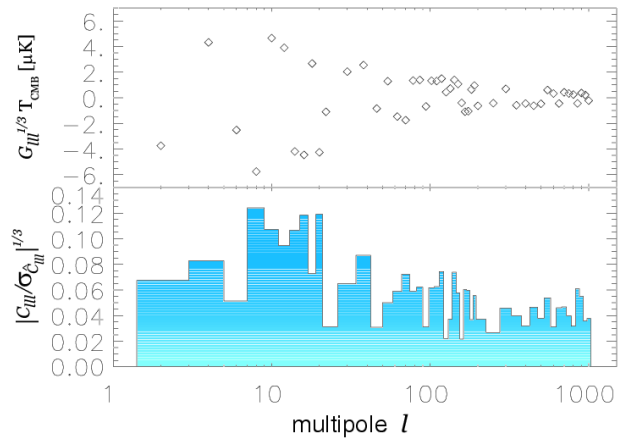} 
%\epsffile{levo2figs3.eps} 
\end{center} \caption{{\sl The CMB
angular bispectrum in the `diagonal' case ($ G_{lll}^{1/3}$) from
wiggly cosmic strings in a spatially flat model with cosmological
parameters $ \Omega _{\rm CDM}=0.3$, $ \Omega
_{\rm baryon}=0.05$, $ \Omega _{\Lambda }=0.65$, and Hubble
constant $ H=0.65{\textrm{km}}{\textrm{s}}^{-1}{\textrm{Mpc}}^{-1}$
{[}upper panel{]}. In the lower panel we show the ratio of the signal
to theoretical noise $ |C_{lll}/\sigma _{\hat{C}_{lll}}|^{1/3}$ for
different multipole indices.  Normalization follows from fitting the
power spectrum to the BOOMERANG and MAXIMA data.}}
\label{fig:1}\end{figure}

In Fig.~\ref{fig:1} we show the results for $G_{lll}^{1/3}$
{[}cf.~Eq.~(\ref{eq:QtP}){]}. It was calculated using the string model
with $800$ consolidated segments in a flat universe with cold dark
matter and a cosmological constant. Only the scalar contribution to
the anisotropy has been included. Vector and tensor contributions are
known to be relatively insignificant for local cosmic strings and can
safely be ignored in this model \cite{ABR97,PV99}\footnote{The
contribution of vector and tensor modes is large in the case of global
strings [Turok, Pen \& Seljak, 1998; Durrer, Gangui \& Sakellariadou,
1996].}. The plots are produced using a single realization of the
string network by averaging over $720$ directions of ${\mathbf
k}_{i}$. The comparison of $G_{lll}^{1/3}$ (or equivalently
${C}_{lll}^{1/3}$) with its cosmic variance
{[}cf.~Eq.~(\ref{eq:sigma}){]} clearly shows that the bispectrum (as
computed from the present cosmic string model) lies hidden in the theoretical
noise and is therefore undetectable for any given value of $l$.

Let us note, however, that in its present stage the string code
employed in these computations describes Brownian, wiggly long strings
in spite of the fact that long strings are very likely not Brownian on
the smallest scales, as recent field--theory simulations indicate.  In
addition, the presence of small string loops [Wu, \etal, 1998] and
gravitational radiation into which they decay were not yet included in
this model. These are important effects that could, in principle,
change the above predictions for the string-generated CMB bispectrum
on very small angular scales.

The imprint of cosmic strings on the CMB is a combination of different
effects. Prior to the time of recombination strings induce density and
velocity fluctuations on the surrounding matter. During the period of
last scattering these fluctuations are imprinted on the CMB through
the Sachs-Wolfe effect, namely, temperature fluctuations arise because
relic photons encounter a gravitational potential with spatially
dependent depth. In addition to the Sachs-Wolfe effect, moving long
strings drag the surrounding plasma and produce velocity fields that
cause temperature anisotropies due to Doppler shifts. While a string
segment by itself is a highly non-Gaussian object, fluctuations
induced by string segments before recombination are a superposition of
effects of many random strings stirring the primordial plasma. These
fluctuations are thus expected to be Gaussian as a result of the
central limit theorem.

As the universe becomes transparent, strings continue to leave their
imprint on the CMB mainly due to the Kaiser \& Stebbins [1984] effect.
As we mentioned in previous sections, this effect results in line
discontinuities in the temperature field of photons passing on
opposite sides of a moving long string.\footnote{The extension of
the Kaiser-Stebbins effect to polarization will be treated 
below. In fact, Benabed and Bernardeau [2000] have
recently considered the generation of a B-type polarization field out
of E-type polarization, through gravitational lensing on a cosmic
string.}  However, this effect can result in non-Gaussian
perturbations only on sufficiently small scales. This is because on
scales larger than the characteristic inter-string separation at the
time of the radiation-matter equality, the CMB temperature
perturbations result from superposition of effects of many strings and
are likely to be Gaussian. Avelino \textit{et al.} [1998] 
applied several non-Gaussian tests to the perturbations seeded by
cosmic strings. They found the density field distribution to be close
to Gaussian on scales larger than $1.5 (\Omega _M h^2)^{-1}$ Mpc,
where $\Omega _M$ is the fraction of cosmological matter density in
baryons and CDM combined. Scales this small correspond to the
multipole index of order $l \sim 10^4$. 

%%-------------------------------------------------------------
\subsection{CMB polarization}
\label{sec-poladeff}    

The possibility that the CMB be polarized was first discussed by
Martin~Rees in 1968, in the context of anisotropic Universe models.
In spite of his optimism, and after more than thirty years, there is
still no positive detection of the polarization field.
Unlike the BOOMERANG[G experiment, MAP will have the capability to
detect it and this to a level of better than 10 $\mu$K in its low
frequency channels.  Polarization is an important probe both for
cosmological models and for the more recent history of our nearby
Universe. It arises from the interactions of CMB photons with free
electrons; hence, polarization can {\it only} be produced at the last
scattering surface (its amplitude depends on the duration of
the decoupling process) and, unlike temperature fluctuations, it is
unaffected by variations of the gravitational potential after
last scattering.  Future measurements of polarization will thus
provide a clean view of the inhomogeneities of the Universe at about
400,000 years after the Bang.

For understanding polarization, a couple of things should be
clear. First, the energy of the photons is small compared to the mass
of the electrons. Then, the CMB frequency does not change, since the
electron recoil is negligible. Second, the change in the CMB
polarization (i.e., the orientation of the oscillating electric field
$\vec{\rm E}$ of the radiation) occurs
due to
a certain transition, called {\it Thomson scattering}.  The transition
probability per unit time is proportional to a combination of the old
($\hat\epsilon^{\rm ~in}_\alpha$) and new ($\hat\epsilon^{\rm
~out}_\alpha$) directions of polarization in the form
$|\hat\epsilon^{\rm ~in}_\alpha\cdot\hat\epsilon^{\rm
~out}_\alpha|^2$. In other words, the initial direction of
polarization will be favored.  Third, an oscillating $\vec{\rm E}$
will push the electron to also oscillate; the latter can then be seen
as a dipole (not to be confused with the CMB dipole), and dipole
radiation emits preferentially perpendicularly to the direction of
oscillation. These `rules' will help us understand why the CMB
should be linearly polarized.            

Previous to the recombination epoch, the radiation field is
unpolarized.  In unpolarized light the electric field can be
decomposed into the two orthogonal directions (along, say, $\hat x$
and $\hat z$) perpendicular to the line of propagation ($\hat y$). The
electric field along $\hat\epsilon^{\rm ~in}_{\hat z}$ (suppose $\hat
z$ is vertical) will make the electron oscillate also
vertically. Hence, the dipolar radiation will be maximal over the
horizontal $xy$-plane.  Analogously, dipole radiation due to the
electric field along $\hat x$ will be on the $yz$-plane.  If we now
look from the side (e.g., from $\hat x$, on the horizontal plane and
perpendicularly to the incident direction $\hat y$) we will see a
special kind of scattered radiation.  From our position we cannot
perceive the radiation that the electron oscillating along the $\hat
x$ direction would emit, just because this radiation goes to the
$yz$-plane, orthogonal to us. Then, it is {\it as if} only the
vertical
component ($\hat\epsilon^{\rm ~in}_{\hat z}$) of the incoming electric
field would cause the radiation we perceive.  From the above rules we
know that the highest probability for the polarization of the outgoing
radiation $\hat\epsilon^{\rm ~out}_\alpha$ will be to be aligned with
the incoming one $\hat\epsilon^{\rm ~in}_{\hat z}$, and therefore it
follows that the outgoing radiation will be {\it linearly} polarized.
Now, as both the chosen incoming direction and our position as
observers were arbitrary, the result will not be modified if we change
them. Thomson scattering will convert unpolarized radiation into
linearly polarized one.

This however is not the end of the story. To get the total effect we
need to consider all possible directions from which photons will come
to interact with the target electron, and sum them up. We see easily
that for an initial isotropic radiation distribution the individual
contributions will cancel out: just from symmetry arguments, in a
spherically symmetric configuration no direction is privileged,
unlike the case of a net linear polarization which would select one
particular direction.                                            

Fortunately, we know the CMB is {\it not exactly} isotropic; to the
millikelvin precision the dominant mode is dipolar.  So, what about a
CMB dipolar distribution~?  Although spatial symmetry does not help us
now, a dipole will not generate polarization either.  Take, for
example, the radiation incident onto the electron from the left to be
more intense than the radiation incident from the right, with average
intensities above and below (that's a dipole); it then suffices to sum
up all contributions to see that no net polarization survives.
However, if the CMB has a {\it quadrupolar} variation in temperature
(that {\it it has}, first discovered by COBE, to tens of $\mu$K
precision) then there will be an excess of vertical polarization from
left- and right-incident photons (assumed hotter than the mean) with
respect to the horizontal one from top and bottom light (cooler). From
any point of view, orthogonal contributions to the final polarization
will be different, leaving a net linear polarization in the scattered
radiation.

\begin{figure}[t]\begin{center}\leavevmode \epsfxsize = 9cm
\epsffile{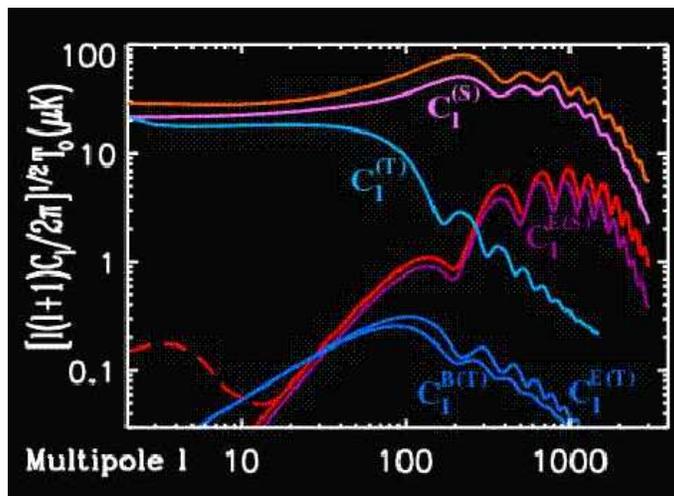} 
%\epsffile{polarizacion8css.ps} 
%\epsffile{polarizacion8c.eps} 
\end{center} 
\vspace{-1cm}
\caption{{\sl
CMB Polarization for two different models.  Red and orange (unlabeled)
curves are the angular spectra derived for a $\Lambda$CHDM model, 
both with (red dashed line) and without (red full line)
reionization.  The temperature anisotropy spectrum from scalar
perturbations (proportional to $[ C_\ell ]^{1/2}$, orange curve) is
virtually unchanged for both ionization histories. The polarization
spectrum ($\propto [ C^{\rm E(S)}_\ell ]^{1/2}$, red curves), although
indistinguishable for $\ell \gsim 20$, dramatically changes for small
$\ell$'s; in this model the Universe is reionized suddenly at low
redshift with optical depth $\tau_{\rm c}=0.05$. 
%Much higher values
%for $\tau_{\rm c}$ are already strongly constrained from reported
%excess (over COBE) of sub-degree scale temperature anisotropies
%(reionization {\it erases} primary fluctuations and generates
%secondary ones, but only weakly). 
Blue and violet curves represent a
SCDM model {\it but} with a high tensor-mode amplitude, T/S=1 at the
quadrupole ($\ell=2$) level, with scale-invariant spectral indices
$n_{\rm S}=1$ and $n_{\rm T}=0$.  Separate scalar (noted $C^{\rm
(S)}_\ell$) and tensor ($C^{\rm (T)}_\ell$) contributions to
temperature anisotropies are shown (top curves). Scalar modes only
generate E-type polarization ($C^{\rm E(S)}_\ell$), which is smaller
than the corresponding red curve of the $\Lambda$CHDM model both due
to differences in the models (notably $\Lambda\not=0$ for the red
curves) and due to the influence of tensors on the normalization at
small $\ell$.  E- and B-type polarization from tensor modes are also
shown, respectively $C^{\rm E(T)}_\ell$ and $C^{\rm B(T)}_\ell$.
Model spectra were computed with {\sc CMBFAST} and are
normalized to $\delta {\rm T}_{\ell = 10} = 27.9\mu{\rm K}$.  
}}
\label{fig-mipola}
%\vspace{-1cm}
\end{figure}        

Within standard recombination models the predicted level of linear
polarization on large scales is tiny (see Figure \ref{fig-mipola}):
the quadrupole generated in the radiation distribution as the photons
travel between successive scatterings is too small. Multiple
scatterings make the plasma very homogeneous and only wavelengths that
are small enough (big $\ell$'s) to produce anisotropies over the
(rather short) mean free path of the photons will lead to a
significant quadrupole, and thus also to polarization.  Indeed, if the
CMB photons last scattered at $z\sim 1100$, the SCDM model with $h=1$
predicts no more than 0.05 $\mu$K on scales greater than a few
degrees. Hence, measuring polarization represents an experimental
challenge. There is still no positive detection and the best upper
limits a few years ago were around $25 \mu$K, obtained by Edward
Wollack and collaborators in 1993, and now improved to roughly $10
\mu$K on subdegree angular scales by Hedman, \etal,
[2000]\footnote{They actually find upper limits of $14 \mu$K and $13
\mu$K on the amplitudes of the $E$ and $B$ modes, respectively, of the
polarization field -- more below. And, if in their analysis they
assumed there are no $B$ modes, then the limit on $E$ improves to $10
\mu$K (all limits to 95\% confidence level).}.

However, CMB polarization increases remarkably around the degree-scale
in standard models. In fact, for $\theta < 1^\circ$ a bump with
superimposed acoustic oscillations reaching $\sim 5 \mu$K is
generically forecasted. On these scales, like for the temperature
anisotropies, the polarization field shows acoustic
oscillations. However, polarization spectra are sharper: temperature
fluctuations receive contributions from both density (dominant) and
velocity perturbations and these, being out of phase in their
oscillation, partially cancel each other.  On the other hand,
polarization is mainly produced by velocity gradients in the
baryon-photon fluid before last scattering, which also explains why
temperature and polarization peaks are located differently.  Moreover,
acoustic oscillations depend on the {\it nature} of the underlying
perturbation; hence, we do not expect scalar acoustic sound-waves in
the baryon-photon plasma, propagating with characteristic adiabatic
sound speed $c_{\rm S}\sim c/\sqrt{3}$, close to that of an ideal
radiative fluid, to produce the same peak-frequency as that produced
by gravitational waves, which propagate with the speed of light $c$
(see Fig.\ref{fig-mipola}).  

The main technical complication with polarization (characterized by a
tensor field) is that it is not invariant under rotations around a
given direction on the sky, unlike the temperature fluctuation that is
described by a scalar quantity and invariant under such rotations.
The level of linear polarization is conveniently expressed in terms of
the {\it Stokes parameters} Q and U. It turns out that there is a
clever combination of these parameters that results in scalar
quantities (in contrast to the above noninvariant tensor description)
but with different transformation properties under spatial inversions
({\it parity} transformations).  Then, inspired by classical
electromagnetism, any polarization pattern on the sky can be separated
into `electric' (scalar, unchanged under parity transformation) and
`magnetic' (pseudo-scalar, changes sign under parity) components (E-
and B-type polarization, respectively).

%%-------------------------------------------------------------
\subsubsection{CMB polarization from global defects}
\label{sec-polaglob}         

One then expands these different components in terms of spherical
harmonics, very much like we did for temperature anisotropies, getting
coefficients $a^{m}_{\ell}$ for E and B polarizations and, from these,
the multipoles $C^{\rm E,B}_{\ell}$. The interesting thing is that
(for symmetry reasons) scalar-density perturbations will {\it not}
produce any B polarization (a pseudo-scalar), that is $C^{\rm
B(S)}_\ell=0$.  We see then that an unambiguous detection of some
level of B-type fluctuations will be a signature of the existence (and
of the amplitude) of a background of gravitational waves~!  [Seljak \&
Zaldarriaga, 1997] (and, if present, also of rotational modes, like in
models with topological defects).

Linear polarization is a symmetric and traceless 2x2 tensor that
requires 2 parameters to fully describe it: $Q$, $U$ Stokes
parameters.  These depend on the orientation of the coordinate system
on the sky. It is convenient to use $Q+iU$ and $Q-iU$ as the two
independent combinations, which transform under right-handed rotation
by an angle $\phi$ as $(Q+iU)'=e^{-2i\phi}(Q+iU)$ and
$(Q-iU)'=e^{2i\phi}(Q-iU)$.  These two quantities have spin-weights
$2$ and $-2$ respectively and can be decomposed into spin $\pm 2$
spherical harmonics ${}_{\pm 2}Y_{lm}$
\begin{eqnarray}
(Q+iU)(\hat{\bbox{n}})&=&\sum_{lm}
a_{2,lm} \, {}_2Y_{lm}(\hat{\bbox{n}}) %\nonumber 
\\
(Q-iU)(\hat{\bbox{n}})&=&\sum_{lm}
a_{-2,lm} \, {}_{-2}Y_{lm}(\hat{\bbox{n}}). %\nonumber
\end{eqnarray}
                       
Spin $s$ spherical harmonics form a complete orthonormal system for
each value of $s$.  Important property of spin-weighted basis:
there exists spin raising and lowering operators $\edth$ and
$\baredth$.  By acting twice with a spin lowering and raising operator
on $(Q+iU)$ and $(Q-iU)$ respectively one obtains quantities of spin
0, which are {\it rotationally invariant}. These quantities can be
treated like the temperature and no ambiguities connected with the
orientation of coordinate system on the sky will arise. Conversely, by
acting with spin lowering and raising operators on usual harmonics
spin $s$ harmonics can be written explicitly in terms of derivatives
of the usual spherical harmonics.  Their action on
${}_{\pm 2}Y_{lm}$ leads to
\begin{eqnarray}
\baredth^2(Q+iU)(\hat{\bbox{n}})&=&
\sum_{lm}
\left({[l+2]! \over [l-2]!}\right)^{1/2}
a_{2,lm}Y_{lm}(\hat{\bbox{n}})
%\nonumber 
\\
\edth^2(Q-iU)(\hat{\bbox{n}})&=&\sum_{lm}
\left({[l+2]! \over [l-2]!}\right)^{1/2}
a_{-2,lm}Y_{lm}(\hat{\bbox{n}}). %\nonumber
\end{eqnarray}                    
With these definitions the expressions for the expansion coefficients
of the two polarization variables become [Seljak \& Zaldarriaga, 1997]
\begin{eqnarray}
a_{2,lm}&=&\left({[l-2]! \over [l+2]!}\right)^{1/2}
\int d\Omega\; Y_{lm}^{*}(\hat{\bbox{n}})
\baredth^2(Q+iU)(\hat{\bbox{n}})
%\nonumber 
\\
a_{-2,lm}&=&\left({[l-2]! \over [l+2]!}\right)^{1/2}
\int d\Omega\;
Y_{lm}^{*}(\hat{\bbox{n}})\edth^2(Q-iU)(\hat{\bbox{n}}).
%\nonumber
\label{alm}
\end{eqnarray}       
Instead of $a_{2,lm}$, $a_{-2,lm}$ it is convenient to introduce their
linear {\it electric} and {\it magnetic} combinations
\begin{eqnarray}
a_{E,lm}=-{1\over 2}(a_{2,lm}+a_{-2,lm}) \qquad
a_{B,lm}= {i\over 2}(a_{2,lm}-a_{-2,lm}). %\nonumber
\label{aeb}
\end{eqnarray}
These two behave differently under {\it parity} transformation:
while $E$ remains unchanged $B$ changes the sign, in analogy
with electric and magnetic fields.

To characterize the statistics of the CMB perturbations only four
power spectra are needed, those for $X = T, E, B$ and the cross
correlation between $T$ and $E$.  The cross correlation between $B$
and $E$ or $B$ and $T$ vanishes because $B$ has the opposite parity of
$T$ and $E$. As usual, the spectra are defined as the rotationally invariant
quantities
\begin{eqnarray}
C_{Xl}={1\over 2l+1}\sum_m \langle a_{X,lm}^{*} a_{X,lm}\rangle
\qquad
C_{Cl}={1\over 2l+1}\sum_m \langle a_{T,lm}^{*}a_{E,lm}\rangle
%\nonumber
\label{Cls}
\end{eqnarray}
in terms of which on has
\begin{eqnarray}
\langle a_{X,l^\prime m^\prime}^{*} a_{X,lm}\rangle&=&
C_{Xl} \, \delta_{l^\prime l} \delta_{m^\prime m} 
%\nonumber 
\\
\langle a_{T,l^\prime m^\prime}^{*} a_{E,lm}\rangle&=&
C_{Cl} \, \delta_{l^\prime l} \delta_{m^\prime m} 
%\nonumber 
\\
\langle a_{B,l^\prime m^\prime}^{*} a_{E,lm}\rangle&=&
\langle a_{B,l^\prime m^\prime}^{*} a_{T,lm}\rangle=0. %\nonumber
\label{stat}
\end{eqnarray}
                                        
According to what was said above, one expects some amount of
polarization to be present in all possible cosmological
models. However, symmetry breaking models giving rise to topological
defects differ from inflationary models in several important aspects,
two of which are the relative contributions from scalar, vector and
tensor modes and the coherence of the seeds sourcing the perturbation
equations. In the local cosmic string case one finds that in general
scalar modes are dominant, if one compares to vector and tensor modes
in the usual decomposition of perturbations. The situation with global
topological defects is radically different and this leads to a very
distinctive signature in the polarization field.

\begin{figure}[t]\begin{center}\leavevmode \epsfxsize = 9cm
\epsffile{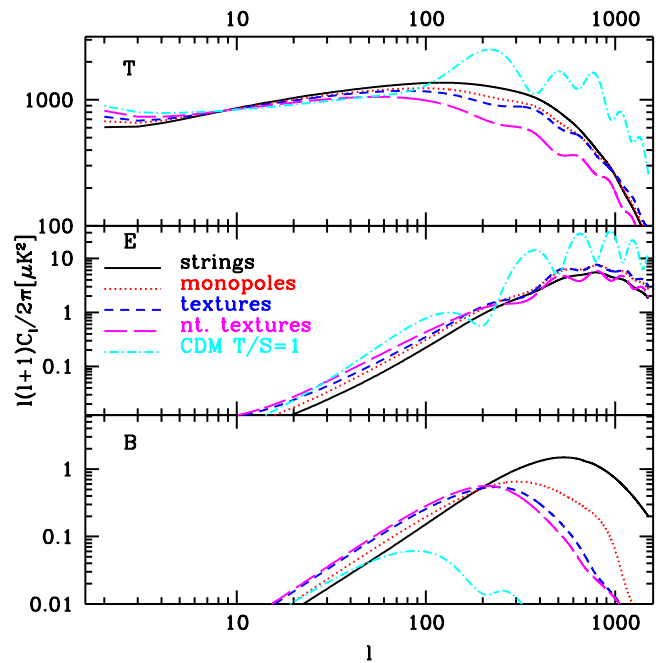} \end{center} \caption{{\sl 
Power spectra of temperature (T), electric type polarization (E) and
magnetic type polarization (B) for global strings, monopoles, textures
and nontopological textures [taken from Seljak. \etal, 1997]. 
The corresponding spectra for a standard CDM model with $T/S=1$
is also shown for comparison. B polarization turns out to be notably 
larger for all global defects considered if compared to the
corresponding predictions of inflationary models on small angular scales.}}
\label{fig-polaglob}\end{figure}           

Temperature and polarization spectra for various symmetry breaking
models were calculated by Seljak, Pen \& Turok [1997] and are shown in
figure \ref{fig-polaglob}. Both electric and magnetic components of
polarization are shown for a variety of global defects. They also plot
for comparison the corresponding spectra in a typical inflationary
model, namely, the standard CDM model ($h=0.5$, $\Omega=1$,
$\Omega_{\rm baryon}=0.05$) but with equal amount of scalars and
tensors perturbations (noted $T/S=1$) which maximizes the amount of B
component from inflationary models. In all the models they assumed a
standard reionization history. The most interesting feature they found
is the large magnetic mode polarization, with a typical amplitude of
$\sim 1 \mu K$ on degree scales [exactly those scales probed by
Hedman, \etal, 2001]. For multipoles below $\ell \sim 100$ the
contributions from $E$ and $B$ are roughly equal.  This differs strongly
from the inflationary model predictions, where $B$ is much smaller than
$E$ on these scales even for the extreme case of $T/S \sim 1$.
Inflationary models only generate scalar and tensor modes, while
global defects also have a significant contribution from vector
modes. As we mentioned above, scalar modes only generate $E$, vector
modes predominantly generate $B$, while for tensor modes $E$ and $B$
are comparable with $B$ being somewhat smaller.  Together this implies
that B can be significantly larger in symmetry breaking models than in
inflationary models.

%%-------------------------------------------------------------
\subsubsection{String lensing and CMB polarization}
\label{sec-bb2000}      

Recent studies have shown that in realistic models of inflation cosmic
string formation seems quite natural in a post-inflationary preheating
phase [Tkachev \etal , 1998, Kasuya \& Kawasaki, 1998].  So, even if
the gross features on CMB maps are produced by a standard (\eg,
inflationary) mechanism, the presence of defects, most particularly
cosmic strings, could eventually leave a distinctive signature. One
such feature could be found resorting to CMB polarization: the lens
effect of a string on the small scale $E$-type polarization of the CMB
induces a significant amount of $B$-type polarization along the
line-of-sight [Zaldarriaga \& Seljak, 1998; Benabed \& Bernardeau
2000]. This is an effect analogous to the Kaiser-Stebbins effect for
temperature maps.

In the inflationary scenario, scalar density perturbations generate a 
scalar polarization pattern, given by \( E \)-type polarization, while
tensor modes have the ability to induce both \( E \) and \( B \) types
of polarization. However, tensor modes contribute little on very small
angular scales in these models. So, if one considers, say, a standard 
\( \Lambda \mathrm{CDM} \) model, only scalar primary
perturbations will be present without defects. But if a few strings
are left from a very early epoch, by studying the patch of the sky
where they are localized, a distinctive signature could come to
light. 

In the small angular scale limit, in real space and 
in terms of the Stokes parameters  \( Q \) and \( U \) 
one can express the \( E \) and \( B \) fields as follows
\begin{eqnarray}
\label{EEBB}
E\equiv \Delta ^{-1}[({\partial x}^{2}-{\partial y}^{2})\,
Q+2{\partial x} {\partial y} \, U], \\
B\equiv \Delta ^{-1}[({\partial x}^{2}-{\partial y}^{2})\,
U-2{\partial x} {\partial y} \, Q]. 
\end{eqnarray}

The polarization vector is parallel transported along the geodesics.
The lens affects the polarization by displacing the apparent position
of the polarized light source.  Hence, the observed Stokes parameters
\( \hat{Q} \) and \( \hat{U} \) are given in terms of the
\emph{primary} (unlensed) ones by:
$\hat{Q} (\vec{\alpha} )=Q(\vec{\alpha} + \vec{\xi} )
 \, {\rm ~and~ } \,
 \hat{U} (\vec{\alpha} )=U(\vec{\alpha} + \vec{\xi} )$.
The displacement \( \vec{\xi} \) is given by the integration of the
gravitational potential along the line--of--sights.
Of course, here the `potential' acting as lens is the cosmic string
whose effect on the polarization field we want to study.

In the case of a straight string which is aligned along the $y$ axis,
the deflection angle (or half of the deficit angle) is \(4\pi G\mu \)
[Vilenkin \& Shellard, 2000] and this yields a displacement 
\(\xi_{x}=\pm \xi_{0}\) with
\be
\xi_{0}=4\pi G\mu
{\mathcal{D}}_{\textrm{lss,s}}/{\mathcal{D}}_{\textrm{s,us}}
\ee
with no displacement along the $y$ axis.
${\mathcal{D}}_{\textrm{lss,s}}$ and 
${\mathcal{D}}_{\textrm{s,us}}$  
are the cosmological angular distances
between the last scattering surface and the string, and
between the string and us, respectively. They can be computed, in an
Einstein-de Sitter universe (critical density, just dust and no
$\Lambda$), from
\be
{\mathcal{D}}_{\textrm{}}(z_1 , z_2)= 
{2c\over H_0}{1\over 1+z_2}[(1+z_1)^{-1/2}-(1+z_2)^{-1/2}]
\ee
by taking $z_1 = 0$ for us and $z_2\simeq 1000$ for the last
scattering surface; see [Bartelmann \& Schneider, 2001]. For the usual
case in which the redshift of the string $z_{\rm s}$ is well below the
$z_{\textrm{lss}}$ one has
${\mathcal{D}}_{\textrm{lss,s}}/{\mathcal{D}}_{\textrm{lss,us}}\simeq
1/\sqrt{1+z_{\rm s}}$. Taking this ratio of order 1/2 (\ie, distance
from us to the last scattering surface equal to twice that from the
string to the last scattering surface) yields $z_{\rm s}\simeq
3$. Plugging in some numbers, for typical GUT strings on has $G\mu
\simeq 10^{-6}$ and so the typical expected displacement is about less
than 10 arc seconds.  Benabed \& Bernardeau [2000] compute the
resulting \( B \) component of the polarization and find that the
effect is entirely due to the discontinuity induced by the string,
being nonzero just along the string itself. This clearly limits the
observability of the effect to extremely high resolution detectors,
possibly post-Planck ones.

The situation for circular strings is different.  As shown by de Laix
\& Vachaspati [1996] the lens effect of such a string, when facing the
observer, is equivalent to the one of a static linear mass
distribution. Considering then a loop centered at the origin of the
coordinate system, the displacement field can be expressed very simply:
observing in a direction through the loop, $\vec{\xi}$ has to vanish, 
while outside of the loop the displacement decreases as \( {\alpha
_{l}}/{\alpha } \), \ie, inversely proportional to the angle.  
One then has [Benabed \& Bernardeau, 2000]
\be 
\vec{\xi}(\vec{\alpha})= -2\xi_{0}
{\alpha_{l} \over \alpha^{2}}
\vec{\alpha} \quad {\rm ~with~} \quad \alpha >\alpha_{l} ,
\ee
where $\alpha_{l}$ is the loop radius. 

\begin{figure}[t]
%\vspace{-2cm}
\begin{center}
\leavevmode
{\hbox %
{\epsfxsize = 6cm \epsffile{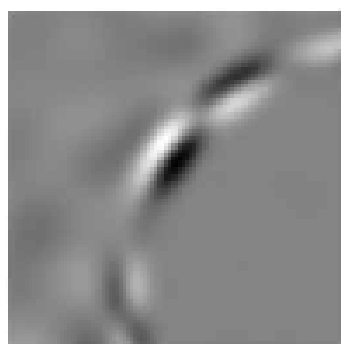} }
{\epsfxsize = 6cm \epsffile{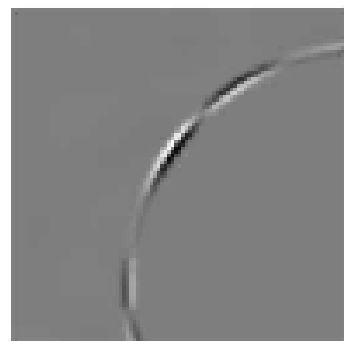} } }
%{\epsfxsize = 7cm \epsffile{DelB.corG.eps} }
%{\epsfxsize = 7cm \epsffile{DelB.corG2.eps} } }
\end{center}
%\vspace{-3cm}  
\caption{{\sl 
Simulations for the \(B\) field in the case of a circular loop.  
The angular size of the figure is \(50'\times 50'\). 
The resolution is 5' (left) and 1.2' (right).
The discontinuity in the \(B\) field is sharper the better the
resolution. Weak lensing of CMB photons passing relatively apart from
the position of the string core are apparent as faint patches outside
of the string loop on the left panel. [Benabed \& Bernardeau 2000].}}
\label{fig-bb2000}\end{figure}   

This ansatz for the displacement, once plugged into the above
equations, yields the \(B\) field shown in both panels of Figure
\ref{fig-bb2000}. A weak lensing effect is barely distinguishable
outside the string loop, while the strong lensing of those photons
traveling close enough to the string is the most clear signature,
specially for the high resolution simulation.  One can check that the
hot and cold spots along the string profile have roughly the same size
as for the polarization field in the absence of the string loop.  The
simulations performed show a clear feature in the maps, although
limited to low resolutions this can well be confused with other
secondary polarization sources. It is well known that point radio
sources and synchrotron emission from our galaxy may contribute to the
foreground [de Zotti \etal\ 1999] and are polarized at a 10 \%
level. Also lensing from large scale structure and dust could add to
the problem.

%%%%-----------------------------------------------------------
\section{Varia}
\label{sec-variouss}  

%%-------------------------------------------------------------
\subsection{Astrophysical footprints}
\label{sec-astrofoot}      

Cosmic strings, with or without current carrying capabilities, are
predicted by many theories of high energy physics, and they have been
postulated {\sl ad hoc} as a possible explanation of various
phenomena, many of which we have explained above. 
If indeed present in our universe, cosmic strings could help in the
reconciliation between theory and observations in many cases, as well
as lead to interesting and testable predictions in others.   
These areas include galactic magnetic fields, stable string loops
(vortons) as a possible dark matter candidate, gravitational waves
from strings, etc.

%%-------------------------------------------------------------
\subsubsection{Strings and galactic magnetic fields}
\label{sec-galacstrings}      

There are many outstanding astrophysical problems that may perhaps be
explained with the help of superconducting cosmic strings. One of
these concerns how galactic magnetic fields are generated. In the most
commonly held scenario, the magnetic fields possessed by galaxies
today arose from smaller seed fields that already existed before
galaxies themselves formed. These seed fields would have only a small
coherence length --the average size of a region with a roughly uniform
field-- but standard magnetohydrodynamic theory allows both the
strength of the field and its coherence length to grow to galactic
scales.

A field incorporated into a protogalactic structure remains trapped as
that structure grows; in particular, as the protogalaxy shrinks under
its own gravity, the magnetic flux within it is compressed too,
increasing the strength (flux per unit area) of the field. Rotation of
the evolving system may then increase the field strength further,
through a dynamo effect, to the value typical of galactic magnetic
fields, roughly $10^{-6}$ gauss. However, this scenario is not
universally accepted, and other models are being studied that would
produce tiny primordial fields that already have a large coherence
length.

Superconducting cosmic strings may be able to do the job. They carry
electric currents, and in fact fairly large ones. As we saw, Witten
[1985] was the first to suggest that strings could become
superconducting, and he went on to calculate a maximum current based
on the mass and charge of a string's current-carrying fermion: some
${\cal J}_{\rm max}\sim 10^{20}$A for particles on the grand unified
mass scale -- a huge value not so often met even in
astrophysics. Magnetic fields are produced when an electrically
charged object moves in space; theoretically this is precisely what
cosmic strings are and what they do. Calculations suggest that
superconducting strings could generate interesting seed magnetic
fields with strengths of about $10^{-20}$ gauss and with coherence
scales of roughly 100 kiloparsecs.  This corresponds to the size of
protogalaxies, and dynamo effects could then increase the field
strength to the observed values. The string's motion through the
turbulent primordial plasma might induce vorticity that could also
amplify field strengths. Conducting strings could thus easily provide
magnetic fields that would evolve into modern galactic fields [see,
\eg, Martins \& Shellard, 1998].

%%-------------------------------------------------------------
\subsubsection{Cosmic rays from cosmic strings}
\label{sec-crfromcs}       

A second problem is much closer to home. Earth's atmosphere is
constantly assaulted by lots of particles, such as photons, electrons,
protons and heavier nuclei. Recent detections have recorded
astonishingly energetic cosmic-ray events, with energies on the order
of a few hundred exaelectron-volts (1 EeV = $10^{18}$ electron-volts).
This is roughly the kinetic energy of a tennis ball traveling at over
150 kilometers an hour, all concentrated into an atomic
particle. Particles with such energies cannot easily move through
intergalactic space, which, far from being empty, is pervaded by
cosmic background radiation fields, including the already mentioned
microwave background (CMB) as well as diffuse radio backgrounds. From
the perspective of particles moving faster than some critical
velocity, these fields look like bunches of very damaging photons,
which degrade the particle's energy through collisions and
scattering. For example, a proton that reaches Earth's atmosphere with
the necessary energy to explain these ultra--energetic events could
not have come from farther away than about 30 million parsecs,
according to a result known as the Greisen-Zatsepin-Kuz'min (GZK)
cutoff [see, \eg, Bhattacharjee \& Sigl, 2000].

One might therefore conclude that the ultra--high--energy cosmic rays
(uhecr{\sl ons}) must come from sources that are close
(astrophysically speaking) to our galaxy. However, unusual and
energetic objects like quasars and active galactic nuclei are mostly
too far away. The high-energy particles remain a mystery because when
one looks back in the direction they came from, there is nothing
nearby that could have given them the necessary kick. So what are
they, and how did they manage to reach us?

For the time being, standard astrophysics seems unable to answer these
questions, and in fact essentially states that we should not receive
any such rays. As Ludwik Celnikier from the Observatoire de
Paris--Meudon has said, comparing cosmological dark matter to
ultra--high--energy cosmic rays: the former is a form of matter which
{\sl should} exist, but until further notice {\sl doesn't}, whereas the
high--energy rays are particles which {\sl do} exist but perhaps {\sl
shouldn't}. 

This is where topological defects, and in particular superconducting
cosmic strings, can lend a hand. They offer two ways to deliver
extremely energetic particles: they may directly emit particles with
tremendous energies, or, more excitingly, they may send off tiny loops
of superconducting cosmic string which would then be misinterpreted as
ordinary but very energetic particles.

The first mechanism arises because the currents carried by strings can
be thought of as streams of trapped particles, which would in general
be extremely massive and unstable. Like neutrons, however, which decay
in a few minutes when left by themselves but live happily inside
nuclei, these heavy particles can exist indefinitely when confined
within strings. Indeed, cosmic strings are the only objects that could
preserve such particles from their origin to the present time. The
trapped particles can nonetheless emerge occasionally when strings
suffer violent events. A single string may bend sharply to create a
kink or cusp\footnote{Movies of a cosmic string cusp simulation
can be found at {\tt http://cosmos2.phy.tufts.edu/\~{}kdo/}}, 
or a pair of strings may intersect in such a way that
their ends switch partners. In these events some trapped particles can
find their way out of the string, at which time they would almost
instantly decay. They are so massive, however, that the light
particles produced in their decay would be energetic enough to qualify
as ultra--high--energy cosmic rays.

Disintegration of superconducting strings has also been proposed as
the origin of ultra high energy cosmic rays [Hill, Schramm \& Walker,
1987; see however Gill \& Kibble, 1994], with the advantage of getting
round the difficulties of the conventional shock acceleration of
cosmic rays.  This mechanism will also produce neutrinos of up to
$10^{18}$ eV energies.  Horizontal air shower measurements, like the
Akeno Giant Air Shower Array (AGASA) experiment [Yoshida, \etal, 1995],
however, constrain $\nu_e + \bar\nu_e$ fluxes, and current estimates
from superconducting strings seem to exceed these bounds
[Blanco-Pillado, Vazquez \& Zas, 1997].

%%-------------------------------------------------------------
\subsubsection{Vortons as uhecr{\sl ons}}
\label{sec-uhecr}       

\begin{figure}[t]\begin{center}\leavevmode \epsfxsize = 10cm
\epsffile{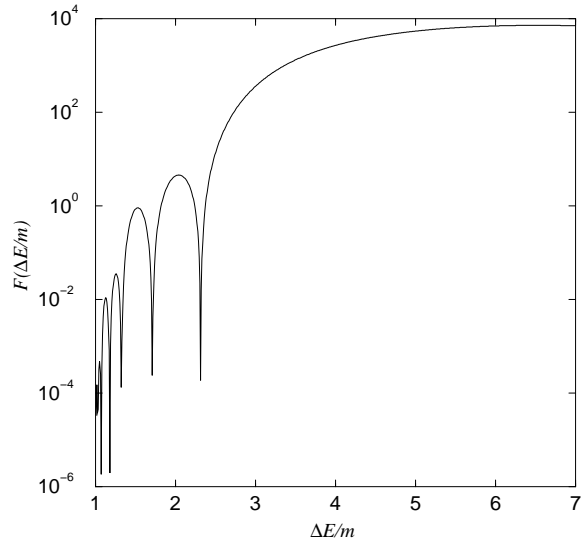} \end{center} 
%\vspace{-1cm}
\caption{{\sl Interaction of a vorton with a proton in Earth's
atmosphere varies with energy in a way that depends on the interaction
of quarks inside the proton with current-carrying particle states in
the string loop [Bonazzola \& Peter, 1997].  Ultra--high--energy cosmic
rays created in this way might have a characteristic energy spectrum
that would identify vorton collisions as their origin.}}
\label{fig-bonappp}
%\vspace{-1cm}
\end{figure}        

A second possibility was proposed by Bonazzola \& Peter [1997] who
have recently suggested that the high--energy cosmic rays are in fact
vortons. As we saw, vortons typically have more than a hundred times
the charge of an electron, $Q=Ze$, and thus they are efficiently
accelerated along electric field lines in active galactic
nuclei. Their huge mass, moreover, means that compared to protons they
need smaller velocities to attain equivalently high energies, and
these lower velocities mean they can travel enormous distances without
running up against the GZK cutoff. A vorton hitting any air molecule
in the atmosphere would decay as if it were a very energetic but
otherwise ordinary particle. The interaction of the trapped current
carriers in the vorton with the quarks within atmospheric protons
would proceed with a characteristic energy spectrum (Figure
\ref{fig-bonappp}), which would be mirrored by the energy spectrum of
observed high-energy rays. 

Other interesting possibilities in which defects play themselves the
r\^ole of high--energy cosmic rays have been proposed in the literature
in connection with gauge monopoles [\eg, Huguet \& Peter, 1999; 
Wick, Kephart, Weiler \& Biermann, 2000]. 
It is hoped that the enigma of ultra--high--energy cosmic rays will be
clarified in the near future with the data gathered in the very large
Pierre Auger Observatory\footnote{See the internet sites {\tt
http://www-lpnhep.in2p3.fr/auger} and 
{\tt http://www.fisica.unlp.edu.ar/auger/}}.
  
%%-------------------------------------------------------------
\subsection{Cosmology in the Lab}
\label{sec-cosmolab} 

As we mentioned earlier, unlike any other proposed mechanism for the
generation of observable cosmological features, topological defects
can be reproduced in the laboratory! In fact, when all relevant
lengths are uniformly scaled down, experimentalists have within reach
a manageable laboratory experiment that offers a physical equivalent
of the early universe. Some years ago, Zurek [1985] proposed testing
the Kibble mechanism using the transition that the liquefied noble gas
helium-4 makes from its normal state to the superfluid state, which
exists at temperatures lower than about 2 degrees above absolute zero
and in which fluid flow occurs without any friction.

If liquid helium were rapidly pressure--quenched around the critical
temperature, Zurek argued, the rotation of the fluid as a whole would
become trapped in a number of isolated vortices -- tiny tornadoes, in
effect. The vortices, carrying rotation in quantized amounts, would
represent defects closely analogous to cosmic strings, and studying
their formation might offer interesting hints for cosmology. Of
course, although defects in condensed matter systems are topologically
identical to those in field theory, there are also some important
differences. The dynamics of the laboratory system is nonrelativistic,
and friction is the controlling force, whereas in the cosmological
case defects can move at almost the speed of light, and gravity is
important. An additional technical difficulty is that the infinite and
homogeneous nature of the universe before a phase transition cannot be
matched by a laboratory sample of finite size.

Dealing with the superfluid transition of helium turned out to be
hard, requiring extreme laboratory conditions.  Some groups have
demonstrated vortex generation, but it remains unclear how well the
experimental findings match the Kibble-Zurek predictions. However, a
more tractable laboratory analogue has been found, in the form of
organic compounds called liquid crystals. In the second half of the
19th century chemists found several materials that behaved strangely
around their melting point. In 1850, W. Heintz reported on the
peculiarities of stearin, an organic compound used to waterproof paper
and make metal polishes and soap. Heated from about 52 to 62 degrees
Celsius, stearin first changed from a solid to a cloudy liquid, then
took on an opaque coloring, then finally became a clear
liquid. Similar behavior was later observed in other biological
materials, leading eventually to the recognition of liquid crystals as
a new form of matter -- which got their badge of honor with the award of
the 1991 Nobel Prize in Physics to Pierre-Gilles de Gennes for his
accomplishments on order phenomena in liquid-crystal systems.

Liquid crystals are organic compounds with phases intermediate to the
liquid and solid phases: They can flow like liquids while retaining
anisotropic properties of crystalline solids, meaning that their
molecular structure has a spatial alignment or orientation. They can
be imagined as crystals whose molecules are able to move around, as in
a liquid, while maintaining their relative orientation. For example,
nematic liquid crystals consist of rodlike molecules, about 20
angstroms long, which tend to maintain themselves in a parallel
alignment. Their structure endows them with useful optical
properties. Such materials are used in digital displays, where
electrical signals flip the orientation of the crystals, switching
them between opaque and reflective states.

Liquid-crystal transitions occur at temperatures ranging from 10 to
200 degrees Celsius and generate structures easily detectable with the
naked eye or with a microscope. These transitions proceed by the
formation of domains, as different regions within a crystal settle
into different alignments, so once again there is the possibility of
defect formation. Experiments have shown that networks of defects in
nematic crystals evolve in a self-similar manner, meaning that
although the characteristic scale of the pattern changes, its
maintains the same overall appearance.  As we saw in previous
sections, such behavior is needed in a cosmological context for
strings to be harmless cosmologically and, moreover, eventually useful
as progenitors of structure: self-similarity means that the defects
contribute a constant fraction of the universe's total energy density
from small to large length scales.

%FIGURE
\begin{figure}[t]\begin{center}\leavevmode \epsfxsize = 17cm
\epsffile{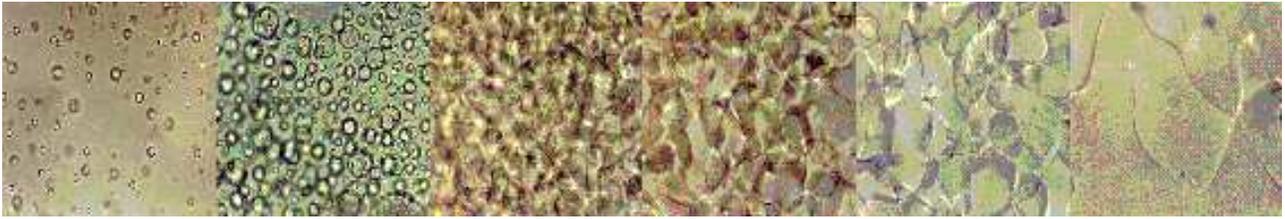} 
%\epsffile{all-Figs2bs.ps} 
%\epsffile{all-Figs2.eps} 
\end{center} 
\caption{{\sl Cosmological-defect formation can be simulated in the
laboratory by observing the transformation of liquid crystal between
phases with different optical properties. In this sequence, bubbles of
a new phase nucleate in an initially uniform liquid. As the bubbles
grow and coalesce, their boundaries develop into structures analogous
to cosmic strings. The scale of the pattern grows similarly to the way
the scale of a network of cosmic strings increases with cosmic
expansion. (Images courtesy of Ajit Srivastava, Institute of Physics,
Bhubaneswar, India.)}}\label{fig-ajit}\end{figure}        

Recently many groups have succeeded in carrying out a variation of
Zurek's original idea using the superfluid transition in another
isotope, helium-3, at temperatures close to 1 millikelvin, rather than
the higher-temperature transition in helium-4 [see, \eg, Bunkov \&
Godfrin, 2000]\footnote{See for instance the internet sites \\
{\tt
http://www-crtbt.polycnrs-gre.fr/ult/superfluid\_{}3He/topo-defects/topo\_{}eng.html}
\\ and 
{\tt http://boojum.hut.fi/research/applied/rotating3he.html}}. 
In 1996, Ruutu and collaborators in Helsinki
succeeded in heating up a volume of superfluid helium-3 with thermal
neutrons to just above the transition temperature, then cooling it
back through the superfluid transition.  They observed copious
production of quantized vortices. The precision in these experiments
is such that the number of vortex lines can be monitored, allowing
quantitative testing of defect-formation theories. Laboratory tests
using both liquid crystals and helium have provided a kind of
experimental confirmation of cosmological topological defect theory,
increasing the credibility of these ideas. 

%%-------------------------------------------------------------
\subsection{Gravitational waves from strings}
\label{sec-gwstrings}      

Next generation of gravitational waves instruments yield a good
prospect of detecting a stochastic GW background generated in the very
early universe. This opens up a brand new window, in some sense
comparable to the advent of radio--astronomy to complement the
existing (and as we know, limited) optical--astronomy, many years ago
now. In fact, if one had to limit oneself to those events accessible
through electromagnetic radiation alone, many of the most interesting
of these events would remain outside our reach. The CMB
provides a snapshot of the universe at about 400,000 years, just as
the universe became transparent to electromagnetic radiation. But what
about those processes that happened before the photon decoupling
`surface'?

\begin{figure}[t]\begin{center}\leavevmode \epsfxsize = 10cm
\epsffile{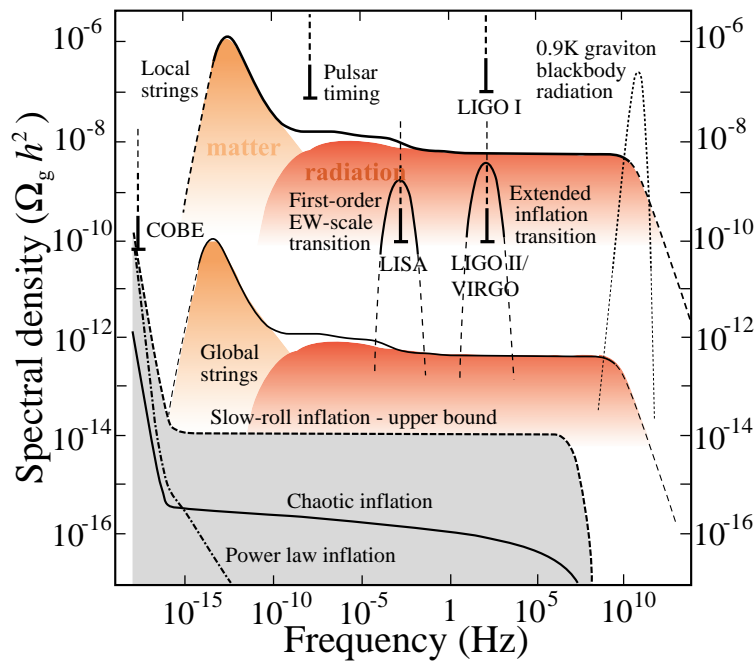} \end{center}\caption{{\sl 
A summary of the spectral density versus frequency for various
potential sources of a stochastic gravitational wave
background. Included in this busy plot are the amplitudes of GW from
different types of inflationary scenarios, from a first--order electroweak
phase transition, and from both gauge and global cosmic strings, 
also including the primordial 0.9 K blackbody spectrum of gravitons. 
[Battye \& Shellard, 1996].}}
\label{fig-bsgwstrings}\end{figure}               

Gravitational waves can penetrate through the electromagnetic surface
of last scattering thanks to the remarkable transparency of the
gravitons and their very weak interactions with ordinary matter. One
can then, by detecting this {\sc relic} background (in `upper case') get
information from the earliest possible times, namely the Planck era
$\sim 10^{-43}$ seconds after the Bang.

For radiation emitted at a time $t_{\rm e}$ before the time of 
equal matter and radiation energy densities, \ie,  
with $t_{\rm e}<t_{\rm eq}\sim 40,000$ years, and 
with a wavelength comparable to the horizon $\lambda(t_{\rm
e})\sim t_{\rm e}$, the GW frequency today is $f \sim
z_{\rm eq}^{-1}(t_{\rm eq}t_{\rm e})^{-1/2}$ where
$z_{\rm eq} \sim 2.3\times10^{4}\Omega_0 h^2$.

In experiments one measures
\[
h_{\rm c}(f)=1.3\times 10^{-20}\sqrt{\Omega_{\rm g}(f)h^2}\left(
{100{\rm Hz} \over f}\right)\,,\quad\Omega_{\rm g}(f)={f\over\rho_{\rm
c}}{\partial\rho_{\rm g} \over \partial f}
\]
with $\Omega_{\rm g}(f)$ giving the energy density in gravitational
radiation in an octave frequency bin centred on $f$, and where 
$h$ is the Hubble parameter in units of 
$100 {\textrm{~km~}}{\textrm{s}}^{-1}{\textrm{Mpc}}^{-1}$ and 
$\rho_{\rm c}$ is the critical density.

We saw above that a network of cosmic strings quickly evolved in a
self similar manner with just a few infinite string segments per
Hubble volume and Hubble time. To achieve this, the generation of
small loops and the subsequent decay of these daughter loops was
required. Both local and global oscillating cosmic string loops are then a
possible cosmological source of gravitational waves (see Figure
\ref{fig-bsgwstrings}) with local strings producing the strongest
signal, as GW emission is their main decay channel (there is also
the production of Goldstone bosons in the global case) 
[Caldwell \& Allen, 1992; Battye \& Shellard, 1996].

\begin{figure}[t]\begin{center}\leavevmode\epsfxsize = 17.5cm 
\epsffile{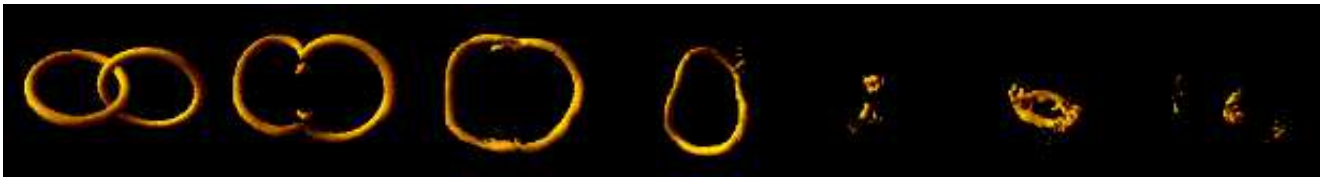} \end{center} \caption{{\sl A series of
snapshots from a two interlocked cosmic string loop decay process
[courtesy of R.~Battye and P.~Shellard].  Loops disintegrate through
the emission of (mainly gravitational) radiation.  However, if endowed
with currents, the loops may eventually reach equilibrium
configurations ({\it vortons}) which will prevent their radiative
decay.  Such a population of vortons would jeopardize the so far
successful standard model, unless it is produced at low enough
energies.}}
\label{fig-dloops}\end{figure}                    

%%-------------------------------------------------------------
\subsection{More cosmological miscellanea}
\label{sec-miscela}      

Regarding vortons, their presence and evolution was recently the
subject of much study, and grand unified models producing them were
confronted with standard cosmological tests, as the primordial
nucleosynthesis bounds and the dark matter content in the universe
today [Carter \& Davis, 2000]. Without entering into too much detail,
in order for the density of vortons at temperatures roughly around 10
MeV not to affect nucleosynthesis results for the light elements, the
maximum energy scale for current condensation cannot exceed $10^{5}$
GeV. This is a limit for approximately chiral vortons, where the
velocity of the carriers approaches that of light, and constitutes a
much more stringent bound that for nonchiral states. For this result,
the analysis demanded just that the universe be radiation dominated
during nucleosynthesis. For long--lived vortons the requirement is
stronger, in the sense that this hypothetical population of stable
defects should not overclose the present universe. Hence, present dark
matter bounds also yield bounds on vortons and these turn out to be
comparable to the nucleosynthesis ones. Although these results are
preliminary, due to the uncertainties in some of the relevant
parameters of the models, grand unified vortons seem to be in
problems. On the other hand, vortons issued from defects formed during
(or just above) the electroweak phase transition could represent today
a significant fraction of the nonbaryonic dark matter of the universe.

Fermionic zero modes may sustain vorton configurations.  In grand
unified models, like SO(10), where the symmetric phase is restored in
the interior of the string, there will be gauge bosons in the core. If
vortons diffuse after a subsequent phase transition these bosons will
be released and their out--of--equilibrium decay may lead to a baryon
asymmetry compatible with nucleosynthesis limits [Davis \& Perkins,
1997; Davis, \etal, 1997]. Another recent mechanism for the generation
of baryon asymmetry, this time at temperatures much lower than the
weak scale, takes advantage of the fact that superconducting strings
may act like baryon charge `bags', protecting it against sphaleron
effects [Brandenberger \& Riotto, 1998].

\begin{figure}[t]\begin{center}\leavevmode\epsfxsize = 3.3in
\epsffile{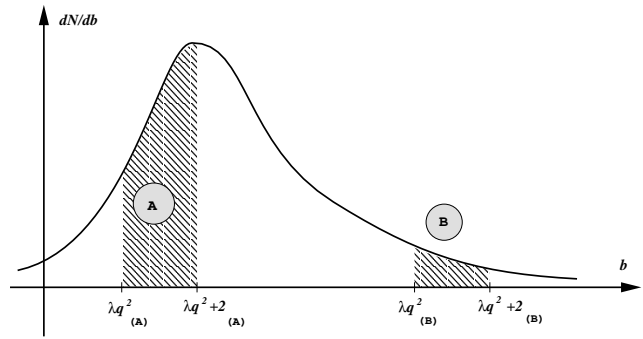} \end{center} 
\caption{{\sl A
possible way out of the vorton excess problem: a sketch of a
distribution of loops with $b\simeq N/Z$, and vorton--forming
intervals for different values of the electromagnetic correction
$\lambda q^2$ to the vorton equation of state. It is clear that the
actual number density of ensuing vortons, at most proportional to the
shaded areas, will depend quite strongly on the location of this
interval. Note also that this electromagnetic correction may reduce
drastically the available phase space for vorton formation since the
maximum of the $dN/db$ distribution is usually assumed to be peaked
around $b=1$.  [Gangui, Peter \& Boehm, 1998].}}
\label{fig-celine6}\end{figure}              

The above mentioned bounds on vortons can be considerably weakened if,
as we showed before, electromagnetic corrections to the string
equation of state are properly taken into account. In other words, a
proto--loop can become a vorton only provided certain relations
between the values of the conserved parameters characterizing the
vorton hold. We have seen that these relations (embodied in the
relevant master function of the string) change whenever
electromagnetic self couplings are considered. 
A given distribution of vortons will be characterized by the ratio of
the conserved numbers $b\simeq N/Z$.  As it turns out, increasing the
electromagnetic correction is equivalent to reducing the available
phase space for vorton formation, as $b$ of order unity is the most
natural value [see, \eg, Brandenberger, Carter, Davis \& Trodden,
1996] situation that we sketch in Figure \ref{fig-celine6}.  On this
figure, we have assumed a sharply peaked $dN/db$ distribution centered
around $b=1$; with $\lambda q^2 =0$, the available range for vorton
formation lies precisely where the distribution is maximal, whereas
for any other value, it is displaced to the right of the
distribution. Assuming a Gaussian distribution, this effect could
easily lead to a reduction of a few orders of magnitude in the
resulting vorton density, the latter being proportional to the area
below the distribution curve in the allowed interval.  This means that
as the string loops contract and loose energy in the process, they
keep their `quantum numbers' $Z$ and $N$ constant, and some sets of
such constants which, had they been decoupled from electromagnetism,
would have ended up to equilibrium vorton configurations, instead
decay into a bunch of Higgs particles, themselves unstable. This may
reduce the cosmological vorton excess problem if those are
electromagnetically charged.
          
The cosmic microwave background radiation might also be used as a
charged string loop detector. In fact vortons are like point masses
with quantized electric charge and angular momentum.  They are
peculiar for, if they are formed at the electroweak scale, their
characteristic size cannot be larger than a hundredth the classical
electron radius, while their mass would be some five orders of
magnitude heavier than the electron.  They can however contain up to
$\alpha^{-1}\sim 137$ times the electron charge, and hence Thomson
scattering between vortons and the cosmic background radiation at
recombination would be (we are admittedly optimistic in here) just nearly at
the same level as the standard one, with important consequences for,
\eg, the polarization of the relic radiation. The signature would
depend on the actual distribution of relic vortons at $z\sim 1000$, an
input that is presently largely unknown. According to current
estimates [\eg, Martins \& Shellard, 1998b], electroweak vortons could
contribute non--negligibly to the energy density. However, current
figures are still well below what is needed to get a distinguishable
signal from them and thus their CMB trace would be hidden in the `noise'
of the vastly too numerous electrons.

%$~$

%$~$

%\indent
%{\bf End of the draft.}

%%%%%-----------------------------------------------------------   
\section*{Acknowledgments}

For these lectures I've drawn freely from various sources. I thank the
people who kindly provided figures and discussions. Among them, I owe
special debts to Brandon Carter, J\'er\^ome Martin, Patrick Peter,
Levon Pogosian, and Serge Winitzki for very enjoyable recent
collaborations.  
Thanks also to the other speakers and students for the many
discussions during this very instructive study week we spent together,
and to the members of the L.O.C. for their superb job in organizing
this charming school.
A.G. thanks {\sc CONICET}, {\sc UBA} and {\sc Fundaci\'on Antorchas}
for financial support.

%%%%-----------------------------------------------------------   

%%%%%%%%%%%%%%%%%%%%%%%%%%%%%%%%%%%%%%%%%%%%%%%%%%%%%%%%%%%%%%%

\end{document}